\documentclass{jfm}

\usepackage{natbib}

\usepackage{amsmath,amsfonts,amssymb}
\usepackage{latexsym}
\usepackage{epsfig,overpic}
\usepackage{subfigure, graphicx}
\usepackage{epstopdf}
\usepackage{color}
\usepackage{ulem}
\usepackage[utf8]{inputenc}

\DeclareGraphicsExtensions{.pdf,.eps,.ps,.eps.gz,.ps.gz,.eps.Y}

\def\eps{\varepsilon}

\newcommand{\ignore}[1]{}

\numberwithin{equation}{section}

\begin{document}
\title[Dynamical contact angle hysteresis]{Effective boundary conditions for dynamic contact angle hysteresis on chemically inhomogeneous surfaces
}
\author[Z. Zhang and X. Xu]
{
  Zhen Zhang \aff{1} 
  \and \,
  Xianmin XU \aff{2}\corresp{\email{xmxu@lsec.cc.ac.cn}}
}

\affiliation{
\aff{1} Department of Mathematics,
Guangdong Provincial Key Laboratory of Computational Science and Material Design, Southern University of Science and Technology (SUSTech), Shenzhen 518055, P.R. China.
\aff{2}
	NCMIS \& LSEC, Institute of
	Computational Mathematics and Scientific/Engineering
	Computing, Academy of Mathematics and Systems
	Science, Beijing 100190, P.R. China\\
	School of Mathematical Sciences,
	University of Chinese Academy of Sciences, Beijing
	100049, P.R. China
}

\maketitle
\begin{abstract}
Recent experiments \textcolor{black}{\citep{guan2016asymmetric,guan2016simultaneous}} showed many interesting phenomena on dynamic contact angle hysteresis while there is still a lack of
complete  theoretical interpretation.
In this work, we study the time averaging of the apparent advancing and receding contact angles on surfaces with
periodic chemical patterns.
We first derive a Cox-type boundary condition for the apparent dynamic contact angle on homogeneous surfaces using Onsager variational principle.
Based on this condition, we propose a reduced model for some typical moving contact line problems on chemically inhomogeneous surfaces in two dimensions.
Multiscale expansion and averaging techniques are employed to approximate the model for asymptotically small chemical patterns.
We obtain a quantitative formula for the averaged dynamic contact angles.
It gives explicitly how the advancing and receding contact angles depend on the velocity and
the chemical inhomogeneity of the substrate.
The formula is a coarse-graining version of the Cox-type boundary condition on inhomogeneous surfaces.
Numerical simulations are presented to validate the analytical results. The numerical results also show that the formula characterizes very well
the complicated behaviour of dynamic contact angle hysteresis observed in the experiments.

\end{abstract}



\section{Introduction}

Wetting is a fundamental process in nature and our daily life (\cite{de1985wetting, bonn2009wetting}). It is of critical importance in many
industrial applications, e.g., coating, printing, chemical engineering, oil industry, etc.
The wetting property is mainly characterized by the contact angle between the liquid-vapor interface
and the solid surface. When a liquid drop is in equilibrium  on a homogeneous smooth surface,
the  contact angle is described by the famous Young's equation \citep{Young1805}. The equilibrium contact angle, also named Young's angle, depends on the surface tensions and reflects the material properties of the substrate. However, the real surface is usually neither homogeneous nor smooth
and  the chemical and geometric inhomogeneity may affect the wetting property dramatically. This
makes the wetting phenomena very complicated in real applications, especially for dynamic problems.

For a liquid drop on an inhomogeneous surface, the apparent contact angle is usually  different from
the microscopic contact angle near the contact line even in equilibrium state.
The equilibrium state of a droplet is determined by minimizing the total free energy
of the system. When a global minimum is obtained, the apparent contact angle
of a liquid can be described either by the Wenzel equation (\cite{Wenzel36}) or by the Cassie-Baxter equation (\cite{Cassie44}).
In reality, there exist many local minimizers that can not be described by the two equations (c.f. \cite{Gao07,Marmur09}).
One can observe many equilibrium contact angles in experiments.
The largest contact angle is called the advancing contact angle and
the smallest one is  the receding contact angle.
The difference betweene the advancing angle and the receding one  is usually referred to
the (static)  contact angle hyteresis (CAH).

The static contact angle hysteresis  has been studied a lot in the literature (see e.g. \cite{Johnson1964,Neumann1971,cox1983spreading,joanny1984model,Schwarts1985,Extrand02,priest2007asymmetric,Whyman08} among many others).
For a two dimensional problem, the contact line is reduced to a point.
When the surface is chemically composed of two or more materials with different Young's angles, it is found
that the advancing contact angle is equal to the largest Young's angle in the system and the receding
contact angle equals the smallest Young's angle(see \cite{Johnson1964, XuWang2011}).
For a three dimensional problem,
the situation becomes more complicated. The contact angle hysteresis
due to a single defect on a homogeneous solid surface was analysed in \cite{joanny1984model}.
The analysis can be generalized to surfaces with dilute defects.
Recently,  some modified
Wenzel and Cassie equations  are proposed to  characterize quantitatively the local equilibrium
 contact angle and the contact angle hysteresis in \cite{Choi09,Raj12,XuWang2013,xu2016modified}.
By these equations, the apparent contact angle can be computed once the position of the contact line is
 given. However, since the actual position of a contact line usually depends on the dynamic process (see \cite{iliev2018contact}),
 we need study the dynamic wetting problem for real applications.

Dynamic wetting  is much more challenging than the static case due to the motion of the contact line,
which is still an unsolved problem in fluid dynamics (see \citet{pismen2002mesoscopic,blake2006physics,snoeijer2013moving,sui2014numerical}).
The standard no-slip boundary condition may lead to a non-physical non-integrable
stress in the vicinity of the contact line~\citep{huh1971hydrodynamic,dussan1979spreading}.
To cure {this} paradox, many models were developed.
A natural way is to explicitly adopt the Navier slip boundary condition instead of the no-slip condition ~\citep{huh1971hydrodynamic,zhou1990dynamics,haley1991effect, spelt2005level,ren2007boundary} or implicitly impose the slip effect by numerical methods~\citep{renardy2001numerical,marmottant2004spray}.
Some other approaches  include: to assume a precursor thin film and a disjoint pressure~\citep{schwartz1998simulation,pismen2000disjoining,eggers2005contact};
to introduce a new thermodynamics for surfaces~\citep{shikhmurzaev1993moving}; to treat the
moving contact line as  a thermally activated process~\citep{blake2006physics,blake2011dynamics,seveno2009dynamics}, to use a
diffuse interface model for moving contact lines~\citep{seppecher1996moving,gurtin1996two,Jacqmin2000,QianWangSheng2003,yue2011can}, etc.
Most of these models can be regarded as a microscopic model for moving contact lines, due to
the existence of very small parameters, e.g. the slip length and the diffuse interface thickness, etc.
It is  very expensive to use them in quantitative numerical simulations for dynamic wetting problems,
unless the characteristic size of the problem is very small (\cite{gao2012gradient,sui2014numerical}).


To simulate the macroscopic wetting problem efficiently, various effective models have been proposed.
One important model is the so-called Cox's model  developed in  \cite{cox1986} for viscous flows.
The relation between the (macroscopic) apparent contact angle and the contact line velocity (characterized by the capillary number)
is derived by  matched expansions and asymptotic analysis.
The model was further validated and developed in \cite{sui2013validation,Sibley15}, and generalized in \cite{ren2015distinguished, zhang2019distinguished} for distinguished limits in different time regimes.
In real simulations, one can use the macroscopic model directly and there is no need to
resolve the microscopic slip region in the vicinity of the contact line.
This  improves  significantly the efficiency of the numerical methods and make it possible to
quantitatively simulate some macroscopic moving contact line problems (\cite{sui2013efficient}).

To study the dynamic contact angle hysteresis, one needs to consider
the moving contact line problems on (either geometrically or chemically) inhomogeneous surfaces.
The problem has been studied theoretically in \cite{raphael1989dynamics} and \cite{Joanny1990}
for the single defect case.
For the moving contact line problems with chemically patterned substrates, theoretical study is more difficult. Direct numerical simulations
in two dimensions have been done in \cite{QianWangSheng2008} and \cite{RenE2011}.
In these simulations, the authors adopted some standard
microscopic moving contact line models where 
 the inhomogeneity of the substrates are described explicitly in the boundary conditions.
The stick-slip behaviour of the contact lines was observed.
The  contact angle hysteresis was also observed  when the period of
the chemical pattern is small. The main challenge in direct simulations is
that the computational complexity is very large in order to resolve the microscopic
inhomogeneity.
Besides the direct simulations,  more studies 
were done by using phenomenological contact angle hysteresis models(\cite{dupont2010numerical,zhang2020level,yue2020thermodynamically}),
where the advancing and receding contact angles were given a priori and the
contact line can not move unless the dynamic contact angle was beyond the interval bounded by the
two angles.  
The advancing and receding contact angles in these models are effective parameters.
In general, it is not clear how the parameters are related to the chemical inhomogeneity of the substrates.

More recently, some experimental results on the dynamic contact angle hysteresis have been presented in \cite{guan2016asymmetric,guan2016simultaneous}.
The authors showed many interesting properties on the dynamic advancing and receding contact angles. Both
the advancing and receding contact angles can change with the increase of the contact line velocity.
The dependence of the contact angles on the velocity is quite complicated.
Sometimes it seems symmetric while it can be very asymmetric in other cases.
The asymmetric dependence is partially understood from the  distributions of the chemical
patterns(see \cite{XuZhaoWang2019,xu2020theoretical}). But there is still a lack of complete understanding of
all the experimental results.

The motivation of the work is two folds. Firstly, we would like to develop an
averaged Cox-type  boundary condition
for moving contact lines on inhomogeneous surfaces. The boundary condition characterizes quantitatively
 how the macroscopic advancing and receding angles depend on the microscopic inhomogeneity of the substrates.
With this  condition, a macroscopic model may be developed for dynamic contact angle hysteresis.
Secondly,  we would also like to have more theoretical understandings on the complicated phenomena of dynamic contact angle
hysteresis in recent experiments  in \cite{guan2016asymmetric,guan2016simultaneous}.

%

For these purposes, we conduct our study in two steps. First, we derive a simplified Cox-type
boundary condition for moving contact lines on general surfaces. The main tool here is to use
the Onsager variational principle as an approximation tool. Recent studies showed that it is very useful
for  approximately modelling many complicated problems in viscous fluids and in soft matter( c.f. \cite{Doi15,xu2016variational,DiXuDoi2016,ManDoi2016,zhou2018dynamics,guo2019onset,doi2019application,xu2020theoretical}).
In this paper, we show that the principle can be used to derive a full sharp-interface model  and a new
 simplified Cox-type boundary condition  for moving contact
lines. The  boundary condition
is a first order approximation for the original Cox's condition.
With the condition, we can construct a reduced model for some interesting moving contact line problems,
including the one for the experiments in~\cite{guan2016asymmetric,guan2016simultaneous}.
Second, we do asymptotic analysis for the reduced model on chemically inhomogeneous substrates.
By multiscale expansions,
we derive an averaged dynamics for the contact angle and the contact point. This leads to
an explicit formula for the averaged (in time) macroscopic dynamic contact angle on chemically
 inhomogeneous surfaces.
The formula is a coarse-graining boundary condition for dynamic contact angle hysteresis.
Our analysis is validated by numerical experiments. Furthermore,  numerical examples
show that the reduced model and the new boundary conditions can be used to understand the
complicated behaviours of the apparent contact angles observed in the experiments in~\cite{guan2016asymmetric,guan2016simultaneous}.
All the main phenomena can be captured by the reduced model and described by the averaged formula.

Although the averaging analysis is conducted for a reduced model in the paper,
the averaged boundary condition for dynamic contact angle hysteresis  is quite general,
since it does not depends on the specific setup of the problem at all.
The condition is a form of harmonic average of the simplified Cox-type boundary condition.
The main result can also be generalized to the case where the original Cox's boundary condition applies.
We expect that the formulae for the averaged macroscopic dynamic contact angles can be
used as an effective boundary condition for the two-phase Navier-Stokes equation
for moving contact lines on inhomogeneous surfaces.

Although the analysis in this paper is restricted to two-dimensional
problems, the main results can be used to understand some three dimensional contact angle hysteresis problems,
e.g. on an inhomogeneous surface with dilute defects. Nevertheless, quantitative descriptions for
general three dimensional problems are still challenging and will be left for future study.


The structure of the paper is as follows. In Section 2, we introduce the derivations for
a continuum model and a Cox-type boundary condition for moving contact lines in a variational approach.
A reduced model is presented for some specific problems. In Section 3, we do asymptotic analysis
for the reduced model to derive the averaged dynamics on chemically inhomogeneous surfaces. Explicit formulae for the apparent dynamic contact angles are derived. In Section 4, we
validate the asymptotic analysis numerically and do comparisons with experiments.
Finally,  in Section 5 we give some concluding remarks, especially discussions
on the generalization to three-dimensional moving contact line problems.

\section{Variational derivation for  moving contact line models}
In this section, we will derive two types of models for moving contact lines by a variational approach. The first one is a full continuum model consisting of a partial differential equation system and the boundary conditions on the microscopic contact angle. The second one is a Cox-type boundary condition, which describes the dynamics of the apparent contact angle. The Cox-type boundary condition is employed to further reduce the
model for two typical problems with moving contact lines. The reduced model
acts as a model problem to study the  dynamic contact angle hysteresis in the following sections.

\subsection{Derivation for a continuum model for moving contact lines}
\begin{figure}
\vspace{0.5cm}
\centering
	\includegraphics[width=0.4\textwidth]{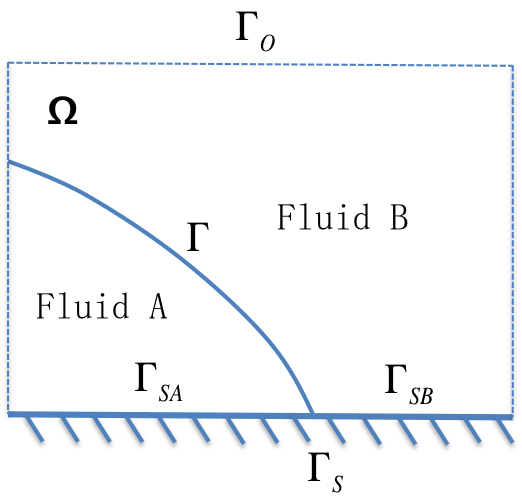}
	\caption{A domain of two-phase flow  with a moving contact line} \label{fig:region0}
\end{figure}

Consider a region $\Omega$ near the contact line as in Figure~\ref{fig:region0}. For simplicity,
we suppose $\Omega$ is a two-dimensional domain. The boundary of $\Omega$
is composed of two parts, the lower solid boundary $\Gamma_S$ and the
outer boundary $\Gamma_{O}$.
On $\Gamma_S$, there exists a moving contact line, which is an intersection
point between the solid boundary and the two-phase flow interface $\Gamma$.
Then $\Omega$ represents an open system near the contact point.
In the following, we will derive a sharp interface continuum model for moving contact lines.
We adopt the Onsager principle to derive the model. The method has
been used to develop the generalized Navier slip boundary condition for
a diffuse interface model  in \cite{QianWangSheng2006}.
In the  derivation, we ignore the gravitational force and the inertial effect, which can be added simply if needed.

To use the Onsager principle, we first compute the rate of change of the total energy in the system.
The total free energy consists of three interface energies,
\begin{equation}\label{eq:energy}
\mathcal{E}=\int_{\Gamma_{SA}}\gamma_{SA} ds + \int_{\Gamma_{SB}}\gamma_{SB} ds
+\int_{\Gamma}\gamma ds,
\end{equation}
where $\Gamma_{SA}$ and $\Gamma_{SB}$ are respectively the parts of solid boundary in contact
with the fluid $A$ and fluid $B$, $\gamma_{SA}$ and $\gamma_{SB}$ are corresponding
surface tensions, and $\gamma$ is the surface tension of the two-phase flow interface.
Direct calculations give
\begin{align}
\dot{\mathcal{E}}&=\gamma(\cos\theta_d-\cos\theta_Y) v_{ct} +\gamma \int_{\Gamma}v_n\kappa ds.
\end{align}
Here $\theta_d$ is the dynamic contact angle with respect to the fluid $A$, $\theta_Y$ is Young's angle, $v_{ct}$ is the velocity of the contact line,  $v_n=\mathbf{v}\cdot\mathbf{n}$ is the normal
velocity, and $\kappa$ is the curvature of the interface.
In the derivation, we have also used the Young's equation $\gamma\cos\theta_Y=\gamma_{SB}-\gamma_{SA}$.

The energy dissipation function, which is defined as half of the total energy dissipation rate in the system, can be written as
\begin{equation}
\Phi=\int_{\Omega_A}\frac{\mu_A}{2} |\nabla \mathbf{v}|^2 dx +\int_{\Omega_B}\frac{\mu_B}{2} |\nabla \mathbf{v}|^2 dx +
\int_{\Gamma_{SA}}\frac{\beta_A}{2} v_{\tau}^2 dx +\int_{\Gamma_{SB}}\frac{\beta_B}{2} v_{\tau}^2 dx +\frac{\xi}{2} v_{ct}^2,
\end{equation}
where $\Omega_A$ and $\Omega_B$ are the regions in $\Omega$ occupied by fluid A and fluid B, respectively,
 $\mathbf{v}$ is the corresponding velocity field, $v_{\tau}$ is
 the slip velocity on the solid boundary, $\mu_A$ and $\mu_B$ are viscosity parameters,
 $\beta_A$ and $\beta_B$ are phenomenological slip coefficients, and $\xi$ is the friction coefficient of the contact line.
 The normal velocity on the solid boundary $\Gamma_{S}$ is zero.

Since the system is an open system, we need also consider the work to the outer fluids at the boundary $\Gamma_O$.
It is given by
\begin{equation}
\dot{\mathcal{E}}^*=-\int_{\Gamma_O}\
\mathbf{F}_{ext}\cdot\mathbf{v}ds=
-\int_{\Gamma_{OA}}\mathbf{F}_{ext}\cdot\mathbf{v}_Ads-\int_{\Gamma_{OB}}\mathbf{F}_{ext}\cdot\mathbf{v}_Bds,
\end{equation}
where $\Gamma_{OA}$ and $\Gamma_{OB}$ are the respective open boundary in contact with fluid A and fluid B.

With the above definitions, the Onsager principle states that \citep{DoiSoftMatter}
the dynamical equation of the system is given by minimizing the  Onsager-Machlup action defined as follows
\begin{equation}
\mathcal{O}=\dot{\mathcal{E}}+\dot{\mathcal{E}}^*+\Phi,
\end{equation}
under the constraint of incompressibility condition
\begin{equation}
\nabla\cdot \mathbf{v}=0.
\end{equation}

Introduce a Lagrangian multiplier $p(x)$  in $\Omega$.
We minimize the following modified functional  with respect to the velocity
\begin{equation}\label{e:ModRay}
\mathcal{R}_{\lambda}=\dot{\mathcal{E}}+\dot{\mathcal{E}}^*+\Phi-\int_{\Omega_A\cup\Omega_B} p \nabla \cdot \mathbf{v} dx.
\end{equation}
Direct calculation for the first variation of $\mathcal{R}_{\lambda}$ gives
\begin{align*}
\delta \mathcal{R}_{\lambda}=&\int_{\Gamma}\gamma\kappa \delta v_n ds +\gamma(\cos\theta_a-\cos\theta_Y) \delta v_{ct}
-\int_{\Gamma_O} \mathbf{F}_{ext}\cdot\delta \mathbf{v} ds +\int_{\Omega_A} \mu_A\nabla \mathbf{v}\cdot \nabla \delta \mathbf{v}dx\\
&+\int_{\Omega_B} \mu_B\nabla \mathbf{v}\cdot \nabla \delta \mathbf{v}dx+\int_{\Gamma_{SA}}\beta_A v_{\tau}\delta v_{\tau} ds
+\int_{\Gamma_{SB}}\beta_B v_{\tau}\delta v_{\tau} ds+ \xi v_{ct}\delta v_{ct} \\
&-\int_{\Omega_A\cup\Omega_B} p \nabla \cdot  \delta \mathbf{v} dx.
\end{align*}
Integration by parts gives
\begin{align*}
&\int_{\Omega_A} \mu_A\nabla \mathbf{v} \cdot \nabla \delta \mathbf{v} dx=
\int_{\partial \Omega_A} \mu_A (\mathbf{n}\cdot \nabla) \mathbf{v} \cdot  \delta \mathbf{v} dx-\int_{\Omega_A}\mu_A\Delta \mathbf{v} \cdot \delta \mathbf{v},\\
&-\int_{\Omega_A} p \nabla \cdot  \delta \mathbf{v} dx=-\int_{\partial \Omega_A} p \mathbf{n} \cdot  \delta \mathbf{v} dx+\int_{\Omega_A} \nabla p \cdot  \delta \mathbf{v} dx.
\end{align*}
Similar calculations can be done in $\Omega_B$.
Combing all this calculations, 
 we immediately derive the corresponding Euler-Lagrange equation for \eqref{e:ModRay},
 \begin{equation}\label{e:Stokes}
\left\{
\begin{array}{ll}
-\mu(x) \Delta \mathbf{v} +\nabla p=0&\\
\nabla \cdot\mathbf{v}=0&
\end{array}
\right.
\qquad\qquad \hbox{in } \Omega,
 \end{equation}
coupled with the interface condition on $\Gamma$
 \begin{equation}
[\mathbf{v}]=0,\quad 
\big[\mu(\mathbf{n}\cdot\nabla)\mathbf{v}-p \mathbf{n}\big]=\gamma\kappa \mathbf{n}
\qquad\qquad \hbox{on } \Gamma,
 \end{equation}
  the Navier slip boundary condition on $\Gamma_S$
 \begin{equation}\label{e:Navier}
v_n=0,\quad 
\beta(x) v_{\tau}=-\mu \frac{\partial v_{\tau} }{\partial n}
\qquad\qquad \hbox{on } \Gamma_S
 \end{equation}
 and the condition for the moving contact line
 \begin{equation}\label{e:BndRenE}
  \xi v_{ct}=-\gamma(\cos\theta_d-\cos\theta_Y).
 \end{equation}
 Here $$\mu(x)=\left\{\begin{array}{ll}
\mu_A, & \hbox{if } x\in\Omega_A;\\
\mu_B, & \hbox{if } x\in\Omega_B;\\
 \end{array}\right.
 \quad
 \hbox{and}
 \quad
 \beta(x)=\left\{\begin{array}{ll}
\beta_A, & \hbox{if } x\in\Gamma_{SA};\\
\beta_B, & \hbox{if } x\in\Gamma_{SB}.\\
 \end{array}\right.
 $$
On the outer boundary $\Gamma_O$, we also have a relation for the external force that
 $$\mathbf{F}_{ext}=\mu(\mathbf{n}\cdot\nabla)\mathbf{v}-p \mathbf{n}.$$

 The  boundary condition \eqref{e:BndRenE} for moving contact lines are
 exactly the  model derived in \cite{ren2007boundary}.
 Specifically, when $\xi=0$, the condition is reduced to
 \begin{equation}
\theta_d=\theta_Y.
 \end{equation}
This implies that the microscopic dynamic contact angle is equal to the Young's angle.
Together with the Navier-Slip boundary condition~\eqref{e:Navier}, this is the model
 used in the asymptotic analysis in  \cite{cox1986}.
%
Therefore, the above analysis indicates that some well-known continuum models for moving contact lines can be derived in a variational
framework of the Onsager principle.
In the following, we will use the variational principle to derive a reduced model for the apparent
dynamic contact angle.

\subsection{Derivation of a Cox type model for the apparent contact angle}
\begin{figure}
\vspace{0.5cm}
\centering
	\includegraphics[width=0.8\textwidth]{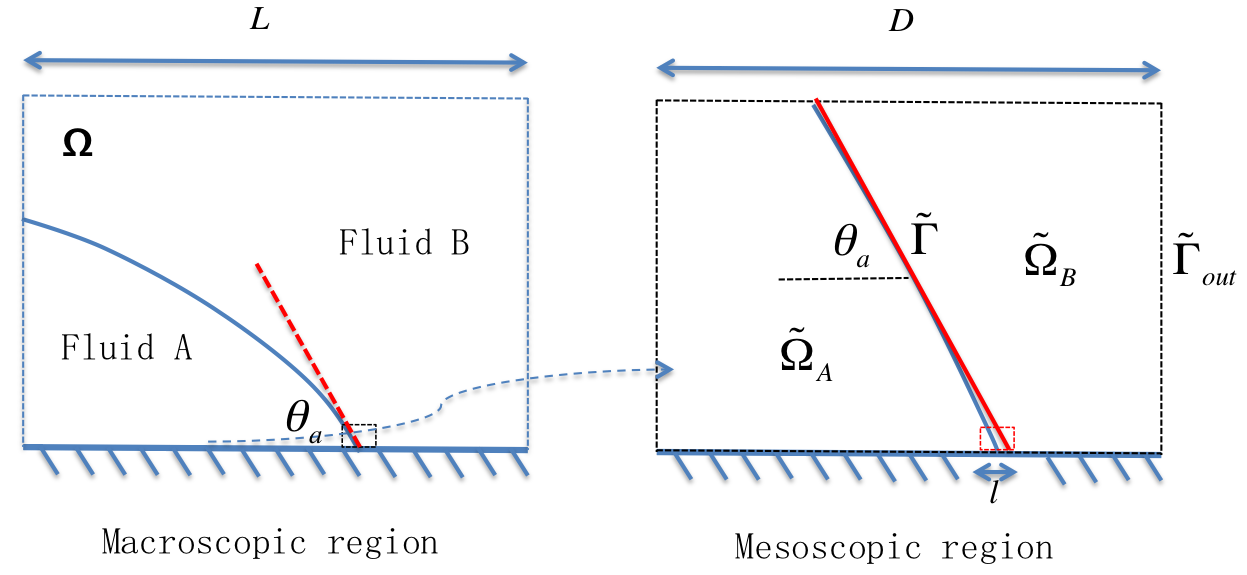}
	\caption{The region near the moving contact point in different scalings($l\ll D\ll L$)} \label{fig:region}
\end{figure}

It is known that the two-phase flow has multiscale property near  moving contact lines.
In general the macroscopic contact angle is different from the microscopic one;
see for example~\cite{cox1986}, where
a dynamic boundary condition for the apparent contact angle was derived by
matched asymptotic analysis.
In the following, we  derive a similar boundary condition by using the Onsager principle
as an approximation tool. The derivation is much simpler but still captures the essential
dynamics of the apparent contact angle.

As shown in Figure~\ref{fig:region}, we separate the domain near the
contact line in three different scales.
The macroscopic region $\Omega$ has a characteristic length $L$.
The microscopic region is of molecular scale with a characteristic length $l$.
In the microscopic scale, the interaction between the liquid molecules and the solid molecules
induces a friction of the contact line \citep{guo2013direct}.
The mesoscopic region has a characteristic length $D\ll L$, but still much larger
than the molecular scale $l$. In this region, the no-slip boundary condition is still a good approximation. 
The  contact angle $\theta_a$ represents the apparent  angle
in macroscopic point of view.

We are interested in the contact line dynamics in the mesoscopic region. For this purpose, we analyze the system by using the Onsager principle as
an approximation tool. We make the ansatz that the interface between the two fluids in this region can
be approximated well by a straight line $\tilde{\Gamma}$ which has a tilting angle equal to
the macroscopic contact angle $\theta_a$. With this assumption, we can use the Onsager principle to derive  a relation between
the apparent contact angle $\theta_a$ and the contact line motion as follows.

We first calculate the rate of change of the total free energy in the system.
Similar to \eqref{eq:energy}, the total approximate free energy $\tilde{\mathcal{E}}$ is composed of
 three surface energies.
The changing rate of the total interface energies with respect to the motion of the contact line is given by
\begin{equation}\label{e:energyReduced}
\dot{\tilde{\mathcal{E}}}=\gamma(\cos\theta_a-\cos\theta_Y)v_{ct} + \int_{\tilde{\Gamma}} \gamma\kappa v_n ds= \gamma(\cos\theta_a-\cos\theta_Y)v_{ct}.
\end{equation}
Here the second term disappears since the curvature is zero for a straight line.
To use the Onsager principle for the open system, we need consider the work  to the exterior region on the outer boundary,
$ \dot{\tilde{\mathcal{E}}}^*=-\int_{\tilde{\Gamma}_{out}} \mathbf{F}_{ext}\cdot \mathbf{v} $.
This is a higher order term in comparison with $\dot{\tilde{\mathcal{E}}}$, since it is of order  $|\mathbf{F}_{ext}| v_{ct} D$
with $D\ll L=O(1)$.
We will ignore this term in the following calculations.

We now compute the  energy dissipations in the wedge region as shown in Figure \ref{fig:region} (right plot).
The total energy dissipations is calculated approximately as
\begin{equation}
\tilde{\Psi}={\xi}v_{ct}^2+\int_{\tilde\Omega_A}\mu_A |\nabla\mathbf{v}|^2 dx+\int_{\tilde\Omega_B}\mu_B |\nabla\mathbf{v}|^2 dx,
\end{equation}
In the above formula, $\tilde{\Omega}_A$ and $\tilde{\Omega}_B$ are the  domains corresponding
to liquids $A$ and $B$ in the mesoscopic region.
The contact line friction term ${\xi}v_{ct}^2$
is due to the dissipations in the microscopic region.
The velocity  $\mathbf{v}$  can be obtained by solving
the Stokes equations in  wedge regions assuming the interface moving in
a steady state.
By careful calculations (see Appendix A),
we can compute the energy dissipation function approximately
\begin{equation}\label{eq:Reduced_dissip}
\tilde{\Phi}=\frac{1}{2}\tilde{\Psi}\approx \frac{1}{2}{\xi}v_{ct}^2+\frac{\mu_A |\ln\zeta|}{2\mathcal{F}(\theta_a,\lambda)}v_{ct}^2\approx \frac{1}{2}\frac{\mu_A |\ln\zeta|}{\mathcal{F}(\theta_a,\lambda)} v_{ct}^2,
\end{equation}
where the dimensionless parameter $\zeta=D/l$ is the ratio between the mesoscopic
size and the microscopic size, and
\begin{align}
&\mathcal{F}(\theta_a,\lambda)\nonumber\\
=&\frac{\lambda(\theta_a^2-\sin^2\theta_a)(\pi-\theta_a+\sin\theta_a\cos\theta_a)+((\pi-\theta_a)^2-\sin^2\theta_a)(\theta_a-\sin\theta_a\cos\theta_a)}{2\sin^2\theta_a\Big(\lambda^2(\theta_a^2-\sin^2\theta_a)
+2\lambda(\sin^2\theta_a+\theta_a(\pi-\theta_a))+((\pi-\theta_a)^2-\sin^2\theta_a)\Big) },\label{eq:F}
\end{align}
with $\lambda=\frac{\mu_B}{\mu_A}$.
In the last approximation in~\eqref{eq:Reduced_dissip}, we have assumed that $\xi$ is much smaller
than the viscous friction coefficient ${\mu_A |\ln\zeta|}/{\mathcal{F}(\theta_a,\lambda)}$.
Actually, the friction coefficient $\xi$ is measured directly in experiments in \cite{guo2013direct},
which is given by $\xi \approx (0.8\pm 0.2)\mu$. Notice that $\ln\zeta$ is generally
chosen as a constant of order $10$ (In~\cite{Gennes03}, it is chosen as $13.6$).
Direct computations also show that $1/\mathcal{F}(\theta_a,\lambda)=O(1)$.
Therefore, $\xi$ can be regarded a small perturbation to the viscous friction coefficient and
 can be ignored.

By using the Onsager principle, we minimize the Rayleighian $\mathcal{R}=\tilde{\Phi}+ \dot{\tilde{\mathcal{E}}}$ with resepct to $v_{ct}$.
This leads to an equation for $v_{ct}$
\begin{equation}\label{e:FU}
\frac{\mu_A |\ln\zeta|}{\mathcal{F}(\theta_a,\lambda)}v_{ct}=-\gamma(\cos\theta_a-\cos\theta_Y).
\end{equation}
This equation basically means that the local dissipative force (due to the viscous dissipation) is balanced by
the effective unbalanced Young force, 
which is the dominant term in the driving force in the mesoscopic region. It gives a relation between the contact line velocity and the apparent contact angle  $\theta_a$.
The relation \eqref{e:FU} can be rewritten in a dimensionless form
%
\begin{equation}\label{e:bndcond}
 |\ln\zeta|Ca=-\mathcal{F}(\theta_a,\lambda)(\cos\theta_a-\cos\theta_Y),
\end{equation}
where $Ca={\mu_A v_{ct}}/{\gamma}$ is the capillary number.
This is a Cox-type formula for the apparent contact angle and the capillary number of the contact line.

\begin{figure}
\vspace{0.5cm}
\centering
	\includegraphics[width=0.5\textwidth]{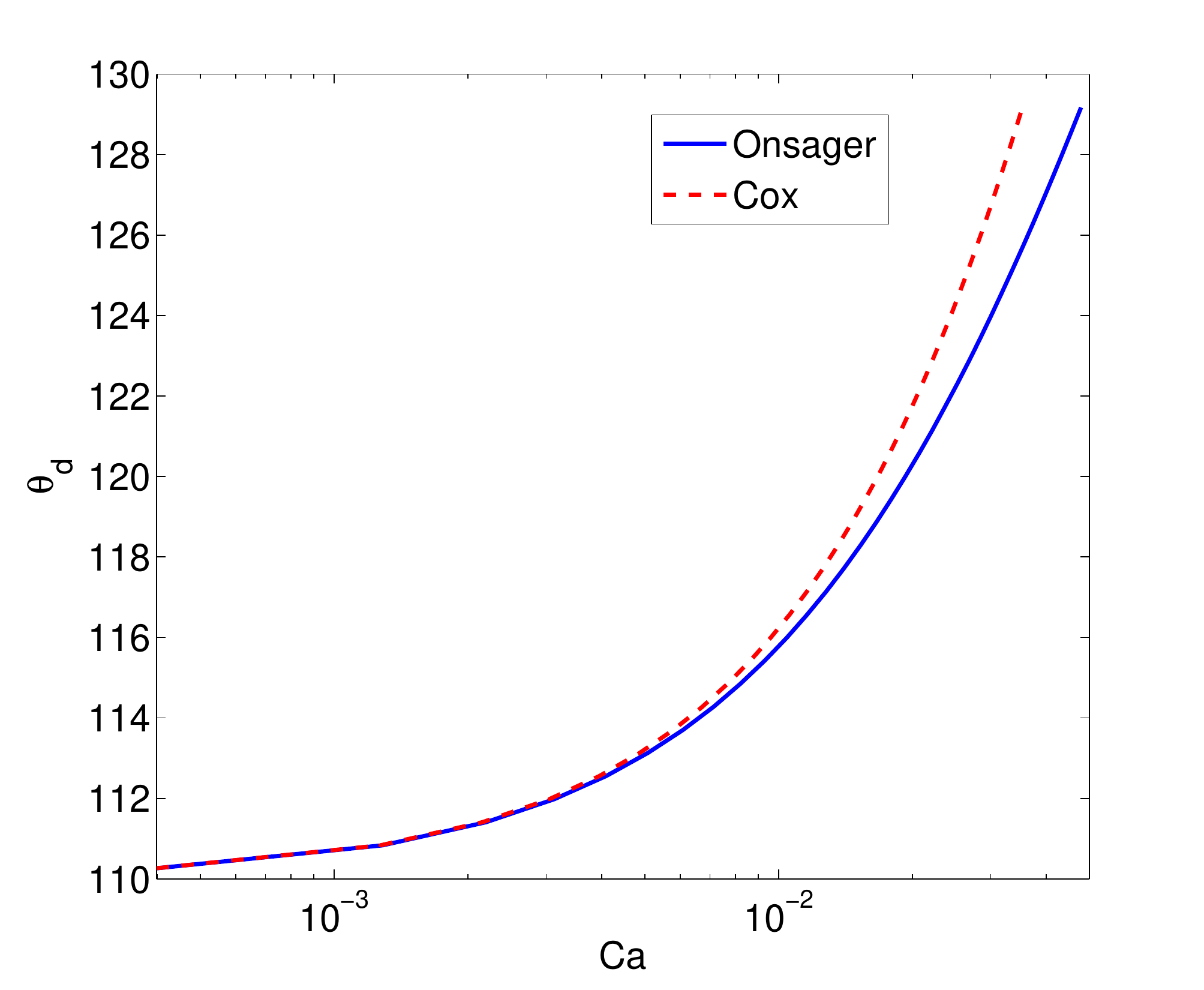}
	\caption{Comparison between  Cox's formula and the equation~\eqref{e:bndcond}. Here we choose $\lambda=0$ and
	$\theta_Y=110^o$. Their difference is small for small capillary number ($Ca\sim O(10^{-2})$).} \label{fig:CompCox}
\end{figure}

This equation~\eqref{e:bndcond} is consistent with Cox's formula in leading order when $Ca$ is small.
This will be discussed as follows.
We recall the Cox's formula  for small $Ca$,
\begin{equation}\label{e:coxformula}
|\ln\zeta|Ca=\mathcal{K}(\theta_a,\lambda)-\mathcal{K}(\theta_Y,\lambda).
\end{equation}
where
\begin{align*}
&\mathcal{K}(\theta,\lambda)\\
&=\int_{0}^{\theta}\frac{\lambda(\beta^2-\sin^2 \beta)(\pi-\beta+\sin \beta \cos \beta)+((\pi-\beta)^2-\sin^2 \beta)(\beta-\sin \beta \cos \beta)}{2\sin \beta \Big( \lambda^2(\beta^2-\sin^2 \beta)
+2\lambda(\sin^2 \beta+\beta(\pi-\beta)+(\pi-\beta)^2-\sin^2 \beta)\Big)}\,\mathrm{d}\beta,\\
&=\int_0^\theta \mathcal{F}(\beta,\lambda)\sin\beta\,\mathrm{d}\beta.
\end{align*}
A first order approximation of Cox's formula leads to
\begin{align*}
 |\ln\zeta|Ca=&\int_{\theta_Y}^{\theta_a}\mathcal{F}(\beta,\lambda)\sin\beta\mathrm{d}\beta\\
\approx &-\mathcal{F}(\theta_a,\lambda )(\cos\theta_a-\cos\theta_Y)+O((\theta_a-\theta_Y)^2).
\end{align*}
This implies that Cox's formula is consistent with our analysis by using the Onsager principle in leading order.
The difference between the two formula is illustrated in Figure~\ref{fig:CompCox}.
We can see that their difference is small for small capillary number case (e.g. $Ca\leq 0.01$).
When $Ca$ is large, the linear approximation of the interface in the mesoscopic region
is not that accurate any more. Then there is a larger difference between the equation~\eqref{e:bndcond} and Cox's formula.

The equation~\eqref{e:bndcond} can  be regarded as a coarse-graining boundary condition for the two-phase Navier-Stokes equation in the
macroscopic region $\Omega$. It gives a relation between the apparent contact angle and
the contact line velocity.
The results are similar to that in \cite{Gennes03}.
One can use~\eqref{e:bndcond} instead of the equation~\eqref{e:BndRenE} to avoid resolving
the mesoscopic region by very fine meshes in numerical simulations.

\subsection{Model problems}\label{sec:examples}
In some problems with small characteristic size, the energy dissipations in the mesoscopic region  may dominate.
Then one can derive a reduced model for such problems. One typical example is
the spreading of a small droplet on hydrophilic surfaces which gives the so-called Tanner's law~\citep{tanner1979spreading}.
Other examples can be found in \cite{Gennes03,chan2020directional,xu2020theoretical}.
In the following, we will introduce two typical examples. Then we show that they can
be described by a unified reduced model, which will be used to study the
dynamic contact angle hysteresis in the next section.


{\it Example 1.  A moving contact line problem in a micro-channel.}
\begin{figure}
\vspace{0.5cm}
\centering
	\includegraphics[width=0.5\textwidth]{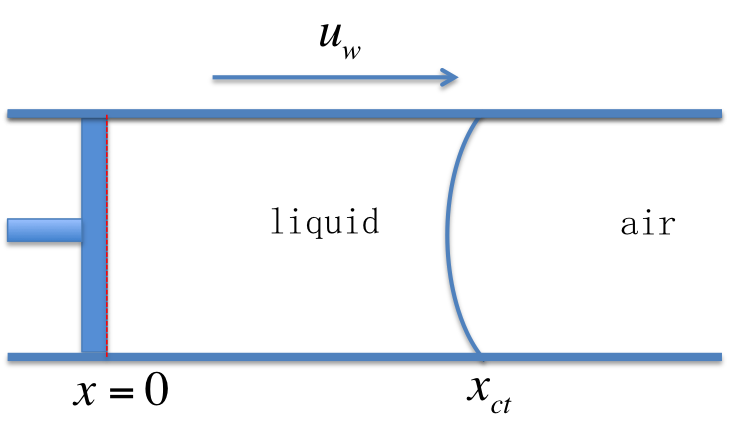}
	\caption{Contact point motion in a micro-channel} \label{fig:tube}
\end{figure}
In the example, we consider a contact line problem in a microfluidics, which is similar to that considered in \cite{joanny1984model}.
As shown in Figure~\ref{fig:tube},
suppose that the two walls of a channel are moving in a velocity $u_{wall}$. There is a bar on the left side of the channel
to keep the liquid from moving out. Suppose that the height $h_0$ of the channel is smaller
than the capillary length. Then we could assume that the interface keeps almost circular
if the velocity  $u_{wall}$ is small. Then the position of the contact point is fully determined
by the dynamic contact angle since the volume of the liquid is conserved.

Suppose the volume of the liquid is $V_0$.  
We suppose the left boundary of the liquid domain is $x=0$ and
 the $x$-coordinate of the contact point is $x_{ct}$.
 Suppose the apparent contact angle is $\theta_a$.
Then the volume of liquid is calculated by
\begin{align*}
V_0&=h_0x_{ct}+\frac{h_0^2}{4}\frac{\theta_a-{\pi}/{2}-\sin(\theta_a-{\pi}/{2})\cos(\theta_a-{\pi}/{2})  }{\sin^2(\theta_a-\pi/2)}\\
&=h_0x_{ct}+\frac{h_0^2}{8}\frac{(2\theta_a-{\pi}+\sin(2\theta_a)  )}{\cos^2\theta_a}.
\end{align*}
This equation gives a relation between ${x}_{ct}$ and the apparent contact angle $\theta_a$.
We do time derivative for the above equation and notice that $\dot{V}_0=0$. Then we have
\begin{equation}\label{eq:kinematic1}
\dot{x}_{ct}=-\frac{h_0}{2}\frac{\cos\theta_a+(\theta_a-\frac{\pi}{2})\sin\theta_a}{\cos^3\theta_a}\dot{\theta}_a.
\end{equation}
On the other hand, since the relative velocity of the contact line with respect to the two walls is  $\dot{x}_{ct} -u_{wall}$,
the equation~\eqref{e:bndcond}  implies
\begin{equation}\label{eq:bndEx1}
\frac{|\ln\zeta|\mu}{\gamma}(\dot{x}_{ct}-u_{wall}) =- \mathcal{F}_1(\theta_a)(\cos\theta_a-\cos\theta_Y),
\end{equation}
where
\begin{equation}\label{eq:F1}
\mathcal{F}_1(\theta_a)=\mathcal{F}(\theta_a,0)=\frac{(\theta_a-\sin\theta_a\cos\theta_a)}{2\sin^2\theta_a}.
\end{equation}

The two equations~\eqref{eq:kinematic1} and \eqref{eq:bndEx1} compose  a complete system of ordinary differential equations (ODEs)  for
 the apparent contact angle and the contact point position.
Denote
$$\mathcal{G}_1(\theta)=-\frac{\cos^3\theta_a}{\cos\theta_a+(\theta_a-\frac{\pi}{2})\sin\theta_a},$$
then the ODE system is written as
\begin{equation}\label{eq:ODEsysEx1}
\left\{\begin{array}{l}
\dot{\theta}_a=\frac{2\mathcal{G}_1(\theta)}{h_0}\Big(-\frac{\gamma}{|\ln\zeta|\mu}\mathcal{F}_1(\theta_a)(\cos\theta_a-\cos\theta_Y)+u_{wall}\Big),
\\
\dot{x}_{ct}= -\frac{\gamma}{|\ln\zeta|\mu}\mathcal{F}_1(\theta_a)(\cos\theta_a-\cos\theta_Y)+u_{wall}.
\end{array}
\right.
\end{equation}

{\it  Example 2.  A capillary problem along a moving thin fiber.}
\begin{figure}
\vspace{0.5cm}
\centering
	\includegraphics[width=0.75\textwidth]{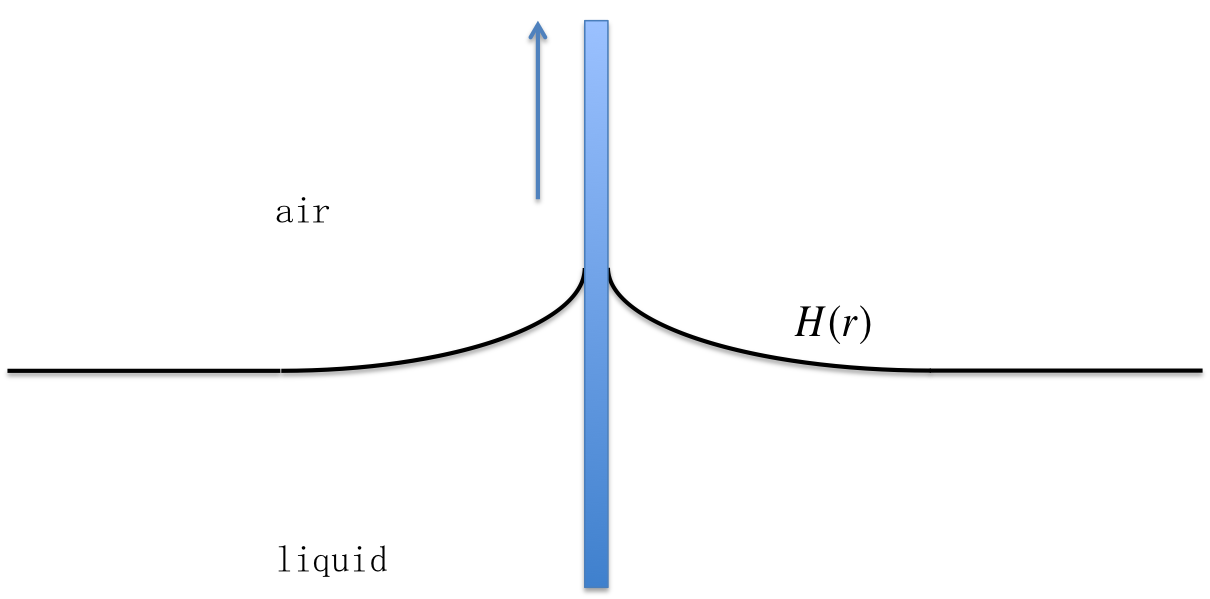}
	\caption{Contact line motion on a moving fiber} \label{fig:fiber}
\end{figure}
Motivated by the recent experiments in \cite{guan2016asymmetric}, we consider a capillary problem along a moving fiber.
As shown in Figure~\ref{fig:fiber}, we suppose a fiber is inserted in a liquid. When the fiber moves up and down,
the contact line will recede and advance accordingly. We assume that the radius $r_0$ of the fiber is much smaller than the
capillary length $l_c$. By the Young-Laplace equation, the radial symmetric liquid-vapor interface
can be described by the following equation (see \cite{Gennes03})
\begin{equation}
x=H(r)=h-r_0\cos\theta_a\ln\frac{r+\sqrt{r^2-r_0^2\cos^2\theta_a}}{r_0\cos\theta_a},\qquad r\geq r_0.
\end{equation}
Here we assume the upper direction is the positive $x$-direction.
There are two parameters $h$ and $\theta_a$ in
this equation. 
By the definition of the capillary length, we can assume that
 the interface intersects with the horizontal surface $x=0$ at $r=l_c$.
Then we have
\begin{equation}
H(l_c)=0.
\end{equation}
This gives a relation between $\theta_a$ and $h$, which is
\begin{equation}
h=r_0\cos\theta_a\ln\frac{l_c+\sqrt{l_c^2-r_0^2\cos^2\theta_a}}{r_0\cos\theta_a}.
\end{equation}

Notice that the $x$-coordinate of the contact line is given by
$$x_{ct}:=H(r_0)=h-r_0\cos\theta_a\ln\frac{1+\sin\theta_a}{\cos\theta_a}=r_0\cos\theta_a\ln\frac{l_c+\sqrt{l_c^2-r_0^2\cos^2\theta_a}}{r_0(1+\sin\theta_a)}.$$
Direct calculation gives
\begin{equation}\label{eq:kinematic2}
\dot{x}_{ct}=r_0 \mathcal{G}^{-1}_2(\theta_a)\dot{\theta}_a,
\end{equation}
where
$$
 \mathcal{G}^{-1}_2(\theta_a)=r_0^{-1}\frac{\partial x_{ct}}{\partial\theta_a}\approx -\Big(
 \sin\theta_a\ln\frac{2l_c}{r_0(1+\cos\theta_a)}+1-\sin\theta_a\Big).
$$
The equation~\eqref{eq:kinematic2} gives a relation between the time derivative of the contact line position and that of
the dynamic contact angle.

In this example, since the relative velocity of the contact line with respect to the fiber surface is $\dot{x}_{ct}-u_{wall}$,
 the equation~\eqref{e:bndcond} turns to be
\begin{equation}
|\ln\zeta|\frac{\mu}{\gamma}(\dot{x}_{ct}-u_{wall})=-\mathcal{F}_1(\theta_a)(\cos\theta_a-\cos\theta_Y).
\end{equation}
The two equations compose a complete system for the capillary rising problem. They can be rewritten as
\begin{equation}\label{eq:ODEsysEx2}
\left\{\begin{array}{l}
\dot{\theta}_a=\frac{\mathcal{G}_2(\theta_a)}{r_0}\Big(-\frac{\gamma}{|\ln\zeta|\mu}\mathcal{F}_1(\theta_a)(\cos\theta_a-\cos\theta_Y)+u_{wall}\Big),
\\
\dot{x}_{ct}=- \frac{\gamma}{|\ln\zeta|\mu}\mathcal{F}_1(\theta_a)(\cos\theta_a-\cos\theta_Y)+u_{wall}.
\end{array}
\right.
\end{equation}
This is the formula given in~\cite{xu2020theoretical}.

We can see that the ordinary differential systems~\eqref{eq:ODEsysEx1} and \eqref{eq:ODEsysEx2} in the two examples have
the same structure. They can be written as a unified form as follows,
\begin{equation}\label{eq:unify}
\left\{\begin{array}{l}
\dot{\theta}_a=\frac{\mathcal{G}(\theta_a)}{l_0}\Big(-\frac{\gamma}{|\ln\zeta|\mu}\mathcal{F}_1(\theta_a)(\cos\theta_a-\cos\theta_Y)+u_{wall}\Big),
\\
\dot{x}_{ct}=- \frac{\gamma}{|\ln\zeta|\mu}\mathcal{F}_1(\theta_a)(\cos\theta_a-\cos\theta_Y)+u_{wall}.
\end{array}
\right.
\end{equation}
The second equation of \eqref{eq:unify} is due to the condition~\eqref{e:bndcond}
where $\mathcal{F}_1(\theta_a)$ is given  in \eqref{eq:F1}.
 Since $\mathcal{F}_1(\theta_a)=\mathcal{F}(\theta_a,0)$, it corresponds to
 the case when $\frac{\mu_B}{\mu_A}=0$ (see in \eqref{eq:F}), i.e. a liquid-vapor system.
 In the first equation of \eqref{eq:unify}, $l_0$ is a characteristic length and
 the function $\mathcal{G}(\theta_a)$ comes from the geometric setup of a problem.
Both $l_0$ and   $\mathcal{G}(\theta_a)$ may change for different problems.
Hereinafter, we will use the equation \eqref{eq:unify} as a model problem to
study the contact angle hysteresis for a liquid-vapor system on chemically patterned surface.

\section{Averaging for dynamic contact angles on inhomogeneous surfaces}
In this section, we consider the case when the solid surface is chemically inhomogeneous. This implies that
the Young's angle $\theta_Y$ is not a constant but a function of the position on the solid substrate. For example, we can assume
that $\theta_Y=\hat{\theta}_Y(x)$, where $x$ is the  position on the solid surface. Then the system \eqref{eq:unify}
can be rewritten as
\begin{equation}\label{eq:unifyPattern}
\left\{\begin{array}{l}
\dot{\theta}_a=\frac{\mathcal{G}(\theta_a)}{l_0}\Big(-\frac{\gamma}{|\ln\zeta|\mu}\mathcal{F}_1(\theta_a)(\cos\theta_a-\cos\hat{\theta}_Y(\hat{x}_{ct}))
+u_{wall}\Big),
\\
\dot{\hat{x}}_{ct}=- \frac{\gamma}{|\ln\zeta|\mu}\mathcal{F}_1(\theta_a)(\cos\theta_a-\cos\hat{\theta}_Y(\hat{x}_{ct})),
\end{array}
\right.
\end{equation}
where $\hat{x}_{ct}$ is the actual position of the contact line on the solid surface
satisfying $\dot{\hat{x}}_{ct}=\dot{x}_{ct}-u_{wall}$.

The system~\eqref{eq:unifyPattern} can be made dimensionless using the Capillary length $l_c$ and the characteristic time scale $l_c/U^*$, where $U^*=\frac{\gamma}{\mu}$.
Using change of variable $ \frac{\hat{x}_{ct}}{l_c}\rightarrow\hat{x}_{ct}$ and $\frac{t}{l_c/U^*}\rightarrow t$
(we still use $\hat{x}_{ct}$ and $t$ for the dimensionless variables for simplicity in notations),
the dimensionless system for \eqref{eq:unifyPattern} is given by
\begin{equation}\label{eq:general}
\left\{\begin{array}{l}
\dot{\theta}_a=g(\theta_a)\Big(f(\theta_a)(\cos\hat{\theta}_Y(\hat{x}_{ct})-\cos\theta_a)+v\Big),
\\
\dot{\hat{x}}_{ct}= f(\theta_a)(\cos\hat{\theta}_Y(\hat{x}_{ct})-\cos\theta_a).
\end{array}
\right.
\end{equation}
where we have denoted
  $f(\theta)=\frac{\mathcal{F}_1(\theta)}{|\ln\zeta|}$,
$g(\theta)=\frac{l_c\mathcal{G}(\theta)}{l_0}$, and $v=\frac{u_{wall}}{U^*}$ for simplicity in notations.
We call the function $f(\theta)$ the dynamic factor, as $f(\theta)$ arises from the force balance equation \eqref{e:bndcond}. We call the function $g(\theta)$ the geometric factor, which describes the geometric relation between the apparent contact angle and contact line velocity. We will see later that $f(\theta)$ plays a very important role in both the dynamic process and the final steady state, while $g(\theta)$ only affects the dynamic process before the steady state is achieved.

\subsection{Time averaging of the ODE system}\label{sec:deterministic}
We first introduce some properties satisfied by  the dynamic factor  $f$ and $g$:
$$ f(\theta)\geq 0, f(0)=0;\quad -M\leq  g(\theta) \leq -m,$$
for some positive numbers $m$ and $M$. In addition, we also assume that $f$ is monotonically increasing in the interval $[0,\pi]$.
 These conditions are quite general and are satisfied by the two examples in the previous section.

We assume that the substrate has periodic chemical patterns with period $\eps$ so that Young's angle at the position $x$ satisfies $\hat{\theta}_Y(x)=\varphi(\frac{x}{\eps})$, where $\varphi$ is a periodic continuously differentiable function with period 1. The minimum and maximum of $\varphi$ are $\theta_A$ and $\theta_B$ respectively with $0<\theta_A<\theta_B<\pi$. One example of such functions is
\begin{equation}\label{eq:sine}
\varphi(z)=\frac{\theta_A+\theta_B}2+\frac{\theta_B-\theta_A}2\sin(2\pi z).
\end{equation}

In the following, we will use the method of averaging to derive the effective dynamics of the contact angle and contact line position. This method has been studied in details (e.g. in \cite{WeinanBook11, pav2008multiscale}) and has been widely used in obtaining the effective boundary conditions (e.g., \cite{miskis1994slip})

We introduce the fast spatial variable $y=\frac{\hat{x}_{ct}}{\varepsilon}$ and fast temporal variable $\tau=\frac{t}{\varepsilon}$. Consider the multiple scale asymptotic expansions
\begin{equation}
\theta_a=\theta_0(t,\tau)+\eps\theta_1(t,\tau)+\cdots,\qquad y=y_0(t,\tau)+\eps y_1(t,\tau)+\cdots.
\end{equation}

Then the system of equations \eqref{eq:general} can be rewritten as
\begin{subequations}
\begin{align*}
  &\frac1{\eps}\frac{\partial\theta_0}{\partial \tau}+(\frac{\partial\theta_0}{\partial t}+\frac{\partial\theta_1}{\partial \tau})+\eps(\frac{\partial\theta_1}{\partial t}+\frac{\partial\theta_2}{\partial \tau})+\cdots\\
  =&g(\theta_0+\eps\theta_1+\cdots)\Big(f(\theta_0+\eps\theta_1+\cdots)(\cos\varphi(y_0+\eps y_1+\cdots)-\cos(\theta_0+\eps\theta_1+\cdots))+v\Big),\\
  &\frac1{\eps}\frac{\partial y_0}{\partial \tau}+(\frac{\partial y_0}{\partial t}+\frac{\partial y_1}{\partial \tau})+\eps(\frac{\partial y_1}{\partial t}+\frac{\partial y_2}{\partial \tau})+\cdots\\
  =&\frac1{\eps}f(\theta_0+\eps\theta_1+\cdots)(\cos\varphi(y_0+\eps y_1+\cdots)-\cos(\theta_0+\eps\theta_1+\cdots)).
\end{align*}
\end{subequations}

\bigskip

\textit{First order equations in $O(\frac1{\eps})$.} We have the two equations in the fast time scale:
\begin{align}
  &\frac{\partial\theta_0}{\partial \tau}=0,\label{eq:theta_0}\\
  &\frac{\partial y_0}{\partial \tau}=f(\theta_0)(\cos\varphi(y_0)-\cos\theta_0) \label{eq:y_0}.
\end{align}
The first equation implies that $\theta_0=\theta_0(t)$ is indeed a slow variable that does not depend on the fast time scale $\tau$.
Then for given $\theta_0$, the second equation is a simple ordinary differential equation for $y_0$ with respect to $\tau$,
that can be easily solved.  
We discuss its solution in two different cases for the parameter $\theta_0$.

\begin{figure}
\centering
  \includegraphics[width=3.6in]{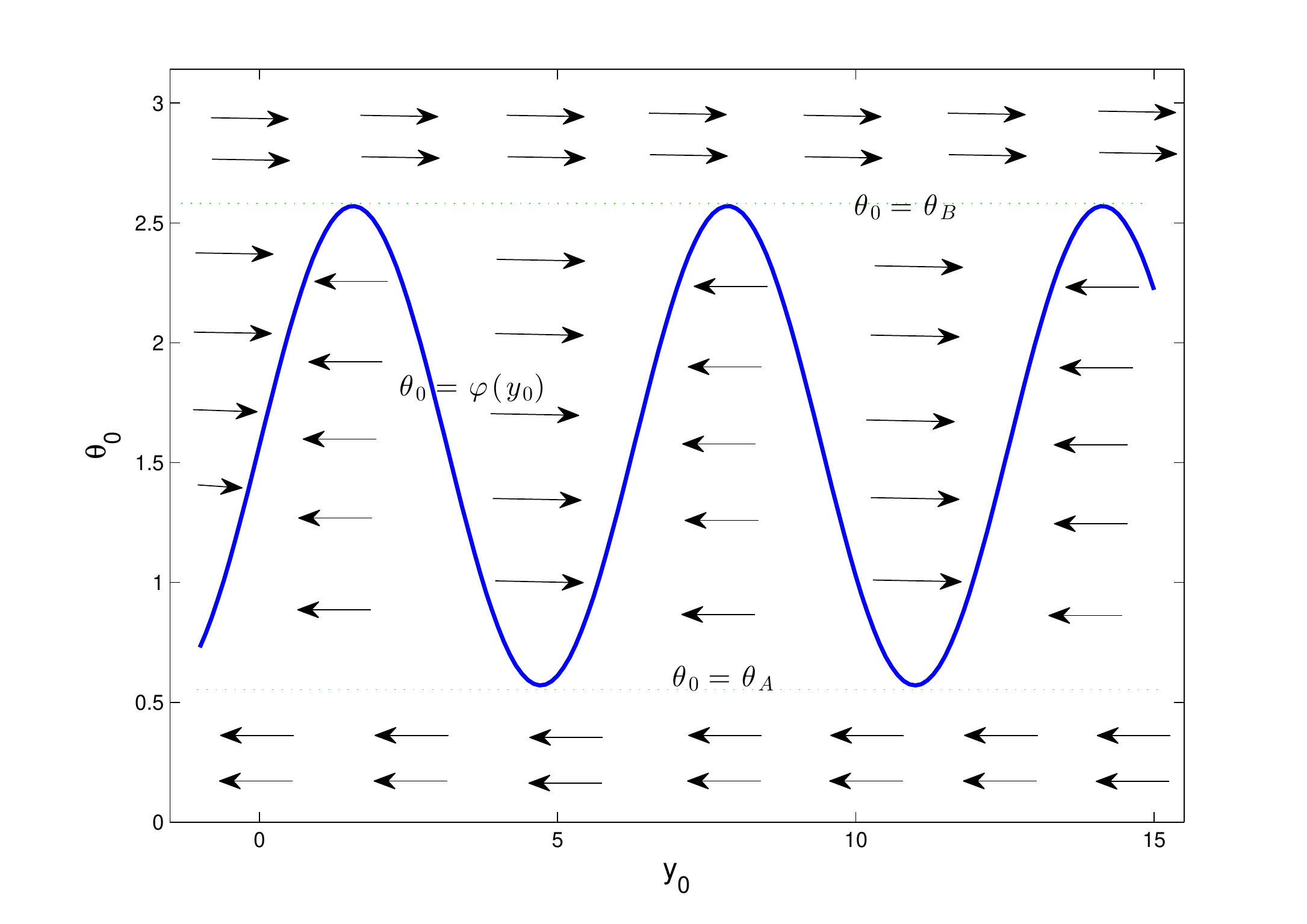}
\caption{Sketch of phase plane analysis for the ordinary differential equations \eqref{eq:theta_0} and \eqref{eq:y_0}.
$\theta_0$ is independent of $\tau$;  $y_0$ increases when $\theta_0>\varphi(y_0)$ and $y_0$ decreases when $\theta_0<\varphi(y_0)$.
}
\label{fig:phase-line}
\end{figure}

In the first case when $\theta_0\in[\theta_A,\theta_B]$, for any given initial value for $y_0$, the solution of \eqref{eq:y_0} approaches to some equilibrium value
$y_{0,\infty}$, which satisfies
 $\varphi(y_{0,\infty})=\theta_0$. By the phase plane analysis (see Figure~\ref{fig:phase-line}), we know that there are three types of equilibrium points: $y_{0,\infty}$ is asymptotically stable, semi-stable, or unstable if $\varphi'(y_{0,\infty})>0$, $\varphi'(y_{0,\infty})=0$ or $\varphi'(y_{0,\infty})<0$ respectively. Every solution path must be attracted to either an asymptotically stable point or a semi-stable point, regardless of the initial value of $y_0$. For instance, if the initial value $y_0^{init}$ satisfies $\varphi(y_0^{init})<\theta_0$, then $y_0$ increases monotonically with $\tau$ until it reaches the nearest equilibrium point $y_{0,\infty}$ which is larger than $y_0^{init}$. This equilibrium point must satisfy $\varphi'(y_{0,\infty})\geqslant0$ and thus becomes asymptotically stable or semi-stable. 
 If the initial value $y_0^{init}$ satisfies $\varphi(y_0^{init})>\theta_0$, then $y_0$ decreases monotonically with $\tau$ until it reaches the nearest equilibrium point $y_{0,\infty}$ that is smaller than $y_0^{init}$. This equilibrium point must also satisfy $\varphi'(y_{0,\infty})\geqslant0$ and thus becomes asymptotically stable or semi-stable. 

In the second case that $\theta_0>\theta_B$ or $\theta_0<\theta_A$, $\cos\varphi(y_0)-\cos\theta_0$ is either always positive or always negative. As a result, $y_0$ is monotonic in $\tau$ and diverges to $\pm\infty$ either increasingly or decreasingly. Moreover, using the method of separable variables, $y_0$ can be solved and represented implicitly as
\begin{equation}\label{eq:sol-y}
  Q(y_0(t,\tau),\theta_0)=Q(y_0(t,0),\theta_0)+\tau,
\end{equation}
where $Q(z,\phi)=\int\frac{\mathrm{d}z}{f(\phi)(\cos\varphi(z)-\cos\phi)}$ is monotonically increasing or decreasing in $z$ and thus can be inverted to give $y_0(t,\tau)$.

\bigskip
\textit{Second order equations in $O(1)$.} We  consider the $O(1)$ order equation for $\theta$, which is given by
\begin{equation}\label{eq:theta0t}
  \frac{\partial\theta_0}{\partial t}+\frac{\partial\theta_1}{\partial \tau}=g(\theta_0)\Big(f(\theta_0)(\cos\varphi(y_0)-\cos\theta_0)+v\Big).
\end{equation}
We can assume that $\theta_1$ is of sublinear growth in $\tau$, i.e., there exists a constant $\alpha\in[0,1)$ such that $|\theta_1(t,\tau)|\leqslant C|\tau|^\alpha$ for some constant $C$ (see also \cite{WeinanBook11}). Define the time averaging operator $<\cdot>_\tau$ as
\begin{equation*}
  <F>_\tau=\lim\limits_{T\rightarrow\infty}\frac1T\int_0^TF(\tau)\mathrm{d}\tau.
\end{equation*}
Then applying this time averaging operator to equation \eqref{eq:theta0t} gives rise to
\begin{equation}\label{eq:averag_theta0t}
  \frac{\partial\theta_0}{\partial t}=g(\theta_0)\Big(<f(\theta_0)(\cos\varphi(y_0)-\cos\theta_0)>_\tau+v\Big),
\end{equation}
where we have used the sublinearity of $\theta_1$ to eliminate $<\frac{\partial\theta_1}{\partial \tau}>_\tau$.

From the previous analysis for the leading order equations, $y_0$ is monotonic in $\tau$. Thus we can simplify the averaged term by
using \eqref{eq:y_0},
\begin{align*}
  <f(\theta_0)(\cos\varphi(y_0)-\cos\theta_0)>_\tau=\lim\limits_{T\rightarrow\infty}\frac1T\int_0^T\frac{\partial y_0}{\partial \tau}\mathrm{d}\tau=\lim\limits_{T\rightarrow\infty}\frac{y_0(t,T)-y_0(t,0)}T.
\end{align*}
We discuss the limit on the right side for different cases on $\theta_0$.
In the first case that $\theta_0\in[\theta_A,\theta_B]$, this limit is zero since every solution path $y(t,\cdot)$ converges to an asymptotically stable or semi-stable equilibrium point (by the analysis for the leading order equations). In the second case that $\theta_0>\theta_B$ or $\theta_0<\theta_A$, we shall show that this limit is $\Big(\int_0^1\frac{\mathrm{d}z}{f(\theta_0)(\cos\varphi(z)-\cos\theta_0)}\Big)^{-1}$, the harmonic average of $f(\theta_0)(\cos\varphi(z)-\cos\theta_0)$ over $z\in[0,1]$.

The argument for the second case is as follows. Let
$$b=\int_0^1\frac{\mathrm{d}z}{f(\theta_0)(\cos\varphi(z)-\cos\theta_0)}=\frac{1}{f(\theta_0)}
\int_0^1\frac{\mathrm{d}z}{(\cos\varphi(z)-\cos\theta_0)}.$$
From \eqref{eq:sol-y} we know that if $\tau=nb$ for any integer $n$, then
\begin{equation*}
  y_0(t,\tau)-y_0(t,0)=n.
\end{equation*}
Writing $T=nb+a$ with $n=\lfloor T/b\rfloor$ (the largest integer no greater than $T/b$) and $0\leqslant a<b$, and letting $n\rightarrow\infty$, we have $\lim\limits_{T\rightarrow\infty}\frac{y_0(t,T)-y_0(t,0)}T=b^{-1}$. This implies that $<f(\theta_0)(\cos\varphi(y_0)-\cos\theta_0)>_\tau$ is exactly equal to the harmonic average of $f(\theta_0)(\cos\varphi(z)-\cos\theta_0)$ over a period $[0,1]$.

By the above calculations, the equation~\eqref{eq:averag_theta0t} is reduced to
\begin{equation*}
  \frac{d \theta_0}{d t}=\begin{cases}g(\theta_0)v,\qquad \qquad \ \theta_0\in[\theta_A,\theta_B];\\
  g(\theta_0)(b^{-1}+v),\quad \theta_0>\theta_B$ or $\theta_0<\theta_A.\end{cases}
\end{equation*}
By this equation, we can discuss the dynamics for the slow variable $\hat{x}_{ct}$. As $\eps\rightarrow0$ we have
\begin{align*}
  \frac{\mathrm{d}\hat{x}_{ct}}{\mathrm{d}t}=&\frac{\partial y_0}{\partial \tau}+\eps(\frac{\partial y_0}{\partial t}+\frac{\partial y_1}{\partial \tau})+O(\eps^2)\\
  \sim&f(\theta_0)(\cos\varphi(y_0)-\cos\theta_0)+O(\eps).
\end{align*}
An application of the time averaging operator $<\cdot>_\tau$, the leading order of $\hat{x}_{ct}$  satisfies
\begin{equation*}
  \frac{\mathrm{d}<\hat{x}_{ct,0}>_{\tau}}{\mathrm{d}t}=
  <f(\theta_0)(\cos\varphi(y_0)-\cos\theta_0)>_\tau=\begin{cases}0,\qquad \theta_0\in[\theta_A,\theta_B],\\b^{-1},\quad \theta_0>\theta_B$ or $\theta_0<\theta_A.\end{cases}
\end{equation*}

We introduce the average of the contact angle and the contact point position  that
$${\Theta}_a:=<\theta_0>_{\tau}=\theta_0,\quad  \hbox{and}\quad \hat{X}_{ct}:=<\hat{x}_{ct,0}>_{\tau}.$$
The above analysis can be summarized as follows. In the first case that $\theta_A\leqslant {\Theta}_a \leqslant\theta_B$, the averaged equations are
\begin{subequations}\label{eq:slave}
\begin{align}
  &\dot{\Theta}_a=g(\Theta_a)v, \label{eq:slave-theta}\\
  &\dot{\hat{X}}_{ct}=0. \label{eq:slave-x}
\end{align}
\end{subequations}
In the second case that ${\Theta}_a >\theta_B$ or ${\Theta}_a<\theta_A$, the averaged equations are 
\begin{subequations}\label{eq:ergodic}
\begin{align}
  &\dot{\Theta}_a=g(\Theta_a)\Big(f(\Theta_a)\Big(\int_0^1\frac{\mathrm{d}z}{\cos\varphi(z)-\cos\Theta_a}\Big)^{-1}+v\Big),\label{eq:ergodic-theta}\\
  &\dot{\hat{X}}_{ct}=f(\Theta_a)\Big(\int_0^1\frac{\mathrm{d}z}{\cos\varphi(z)-\cos\Theta_a}\Big)^{-1}.\label{eq:ergodic-x}
\end{align}
\end{subequations}

\subsection{The effective contact angles}

Based on the above analysis, we can make a prediction for the averaged apparent contact angle
 $\Theta_a$ and the averaged contact line motion $\dot{\hat{X}}_{ct}$. This will lead to
 the formula for the effective advancing and receding contact angles when the contact line moves on a chemically
  patterned surface.

 First consider the case $v>0$, i.e. the wall moves in positive direction.
 This corresponds to a receding contact line, since the fluid moves to the negative direction relative to the substrate.
 We discuss three possible stages while leaving the details to Appendix \ref{sec:ode-averaging}:
\begin{enumerate}
  \item When the initial value of the contact angle $\Theta_a$ lies in the regime $(\theta_B,\pi]$, the averaged contact line dynamics follows \eqref{eq:ergodic}. In this stage, the effective contact angle $\Theta_a$ decreases towards $\theta_B$ exponentially fast. Moreover, the effective contact line position $\hat{X}_{ct}$ moves in the positive direction. The first stage is indeed a transient stage.
  \item When the contact angle $\Theta_a$ reaches $\theta_B$, the averaged dynamics switches to \eqref{eq:slave}. The effective contact angle $\Theta_a$ continues decreasing until it reaches $\theta_A$. However, the effective contact line position $\hat{X}_{ct}$ keeps unchanged.
  \item When the contact angle $\Theta_a$ attains $\theta_A$, the averaged dynamics switches back to \eqref{eq:ergodic}. In this situation, the effective contact angle $\Theta_a$ decreases until it eventually arrives at a stable steady state $\Theta^*<\theta_A$. The steady state $\Theta^*$ satisfies
      \begin{equation}\label{eq:effective-angle}
        f(\Theta^*)\Big(\int_0^1\frac{\mathrm{d}z}{\cos\varphi(z)-\cos\Theta^*}\Big)^{-1}+v=0.
      \end{equation}
      When $\Theta_a$ attains $\Theta^*$, the effective contact line position $\hat{X}_{ct}$ moves in the negative direction at a constant velocity equal to $-v$.
\end{enumerate}

If the initial contact angle lies in the regime $[\theta_A,\theta_B]$ or $[0,\theta_A)$, the above three-stage dynamics
may be reduced to a two-stage process or only  one-stage. In all these situations,  $\Theta_a$ always reaches
the equilibrium contact angle $\Theta^*$. Meanwhile, the average contact line ${\hat{X}}_{ct}$ finally moves with a constant negative velocity $-v$.
Notice that it is the actual contact line position on the solid wall. The equilibrium contact angle  $\Theta^*$ is actually the effective receding contact angle, which is denoted by $\Theta^*=\Theta^*(v)$ as a function of $v>0$.

In the case of $v<0$, a similar analysis can be made to predict the dynamics of the effective advancing contact angle, with the corresponding
velocities in opposite signs. The effective contact angle $\Theta_a$ eventually arrives at a steady state $\Theta^*$, which also satisfies \eqref{eq:effective-angle};
and the actual contact position $\hat{X}$ moves with a constant positive velocity $-v$. The final steady state $\Theta^*$ is called the effective advancing contact angle, which is also denoted by $\Theta^*=\Theta^*(v)$ for $v<0$.

As an interesting but important fact, we remark that: The effective advancing contact angle $\Theta^*(v)>\theta_B$ with $v<0$, and it approaches $\theta_B$ as $v$ increases to zero; the effective receding contact angle $\Theta^*(v)<\theta_A$ with $v>0$, and it approaches $\theta_A$ as $v$ decreases to zero. When $v=0$, the contact line can be pinned for any contact angle between $\theta_A$ and $\theta_B$. Similar ideas have been investigated in many phenomenological moving contact line models in the study of contact angle hysteresis \citep{Prabhala13, yue2020thermodynamically}.

In summary, depending on the initial value of the contact angle, the averaged dynamics of the contact line  and the contact angle  can be
characterized by a process with at most three stages as described above.
 In any cases, one approaches to a steady state where
 the effective advancing angle or receding angle does not change any more.
 The effective contact angles are give by the equation \eqref{eq:effective-angle}.
 It is worth noting that the final effective advancing and receding contact angles only depend on the dynamic factor $f(\theta_a)$ and the dragging velocity $v$. The geometric factor $g(\theta_a)$ only affects the dynamic process that how $\Theta_a$ approaches the effective advancing and receding contact angles. The above results are numerically validated in Section~4.

\subsection{Discussions}\label{sec:discussion}
The main result of the above analysis is that we obtain
an equation~\eqref{eq:effective-angle}
 for the effective advancing and receding contact angles on chemically patterned surface.
 It is easy to see that the dimensionless wall velocity
$v=\frac{u_{wall}}{U^*}$ is opposite to the effective capillary number $Ca$ for the contact line motion.
We replace $v$ by $-Ca$,  the equation~\eqref{eq:effective-angle} can be rewritten as
\begin{equation}\label{eq:effectiveAngleN}
Ca=-{f(\Theta^*)}\left(\int_0^1 \frac{dz}{\cos\Theta^*-\cos\varphi(z)}\right)^{-1}.
\end{equation}
Or equivalently
\begin{equation}\label{eq:effectiveAngleN1}
|\ln\zeta|Ca=-{\mathcal{F}_1(\Theta^*)}\left(\int_0^1 \frac{dz}{\cos\Theta^*-\cos\varphi(z)}\right)^{-1}.
\end{equation}
where $\mathcal{F}_{1}(\theta)=\mathcal{F}(\theta,0)$ is defined in Equation~\eqref{eq:F1}. We can see that the equation is quite similar to
the Cox-type boundary condition~\eqref{e:bndcond} derived in Section 2.
The only difference is the term $(\cos\theta-\cos\theta_Y)$ is replaced by
by its harmonic averaging on chemically inhomogeneous surface (with contact angle pattern $\varphi(z)$).
When $\varphi(z)\equiv \theta_Y$, Equation~\eqref{eq:effectiveAngleN1} will reduce
to Equation~\eqref{e:bndcond}(with $\lambda=0$).
In general cases, it is
 a complicated nonlinear  equation for the dynamic contact angle $\Theta^*$
for given $Ca$ and the chemical pattern function $\varphi$. We will discuss its properties below.

The equation~\eqref{eq:effectiveAngleN1} can be solved simply by numerical methods.
We show some results to see how the effective contact angles depends on
the wall velocity and the chemical patterns.
\begin{figure}
\centering
  \includegraphics[width=2.6in]{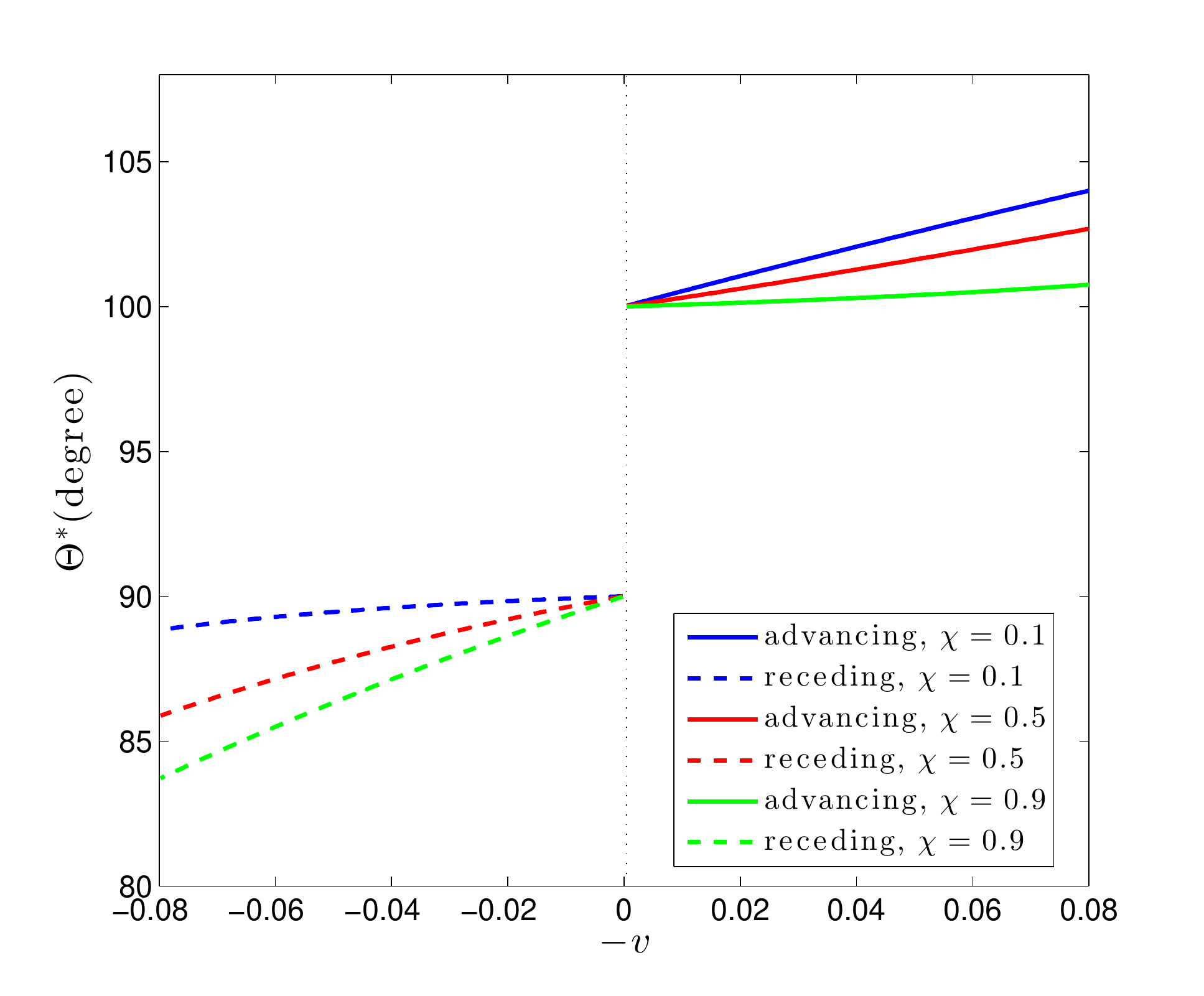}
  \includegraphics[width=2.6in]{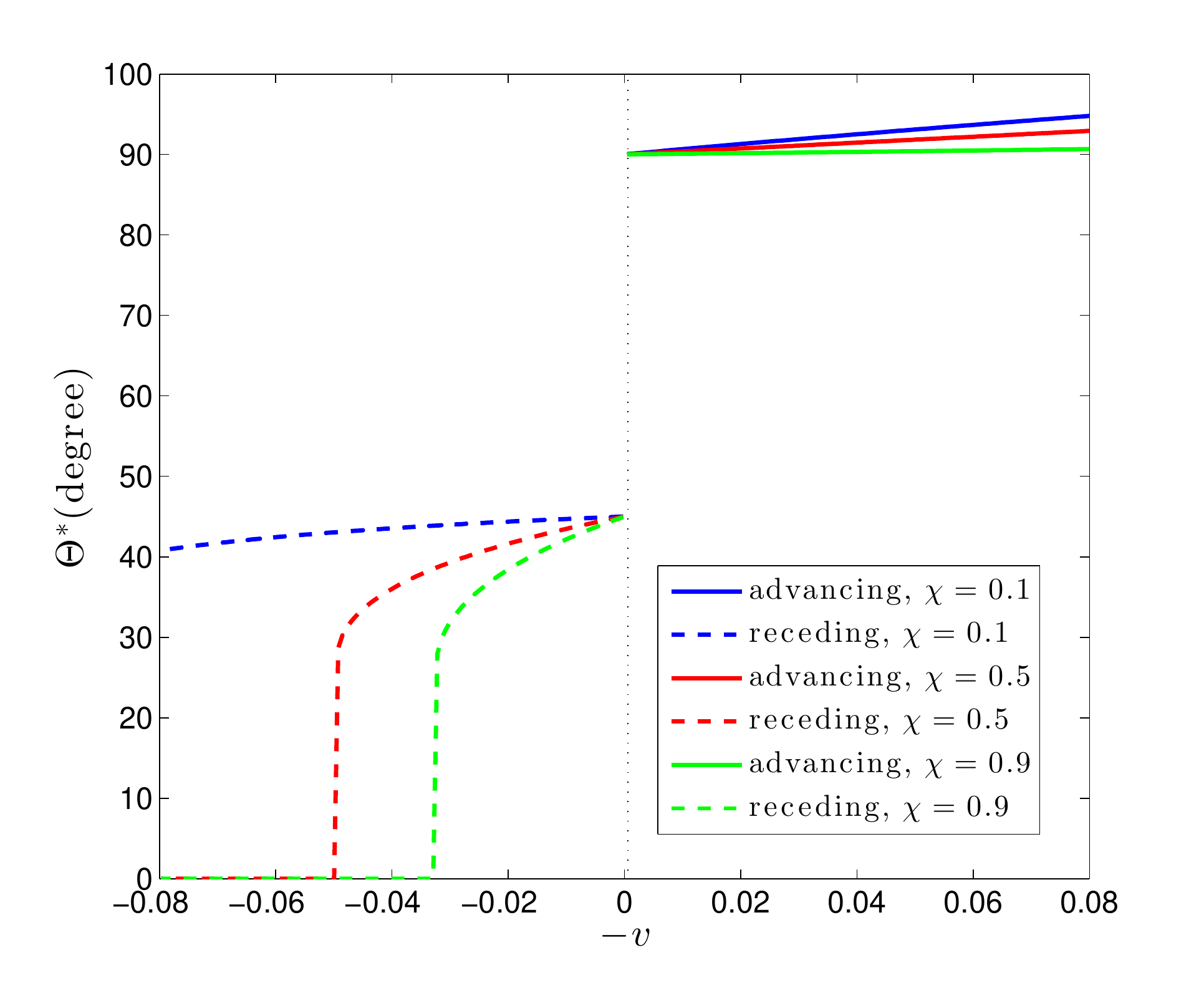}
\caption{The effective contact angles on chemically patterned surfaces. Left  panel: $\theta_A=90^o$, $\theta_B=100^o$.
Right panel: $\theta_A=45^o$, $\theta_B=90^o$.}
\label{fig:anglePattern}
\end{figure}
To show the effect of the chemical inhomogeneity and the velocity on the effective contact angles,
we consider a simple chemically patterned surface.
In this case,  the pattern is described by 
$$
\varphi_{\chi}(z)=  \begin{cases} \theta_A ,\qquad\qquad \quad z\in[0,\chi],
  \\ \theta_B ,\qquad\qquad \quad z\in (\chi,1] .\end{cases}
$$
Here $\chi$ is the fraction of the solid surface occupied by material A.

Figure~\ref{fig:anglePattern} shows the effective advancing and receding contact angles
computed by \eqref{eq:effectiveAngleN}. We could see that the effective contact angles depends
on the two contact angles $\theta_A$ and $\theta_B$, the wall velocity $v$ and also the fraction $\chi$.
Firstly, for given chemical patterns where $\theta_A$, $\theta_B$ and $\chi$ are fixed, the advancing
contact angle increases when the absolute value of the velocity (i.e., $-v$) increases. Meanwhile the receding contact angle decreases
when the velocity $v$ increases.  The dependence of the effective contact angles on
the velocity is affected by both \textcolor{black}{ the chemical properties $\theta_A$, $\theta_B$ and their distribution (represented by $\chi$)}.
For given chemical properties $\theta_A$ and $\theta_B$, the dependence of the
contact angles on the velocity seems more asymmetric for larger $\chi$, which means that
the receding contact angle changes more dramatically than the advancing contact angle when
the absolute value of the velocity increases.
For given distributions (i.e. $\chi$ is fixed),  the dependence seems more symmetric when both $\theta_A$
and $\theta_B$ are close to $90^o$.  These dependence may lead to very complicated
experimental observations~\citep{guan2016asymmetric,guan2016simultaneous}. We will make numerical comparisons in the next section.

From~\eqref{eq:effectiveAngleN1}, we could see that the results depend only on the function
 $\mathcal{F}_1$.
 It is remarkable
that the apparent contact angle does not depend on the specific setup of the problem. The same formula
holds for both a capillary problem in a tube or the forced wetting problem around a fiber. In fact,
the geometric factor $g$ which is determined by the specific setup of a problem  does not appear in the equation.
In this sense, the homogenized formula~\eqref{eq:effectiveAngleN} is valid for general cases.
We can regard it as an effective formula for the advancing contact angle and receding contact angle on
an inhomogeneous surface. It gives explicit relation on how the advancing and receding contact angles
depend on the chemical inhomogeneity of the solid surface. Thus, we expect that
the formula~\eqref{eq:effectiveAngleN1} can be used as a boundary condition for the two-phase
Navier-Stokes equations.

Furthermore,
 we consider only the wetting problem of a liquid-gas system in the above analysis.
The approach can be generalized to the liquid-liquid system. In this case, the function $\mathcal{F}_1$
should be replaced by $\mathcal{F}(\theta,\lambda)$. We get an equation~
\begin{equation}\label{eq:effectiveAngleN2}
|\ln\zeta|Ca=-{\mathcal{F}(\Theta^*,\lambda)}\left(\int_0^1 \frac{dz}{\cos\Theta^*-\cos\varphi(z)}\right)^{-1}
\end{equation}
for two-phase flow with viscous ratio $\lambda$. It is an averaged boundary condition for the apparent contact angle
on chemically inhomogeneous surface, corresponding to the Cox-type boundary condition~\eqref{e:bndcond} on
homogeneous surfaces.
We also remark that the averaging technique also works if we
choose Cox's boundary condition~\eqref{e:coxformula} (instead of the simplified version \eqref{e:bndcond}),
but the analysis should be much more complicated. In this case, the effective boundary condition will be
\begin{equation}\label{eq:effectiveAngleN3}
|\ln\zeta|Ca=-\left(\int_0^1 \frac{dz}{\mathcal{K}(\Theta^*,\lambda)-\mathcal{K}(\varphi(z),\lambda)}\right)^{-1}.
\end{equation}
It is an harmonic average of the force term in Cox's formula. Once again, it reduces to Cox's formula~\eqref{e:coxformula}
when $\varphi(z)\equiv\theta_Y$.
This boundary condition is the averaged version of Cox's formula on a inhomogeneous surface.
It can be used as a coarse graining boundary for moving contact line problems on a chemically inhomogeneous surfaces.
Finally, as we mentioned in Section 2.2, the formula~\eqref{eq:effectiveAngleN3} and \eqref{eq:effectiveAngleN1}
are quite close when the capillary number $Ca$ is small.


\section{Numerical experiments}

In this section, we numerically solve the system \eqref{eq:general}, and its averaged system \eqref{eq:slave}-\eqref{eq:ergodic}. From the numerical comparisons of the dynamics of $\theta_a$ and $\hat{x}_{ct}$ with their averaged dynamics of $\Theta_a$ and $\hat{X}_{ct}$, we can verify our analytical results in the previous section.
We apply the forward Euler scheme to numerically solve the system \eqref{eq:general} and its averaged system \eqref{eq:slave}-\eqref{eq:ergodic}.

\subsection{Verification of the analysis}

We first consider the case of the microscopic channel where the geometric factor is given by $g(\theta)={4\mathcal{G}_1(\theta)}$.
 The two extrema of $\varphi(y)$ are $\theta_A=60^\circ$ and $\theta_B=120^\circ$.

\begin{figure}
\centering
  \includegraphics[width=2.5in]{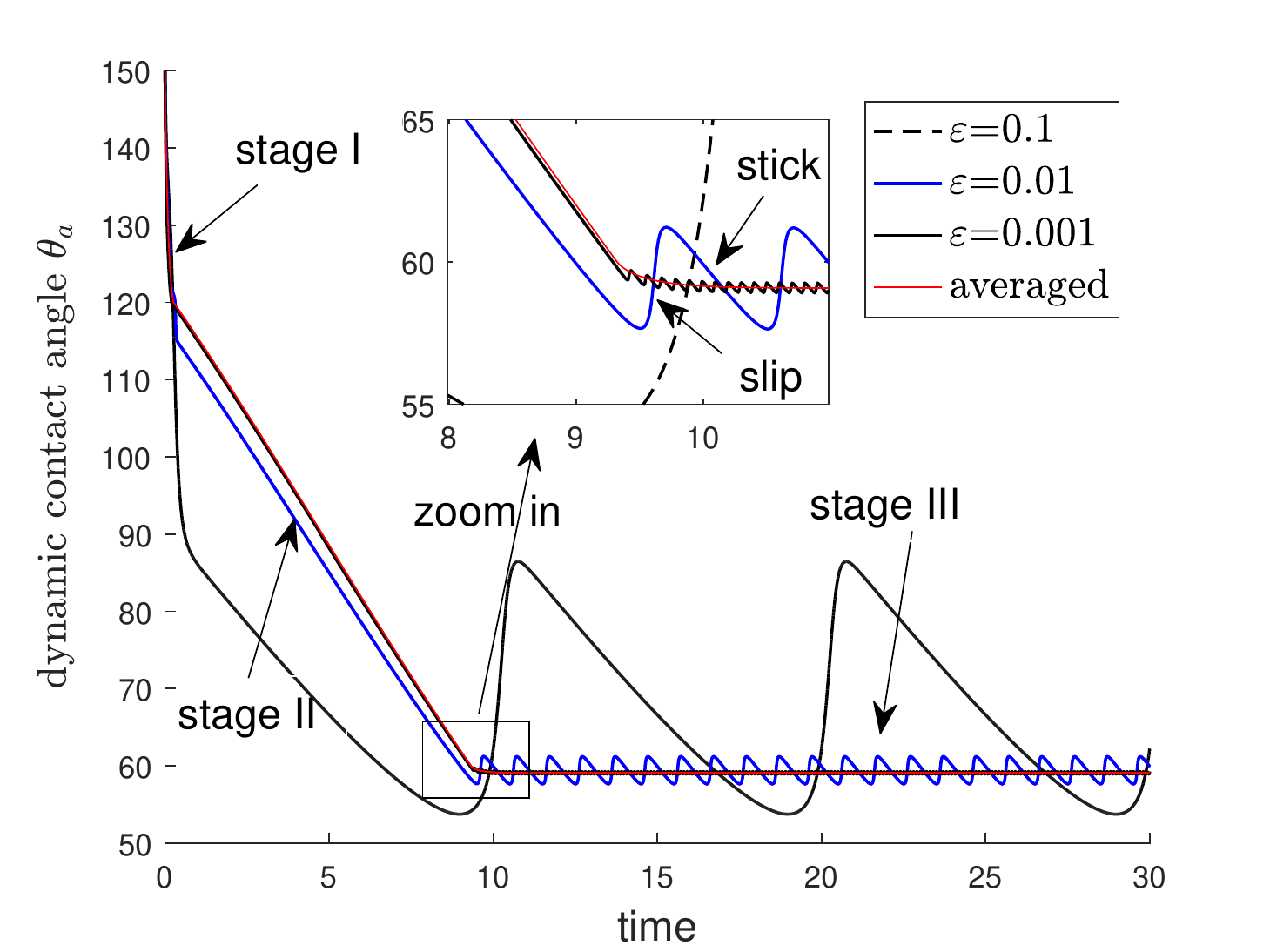}
  \includegraphics[width=2.5in]{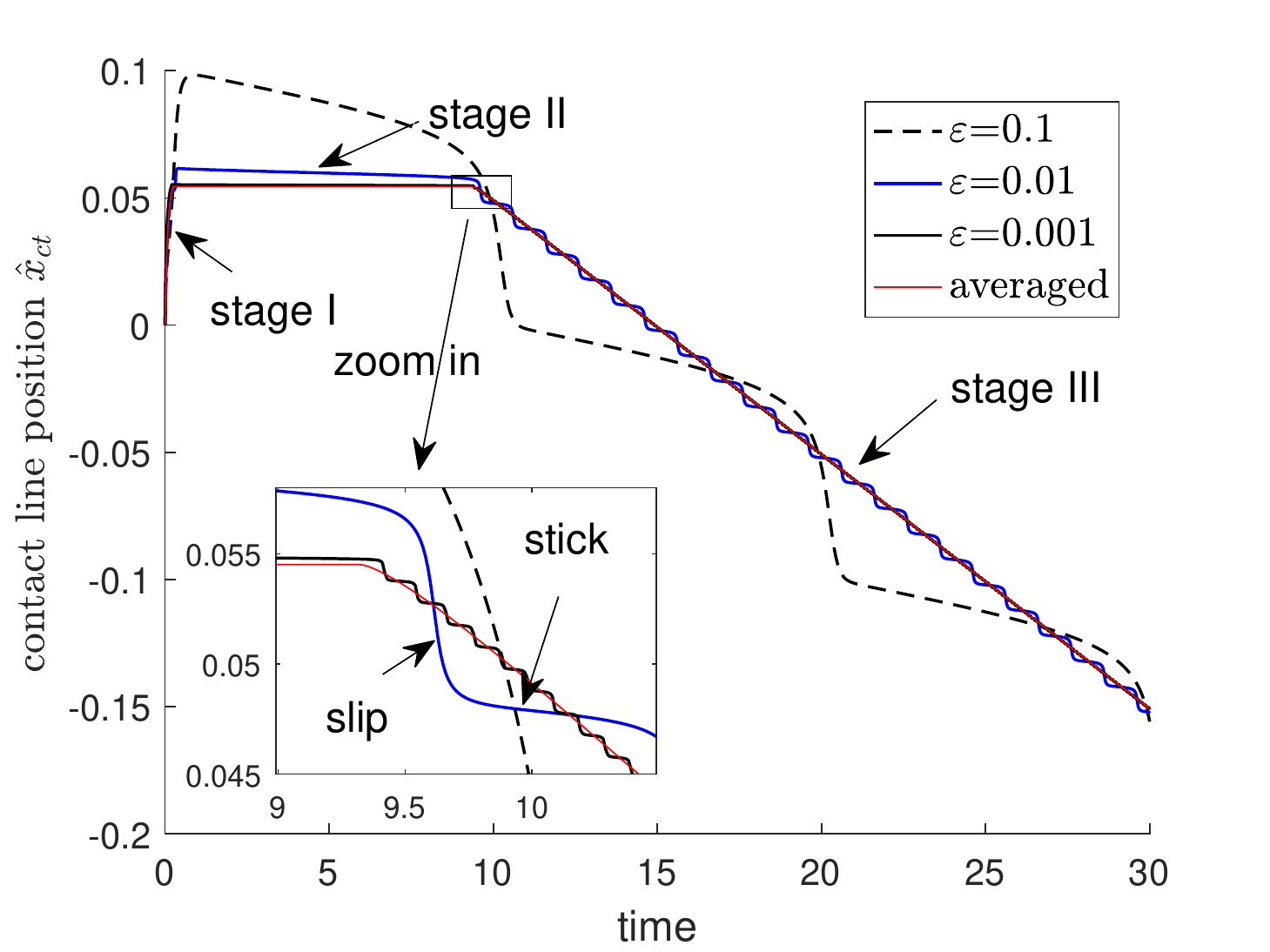}
\caption{Receding dynamics for $\varepsilon=0.1, 0.01, 0.001$ and ${v}=0.01$ given $\theta_A=60^\circ$ and $\theta_B=120^\circ$. Left panel: Dynamical contact angle starting from $\theta_{init}=150^\circ$. Right panel: The contact line position starting from $\hat{x}_{ct}=0$.}
\label{fig:eps-compare}
\end{figure}

 Figure \ref{fig:eps-compare} shows the evolutional behavior of the dynamic contact angle $\theta_a$ and the contact line position $\hat{x}_{ct}$ modeled by \eqref{eq:general} with different periods $\eps=10^{-1}, 10^{-2}$ and $10^{-3}$. The dimensionless velocity is ${v}=0.01$, and the initial contact angle is $\theta_{init}=150^\circ$. It shows  a clear three-stage process.

(i). In the first stage, since the initial contact angle is greater than $\theta_B=120^\circ$ and the wall velocity is positive, the averaged dynamics \eqref{eq:ergodic} predicts that the contact line $\hat{x}_{ct}$ moves to the positive direction and the contact angle $\theta_a$ decreases towards $\theta_B$. This is consistent with the numerical results of the original dynamics \eqref{eq:general} (shown by red curves), in particular for small $\eps$ as depicted by the solid blue and black curves. Moreover, this stage lives very shortly and thus is transient.

(ii). When $\theta_a$ is around $\theta_B=120^\circ$, the second-stage dynamics emerges. The contact line moves very slowly as if it stays almost unchanged, but the contact angle continues to recede.
This behavior is consistent with the prediction by the averaged dynamics \eqref{eq:slave}.

(iii). As the receding angle achieves some value around $\theta_A=60^\circ$, the dynamical process goes into the third stage.
 The contact angle oscillates around the effective receding contact angle given by~\eqref{eq:effective-angle}.
 There are also oscillations for the contact line position $\hat{x}_{ct}$. However, the averaged position of the contact increases linearly with respect to
the time.

In the third stage,  the contact line moves to the negative direction relative to the wall with a stick-slip effect.
Take the $\eps=0.01$ case  for an example (in the zoom-in plot).
As the contact angle achieve about $57.5^\circ$, the contact line starts to move fast (e.g., the zoom-in plot of the $\eps=0.01$ curve when $t=9.5$). This fast movement of $\hat{x}_{ct}$ in turn leads to a fast increasing of $\theta_a$ until $\theta_a$ arrives at about $62^\circ$. This is the slip behavior of the contact line. When $\theta_a$ is around $62^\circ$, the deviation of $\theta_a$ from $\varphi(\hat{x}_{ct})$ is small and so that the contact line moves very slowly as if it sticks there (e.g., $\hat{x}_{ct}\approx 0.047$ in the zoom-in plot of the $\eps=0.01$ curve). The angle $\theta_a$ decreases slowly towards $57.5^\circ$ and another stick-slip period follows. Both the oscillation of the contact angle and the stick-slip of the contact line position share the same period proportional to $\eps$.
The stick-slip motion has also been discussed in details in  \cite{xu2020theoretical}.

Figure~\ref{fig:eps-compare} also shows that the dynamics of the contact angle and the contact line position converges to the averaged dynamics (red curves) modeled by \eqref{eq:slave} and \eqref{eq:ergodic}  as the period $\eps$ decreases. When $\eps=10^{-3}$, the differences between
the original dynamics and the averaged dynamics is quite
small.   

\begin{figure}
\centering
  \includegraphics[width=2.5in]{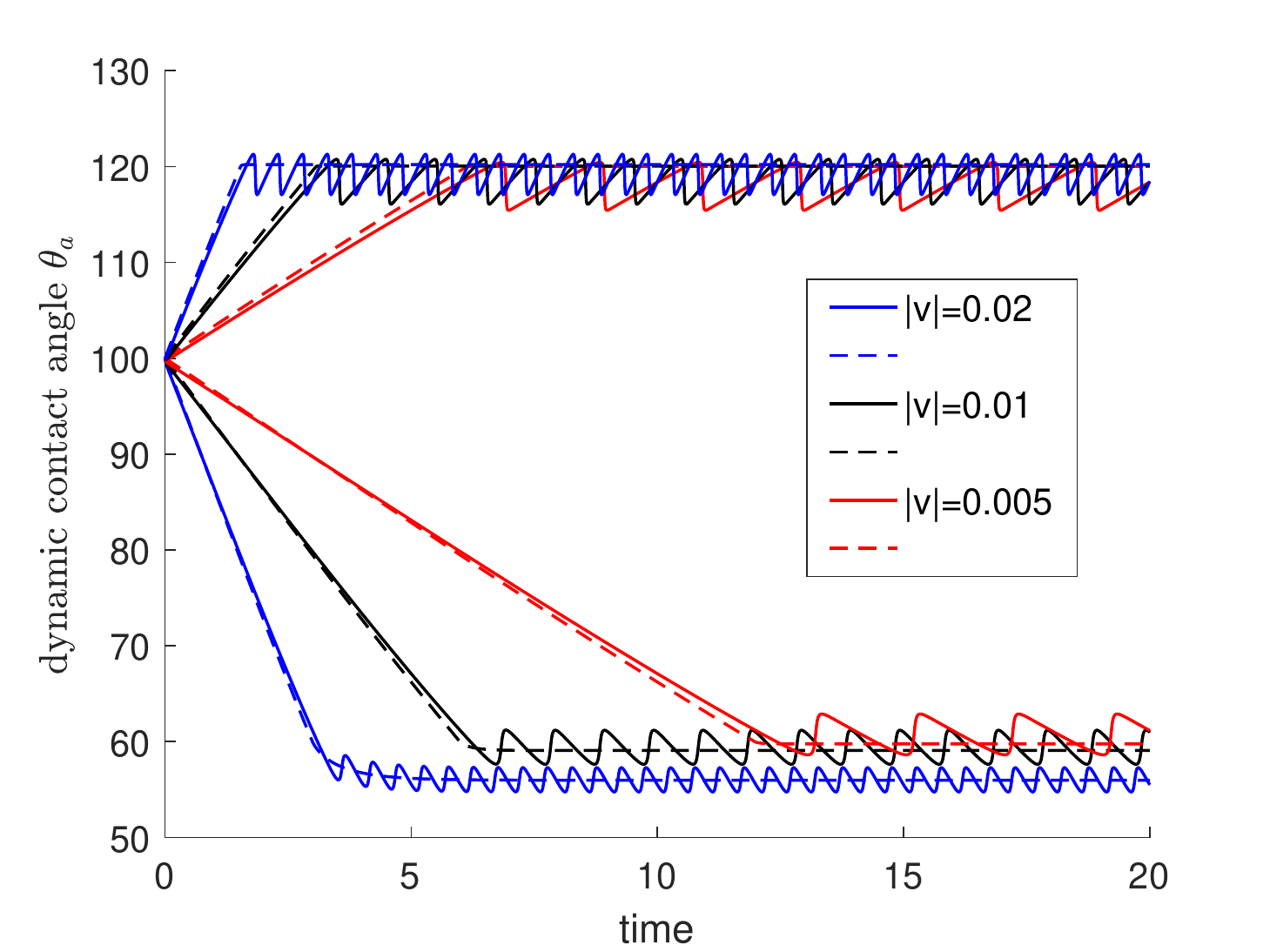}
  \includegraphics[width=2.5in]{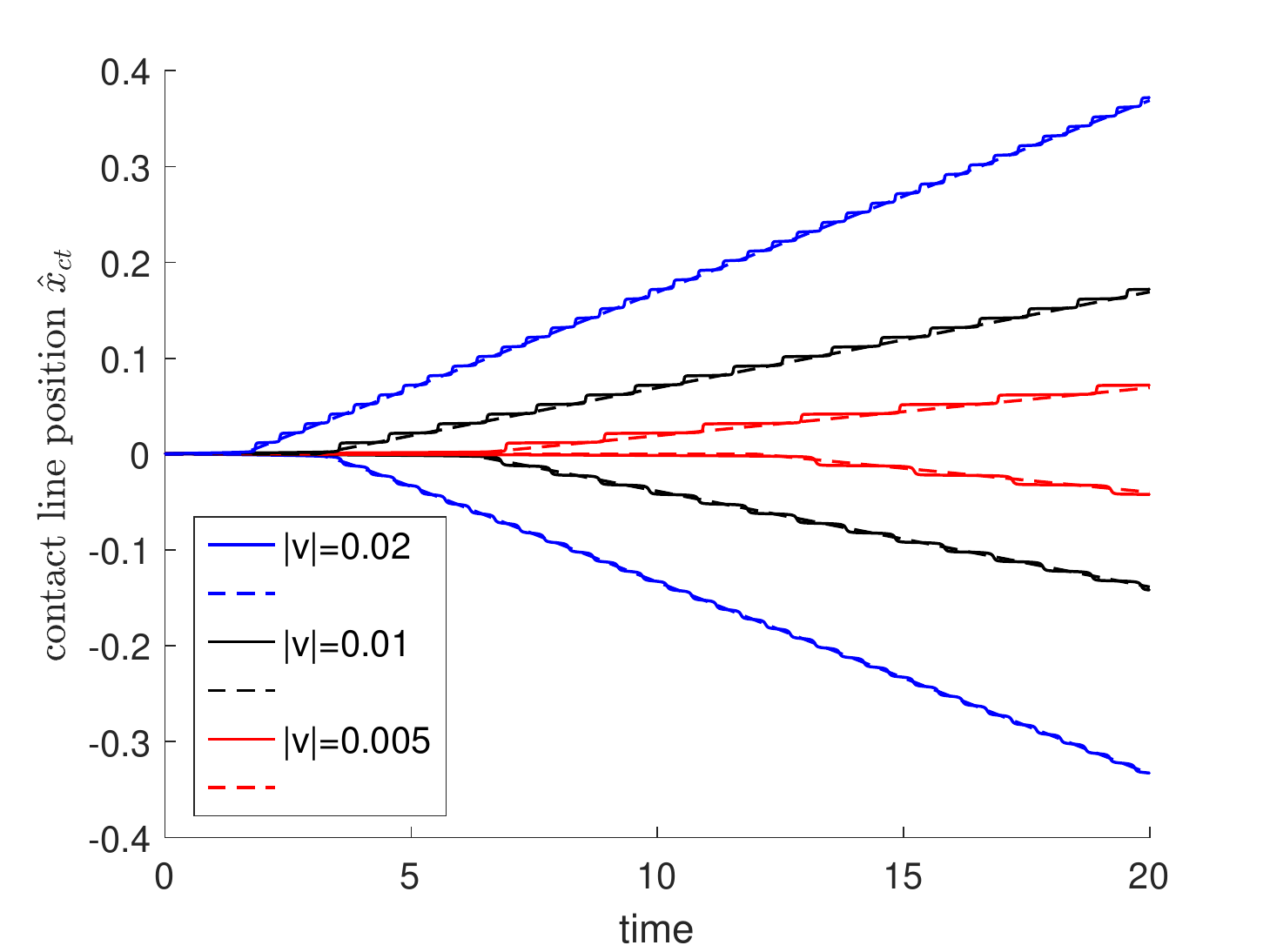}
\caption{Advancing and receding dynamics for $\varepsilon=0.01$ and ${v}=\pm0.02,\pm0.01,\pm0.005$ given $\theta_A=60^\circ$ and $\theta_B=120^\circ$. Left panel: Dynamical contact angle starting from $\theta_{init}=100^\circ$. Right panel: The contact line position starting from $0$. From top to bottom, the solid curves represent the contact angle and contact line motion in the original dynamics. The dashed curves are the corresponding averaged dynamics.}
\label{fig:vel-compare}
\end{figure}

When $v$ is negative, the situation is similar to that shown in Figure~\ref{fig:eps-compare}. In this case, one will observe the effective advancing contact angle in the third stage instead (also see the plots in Figure~\ref{fig:vel-compare}).
For the other choices of $\theta_A$, $\theta_B$ and initial angle $\theta_{init}$, the numerical results are also similar except that the three-stage behavior may be replaced by a two-stage process or a  one-stage process depending on whether $\theta_{init}$ is inside or outside $[\theta_A,\theta_B]$.

Figure~\ref{fig:vel-compare} shows the effect of velocity $v$ on the dynamics.
In these experiments, we fix $\eps=0.01$, $\theta_{init}=100^\circ$, and vary the dimensionless velocity ${v}$ from 0.02 to -0.02.
As shown by the bottom three groups of curves in Figure \ref{fig:vel-compare}, when ${v}>0$, the contact angle $\theta_a$ decreases towards the receding angle and the contact line position $\hat{x}_{ct}$ moves to the negative direction.
In the stick-slip stage, the effective contact line velocity in the averaged dynamic is exactly equal to the wall velocity ${v}$.
The effective receding angle also depends on ${v}$. As $|{v}|$ turns smaller, the effective receding angle is closer to $\theta_A=60^\circ$ from below.
 In the case ${v}<0$, the advancing dynamics is observed and $\theta_a$ increases until it starts to oscillate around the effect advancing angle.
 $\hat{x}_{ct}$ moves to the positive direction with an averaged velocity equal to ${v}$.
 As $|{v}|$ turns smaller, the effective advancing angle is closer to $\theta_B=120^\circ$ from above. This validates our analysis made in Section \ref{sec:deterministic}. Moreover, the sensitivities of the effective receding and advancing contact angles to $|{v}|$ are different and asymmetric in this case.

\begin{figure}
\centering
  \includegraphics[width=2.5in]{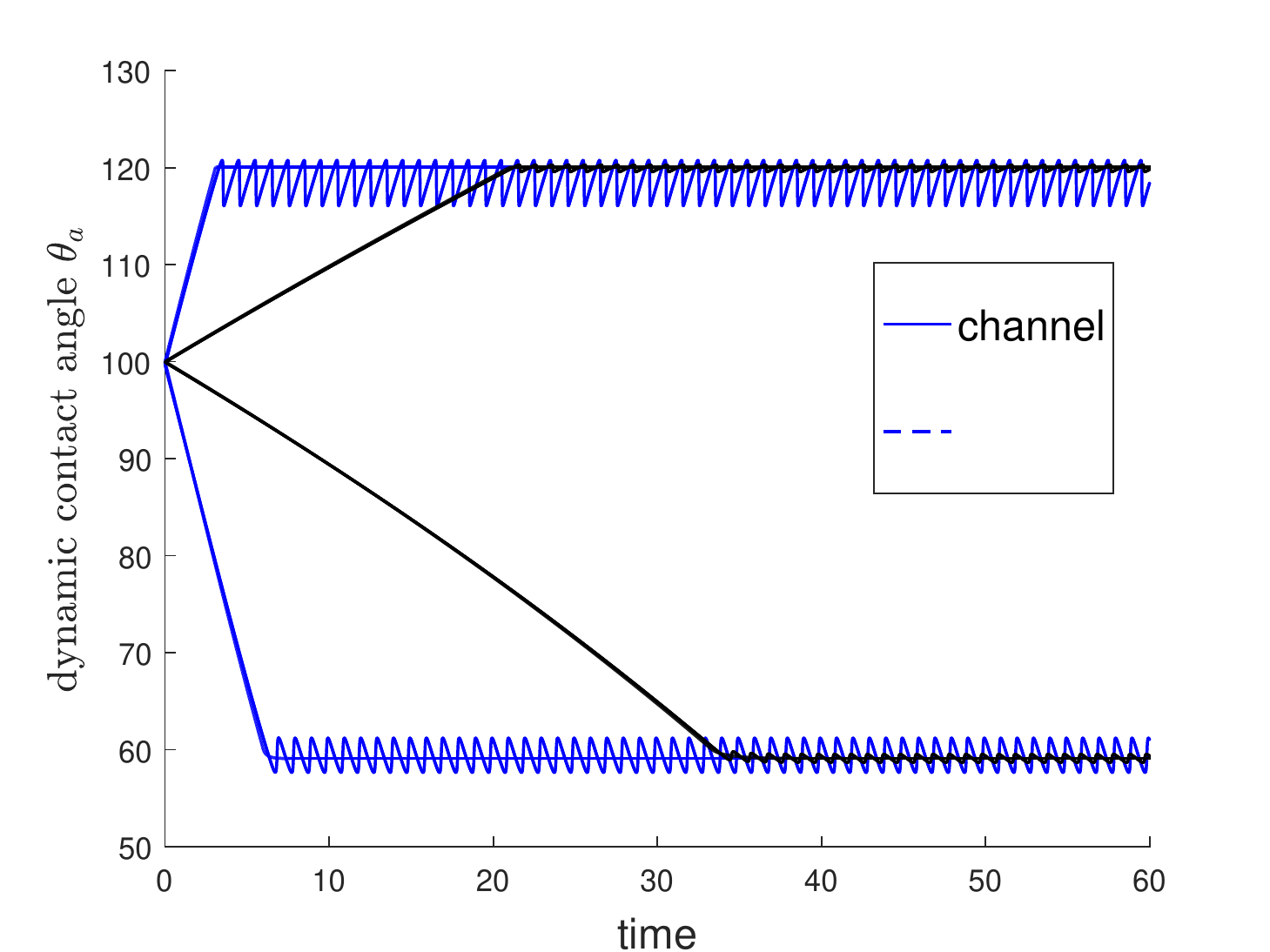}
  \includegraphics[width=2.5in]{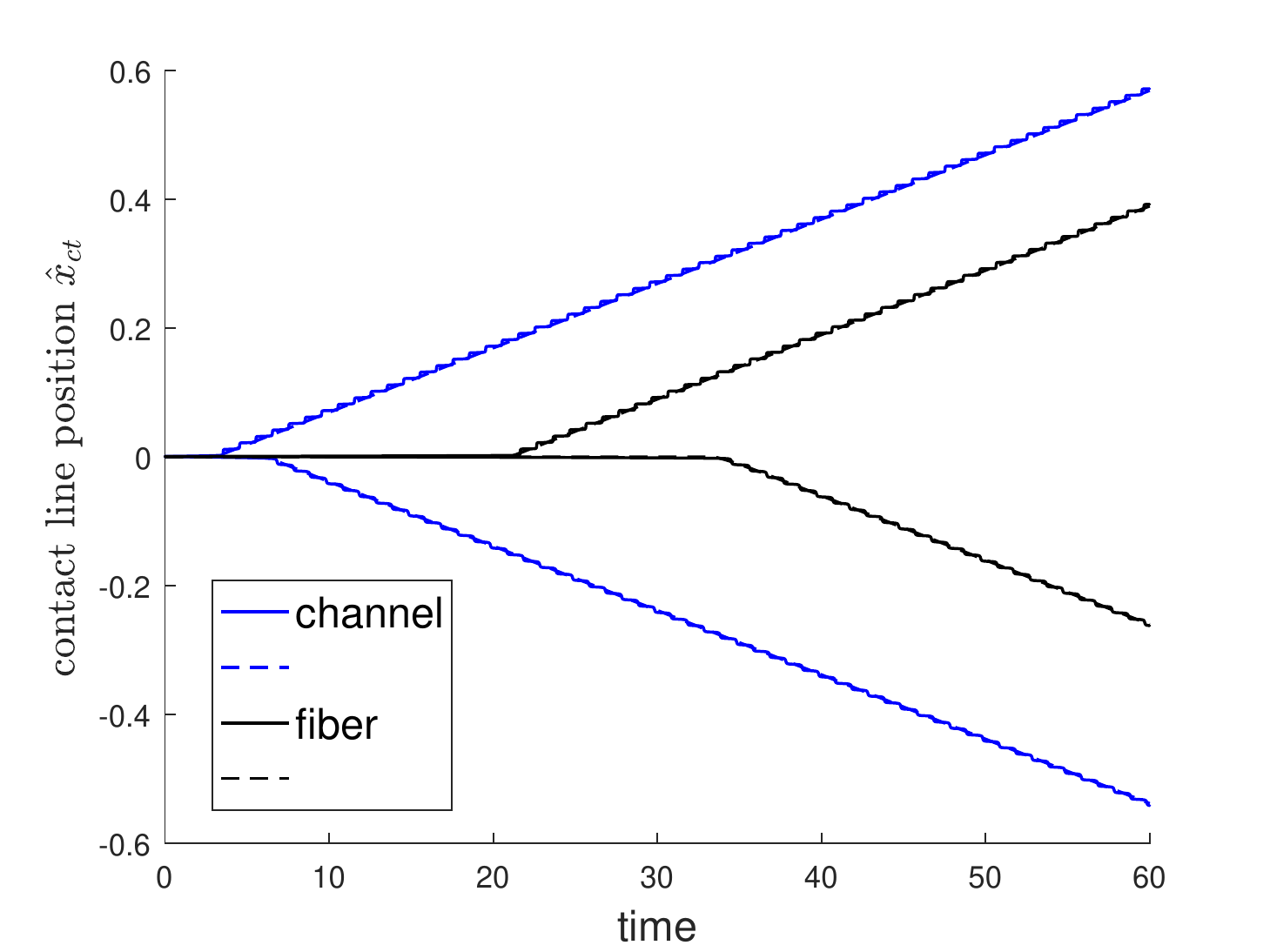}
\caption{Advancing and receding dynamics for $\varepsilon=0.01$ and ${v}=\pm0.01$ given $\theta_A=60^\circ$ and $\theta_B=120^\circ$. Left panel: Dynamical contact angle starting from $\theta_{init}=100^\circ$. Right panel: The contact line position starting from $0$. From top to bottom, the solid curves represent the contact angle and contact line dynamics with respect to ${v}=-0.01$ and ${v}=0.01$. Black curves show the dynamics in the case of fiber while blue curves show the dynamics in the case of microscopic channel. The dashed curves are the corresponding averaged dynamics.}
\label{fig:g-compare}
\end{figure}

Finally, we will study the effect of the geometric factor  $g(\theta_a)$. We solve the problem~\eqref{eq:general} for two different choices of
$g$, which corresponds to the moving contact line problems in a microscopic channel ($g=4\mathcal{G}_1$) and on a moving fiber ($g=4\mathcal{G}_2$).
Figure \ref{fig:g-compare} shows the dynamics of the advancing and receding angles.
It is clear that  the geometric factor affects the dynamic process, especially the first two stages.
 The contact angle approaches to its equilibrium value in the channel case much faster than that it does in the fiber case.
However, the effective advancing and receding angles are the same in the two cases
  and they only depend on the dynamic factor $f(\theta_a)$, just as shown by the equation \eqref{eq:effective-angle}. 

\subsection{Comparison to experimental results}
In this subsection, we will like to use our model and analysis to explain the experimental results of \cite{guan2016asymmetric,guan2016simultaneous}.
There the dynamic contact angle hysteresis is observed by careful design of physical experiments. A very thin glass fiber  with a inhomogeneous surface is inserted into a liquid reservoir. The fiber moves up and down so that a contact line moves on its surface.
The capillary forces on the fiber are measured by AFM and the effective contact angles are computed.
Several liquids with different surface tensions, viscosity and equilibrium angles are tested in their experiments.
Many interesting phenomena are observed in the experiments. It is found that the dependence of the the advancing and receding contact angles
on the fiber velocity is not unified (see Figure 6 in \cite{guan2016simultaneous}).
 In some cases (e.g., the water-air system),
the dependence of the advancing and receding contact angles on the velocity is very asymmetric.
In some cases (e.g., the octanol-air system), this dependence seems symmetric. In some other cases (e.g., the
 FC77-air system), the advancing and receding contact angles are all very small and close to each other.

To compare with the experiments, we use the reduced model \eqref{eq:ODEsysEx2} which is an approximation  for the problem in  \cite{guan2016simultaneous}.
We suppose the surface of the fiber is composed of two different materials $A$ and $B$ with different Young's angles $\theta_A$ and $\theta_B$.
On the surface, the pattern of the Young's angle $\varphi(z)$ is a piecewise constant periodic function  with $\varphi_{\chi}(z)=\theta_A$ if $0\leqslant z<\chi$ and $\varphi_{\chi}(z)=\theta_B$ if $\chi\leqslant z<1$, where $\chi$ is the percentage of material $A$ in a period.
For the convenience in numerical computations, we smooth out this discontinuous pattern by a hyperbolic tangent function:
\begin{equation*}
  \varphi_{\chi}^{\delta}(z)=\frac{\theta_A+\theta_B}{2}+\frac{\theta_B-\theta_A}{2}\tanh\Big(\frac{\sin(2\pi z)-\sin(\chi-1/2)\pi}{\delta}\Big),
\end{equation*}
where $\delta\ll 1$ is chosen to control the thickness of the smooth transition between two patterns. We use $\varphi_{\chi}^{\delta}$ in our simulations.
$\theta_A$ and $\theta_B$ can be chosen approximately according to the receding and advancing contact angles in the experiments with smallest velocity. We choose $\chi$ as a fitting parameter. The dimensionless wall velocity is represented by $v=\frac{\mu u_{wall}}{\gamma}$, where $u_{wall}$ is in the reduced model \eqref{eq:ODEsysEx2}.

\begin{figure}
\centering
  \includegraphics[width=2.5in]{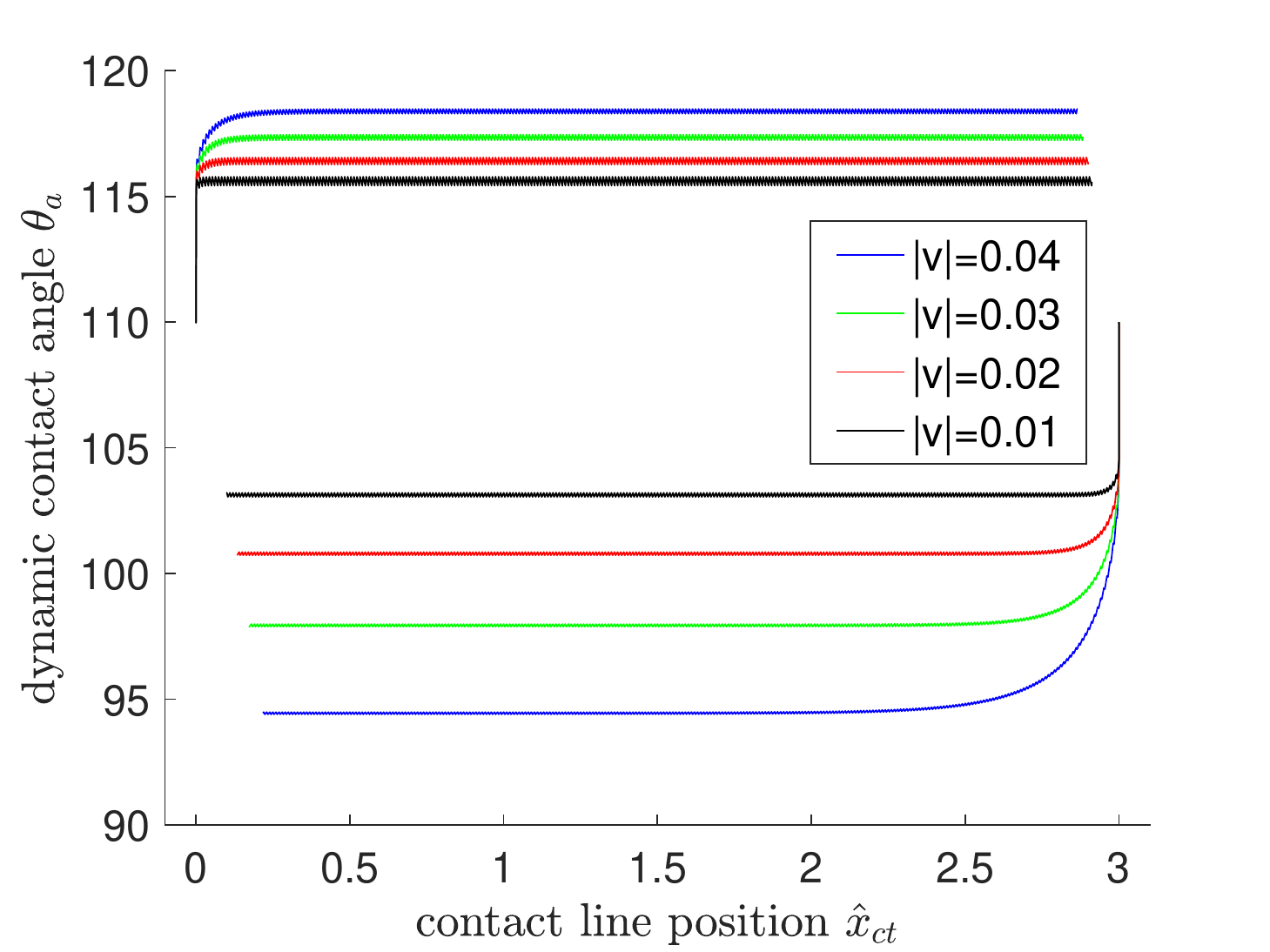}
  \includegraphics[width=2.5in]{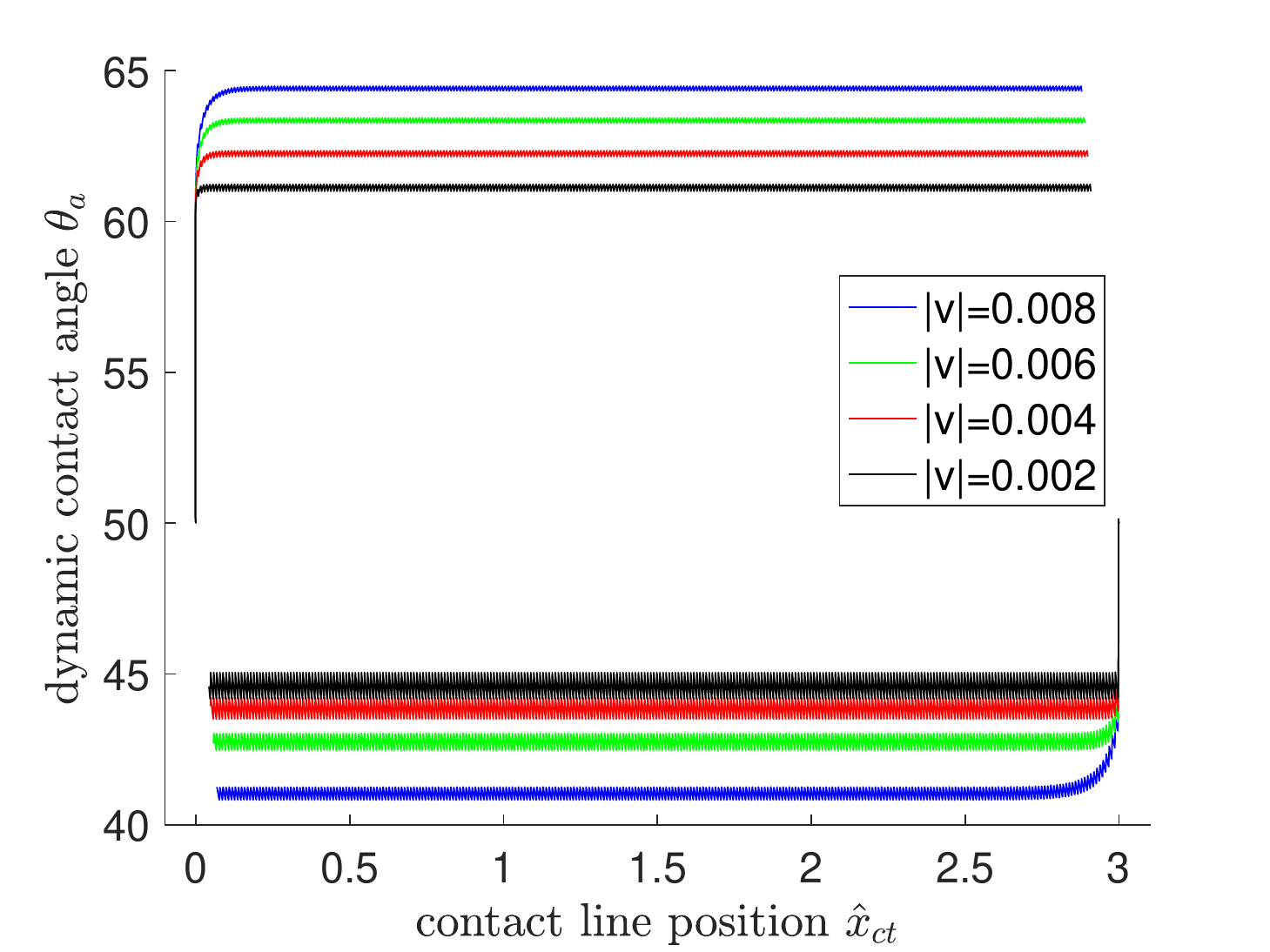}
  \includegraphics[width=2.5in]{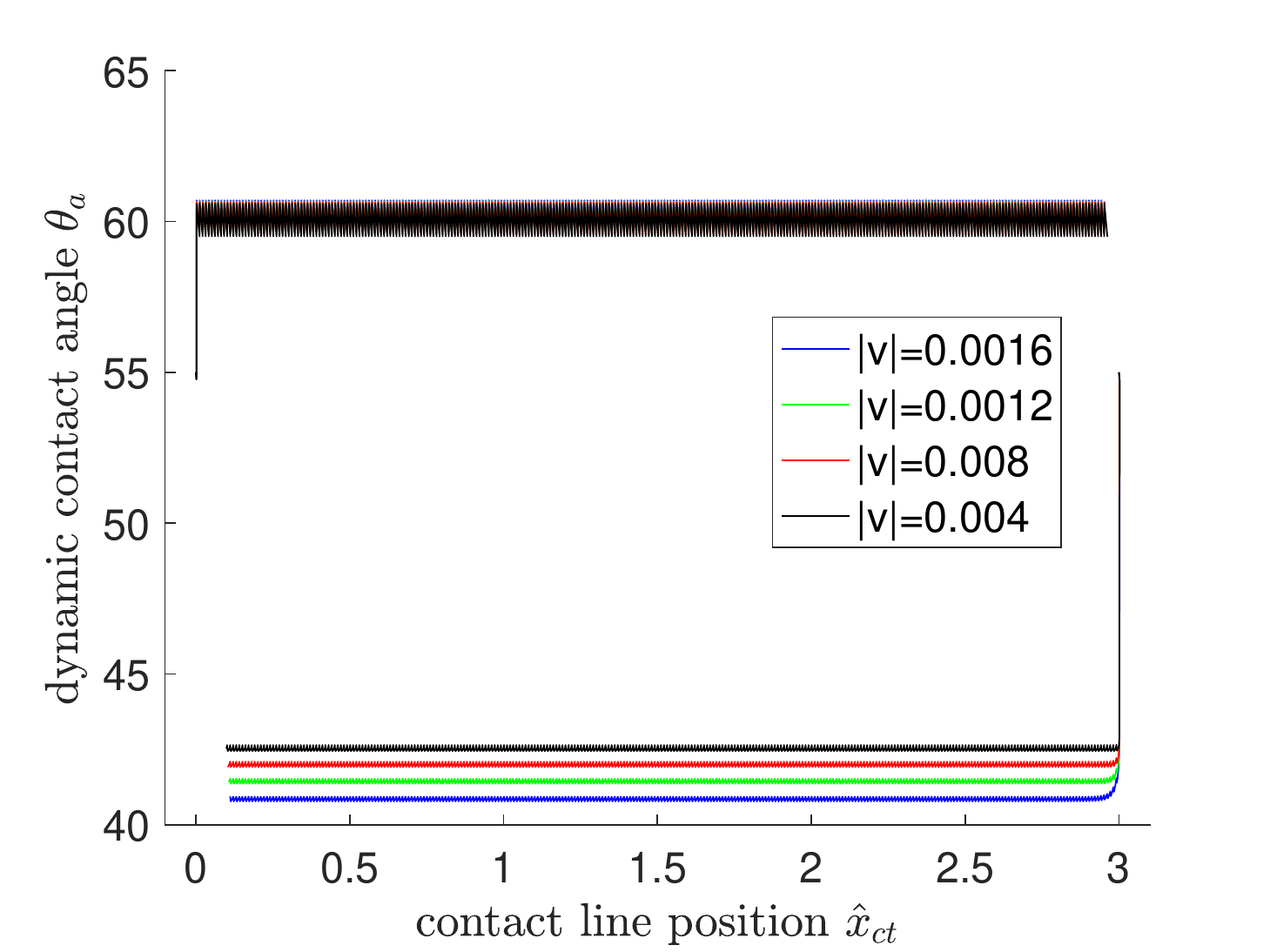}
  \includegraphics[width=2.5in]{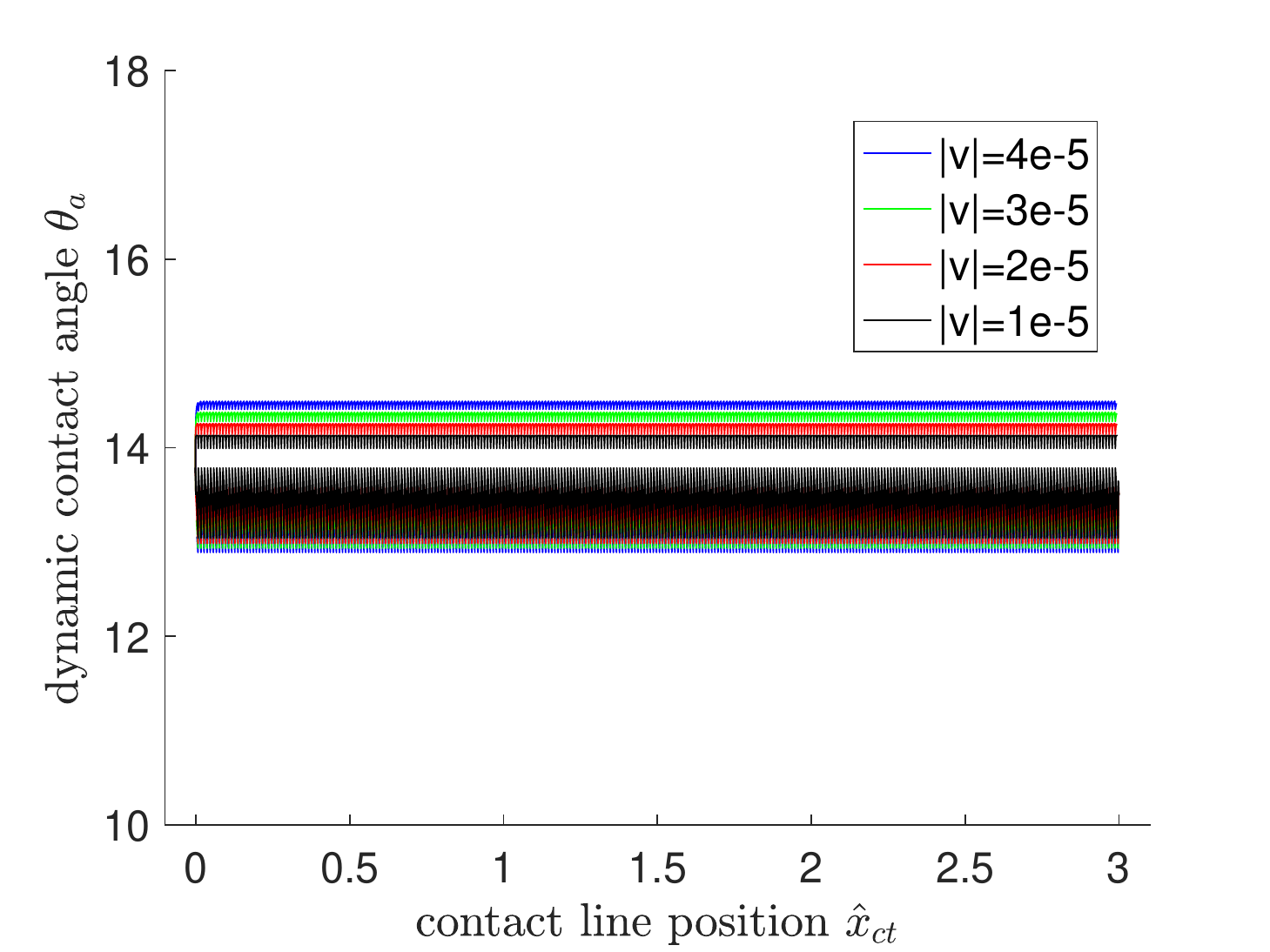}
\caption{Dependence of contact angle hysteresis on different capillary numbers. The period of chemical pattern is set to $\varepsilon=0.01$. Upper left panel: $\theta_A=105^\circ$, $\theta_B=115^\circ$, and $\chi=0.7$. Upper right panel: $\theta_A=45^\circ$, $\theta_B=60^\circ$, and $\chi=0.2$. Lower left panel: $\theta_A=43^\circ$, $\theta_B=63^\circ$, and $\chi=0.9$. Lower right panel: $\theta_A=13^\circ$, $\theta_B=14^\circ$, and $\chi=0.1$. }
\label{fig:experiment-compare}
\end{figure}

The numerical results are given in Figure \ref{fig:experiment-compare}.   We could see that
 the dependence of the contact angle hysteresis on the capillary number is not unified.
For all the four cases, the advancing and receding contact angles and their dependence on the wall velocity are
similar to that in the experiments in \cite{guan2016asymmetric,guan2016simultaneous}.
 In particular, it is found that the asymmetric dependence of the advancing and receding contact angles on the velocity can
 be weakened by  increasing the proportion of the material with smaller contact angle $\theta_A$ in the patterned substrate.
The numerical results also agree with the discussions in Subsection \ref{sec:discussion} based on the formula~\eqref{eq:effective-angle} of the effect contact angle. This also indicates that
the  experimental observations may
be understood from our theoretical analysis in Section 3, especially the equation~\eqref{eq:effective-angle}
 on the  effective contact angles.
There we show that effective advancing and receding
contact angles depend on the capillary number, the Young's angles of the inhomogeneous substrate and
also the spacial distributions of the patterns. All these parameters have important impacts and the interplay among them gives rise to very complicated experimental phenomena.

\begin{figure}
\centering
  \includegraphics[width=2.5in]{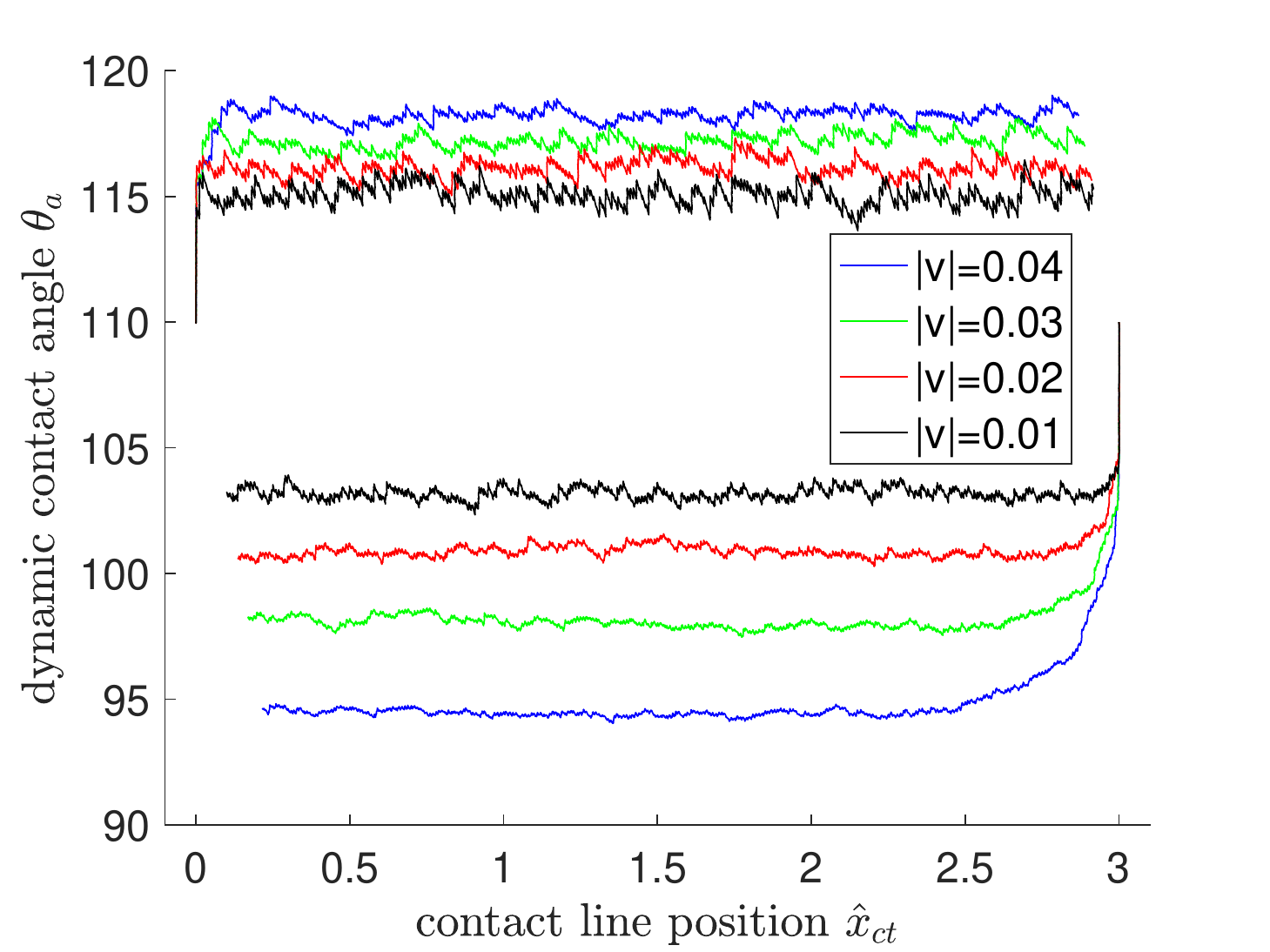}
  \includegraphics[width=2.5in]{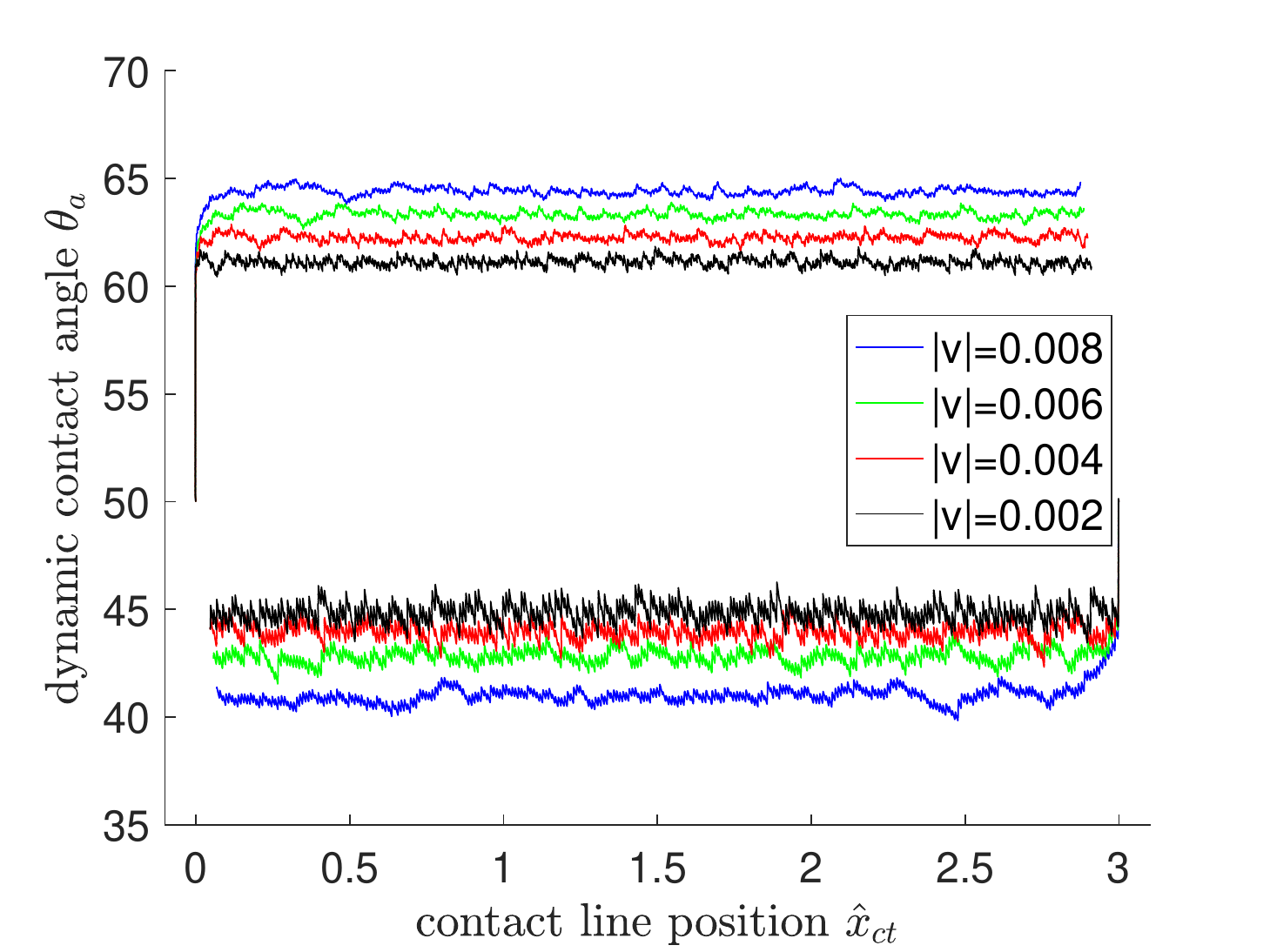}
  \includegraphics[width=2.5in]{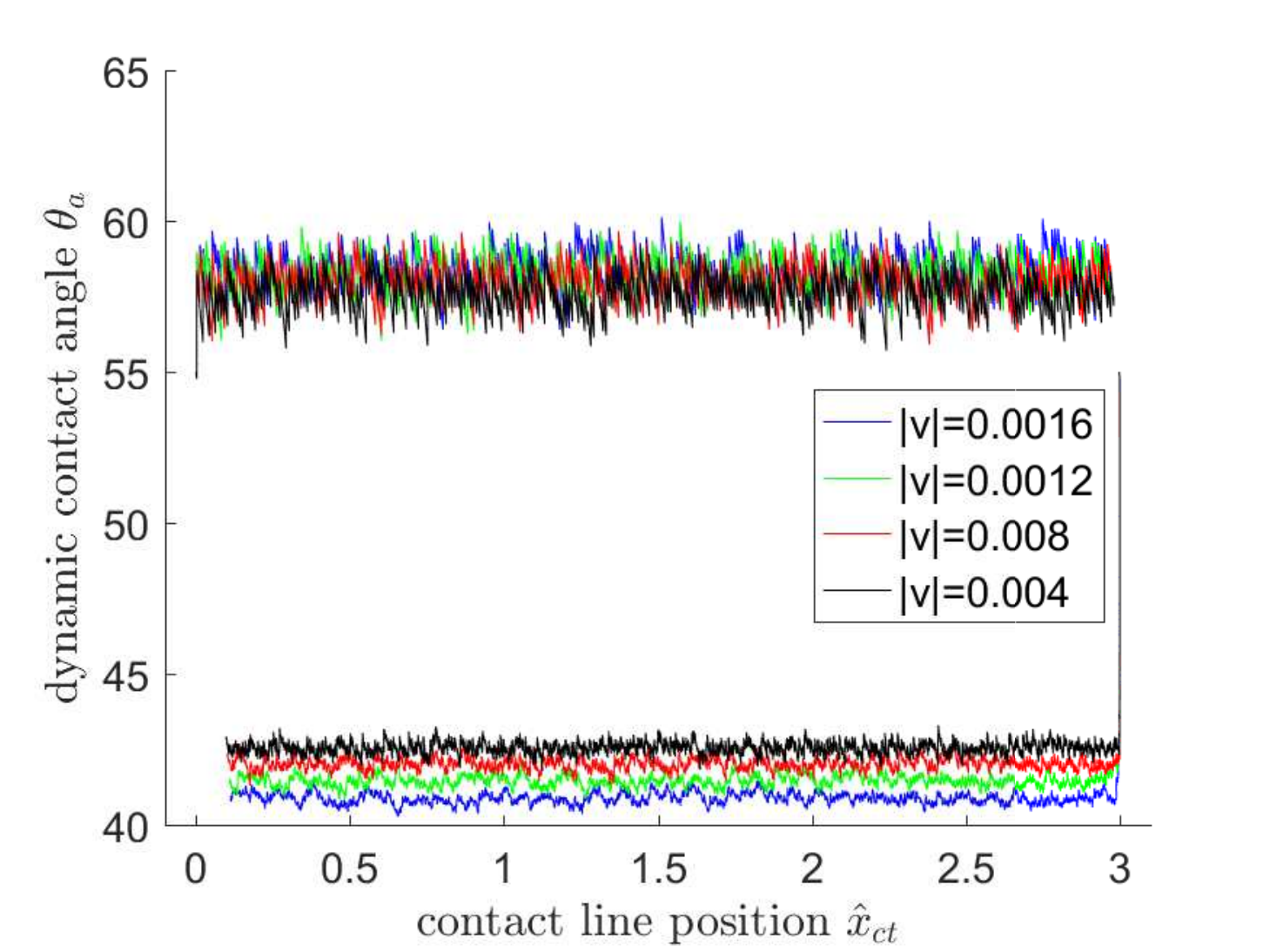}
  \includegraphics[width=2.5in]{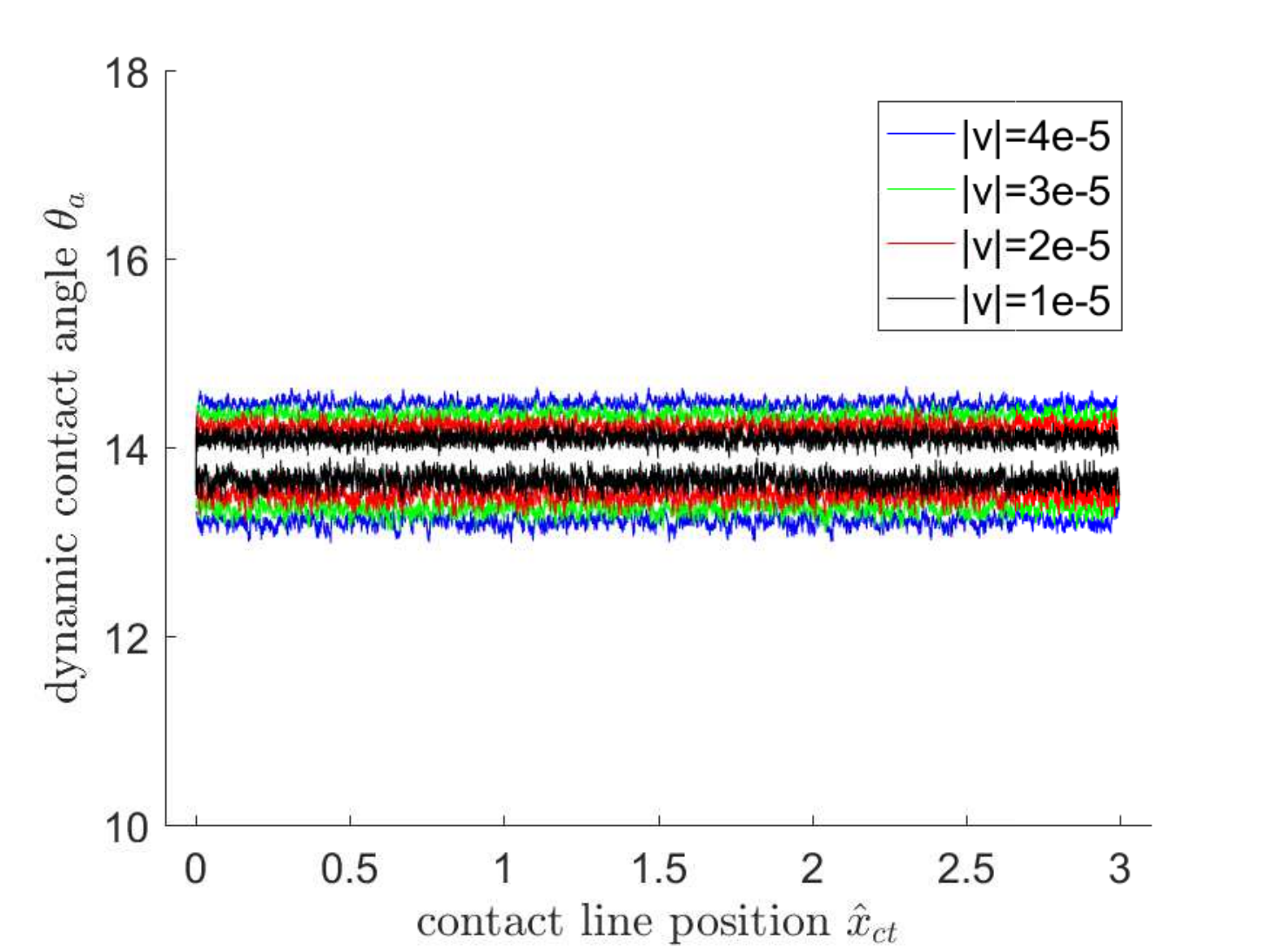}
\caption{Dependence of contact angle hysteresis on different capillary numbers in presence of noise. The period of chemical pattern is set to $\varepsilon=0.01$. The noise level is chosen as $\sigma=0.01$. Upper left panel: $\theta_A=105^\circ$, $\theta_B=115^\circ$, and $\chi=0.7$. Upper right panel: $\theta_A=45^\circ$, $\theta_B=60^\circ$, and $\chi=0.2$. Lower left panel: $\theta_A=43^\circ$, $\theta_B=63^\circ$, and $\chi=0.9$. Lower right panel: $\theta_A=13^\circ$, $\theta_B=14^\circ$, and $\chi=0.1$. }
\label{fig:experiment-compare-noise}
\end{figure}

In real experiments, there are always thermal noises which affect the dynamics of the contact line.
To make the numerical results more comparable with  the experiments, we  add a stochastic force in \eqref{e:bndcond} to model other nondeterministic effects, e.g., thermal noises. The resulting stochastic system for the apparent contact angle $\theta_d$ and contact line position $\hat{x}_{ct}$ reads
\begin{equation}\label{eq:random}
\left\{\begin{array}{l}
\dot{\theta}_a=g(\theta_a)\Big(f(\theta_a)(\cos\hat{\theta}_Y(\hat{x}_{ct})-\cos\theta_a+\sigma\dot{W})+v\Big),
\\
\dot{\hat{x}}_{ct}= f(\theta_a)(\cos\hat{\theta}_Y(\hat{x}_{ct})-\cos\theta_a+\sigma\dot{W}),
\end{array}
\right.
\end{equation}
This system of equations should be understood in the It\^{o} sense. Since the contact angle and the contact line position are linked by the kinematic constraint (e.g., \eqref{eq:kinematic1} and \eqref{eq:kinematic2}) through the geometric factor $g(\theta_a)$, they are driven by the same Brownian motion $W(t)$.
We assume the noise is small with $\sigma=0.01$ so that it does not affect the dynamics too much. We then numerically solve \eqref{eq:random} for one sample path using Euler-Maruyama scheme.
Numerical results by solving the stochastic equation~\eqref{eq:random} are shown in Figure \ref{fig:experiment-compare-noise}.
We could see similar velocity dependence of the contact angle hysteresis as that in Figure~\ref{fig:experiment-compare} in the deterministic case.
Moreover, the dynamic behaviours fit very well with the experiments for all the four cases, i.e, the water-air, octanol-air, decane-air, and FC77-air systems \citep{guan2016simultaneous}.

In the end, we would like to remark that the previous comparisons with experiments are  qualitative rather than quantitative,
since there is a large gap between the capillary numbers chosen in our numerical simulation and those in the experiments.
There are some issues which are not considered in our reduced model~\eqref{eq:ODEsysEx2} (or \eqref{eq:random}).
For example, we assume that the contact line on the fiber is circular and the liquid-air interface is angular symmetric.
The assumptions do not hold in experiments, where the chemical inhomogeneity is  more complex on the  fiber surface.
The relaxation behaviour of the contact line is much more complicated on general surfaces than that in our model.
\textcolor{black}{The contact line hysteresis on general surfaces with chemical and geometrical roughness will be left for future work.}

\section{Conclusion}
In this paper, we derive a formula \eqref{eq:effectiveAngleN} for the apparent contact angles on chemically inhomogeneous surfaces.
It can be regarded as a Cox-type boundary condition for the time averaged apparent contact angle on these surfaces.
The formula characterizes quantitatively how the averaged advancing and receding contact angles depend on the
velocity, the Young's angles and the distributions of the chemical inhomogeneities.
It can be used to understand the complicated behavious for the dynamic contact angle hysteresis
observed in experiments.

The derivation of the above formula is based on a reduced model for the  macroscopic contact angle for moving
contact line problems.
The model is a leading order approximation for the famous Cox's formula for small capillary number and is
easier to analyze in the case with inhomogeneous surfaces.
The reduced model is derived by using the Onsager principle as an approximation tool, which is much simpler
than the standard asymptotic matching methods used in \cite{cox1986}.

Although the main result is obtained by averaging the reduced model for a liquid-vapor system with small size, it can be generalized to
other two-phase flow systems using the same averaging technique. In particular, it is straightforward to do averaging
for the Cox's boundary condition and derive a similar formula~\eqref{eq:effectiveAngleN3}. These formulae can be used coupled with the standard
two-phase Navier-Stokes equation. This will lead to a coarse-graining model for two-phase
flow systems on chemical inhomogeneous surfaces. The dynamic contact angle hysteresis is  given by the formulae and
one does not need to resolve the microscopic inhomogeneity of the solid surfaces as in \cite{yue2020thermodynamically}.

We mainly focus on the two-dimensional problem in the paper. But the results are useful for three dimensional problems when the inhomogeneity is simple. For example, when the defect
is dilute, the static advancing and receding contact angles has been derived by \cite{joanny1984model}.
Then it is possible to reduce the three dimensional problem into a two-dimensional one
by symmetry assumptions. Thus one can apply the results in this paper to these problems.
Some other problems may be handled in a similar way by combing our analysis with the modified Cassie-Baxter equation
 for static contact angle hysteresis in \cite{Choi09,XuWang2013,xu2016modified}.

For more general three-dimensional problems with rough or chemically inhomogeneous surfaces,
the relaxation dynamics of the contact line is very complicated.
the CAH may also depend on the depinning processes of a contact line even in quasi-static processes (\cite{Choi09, iliev2018contact}).
For dynamical problems, the correlations and roughening processes of the contact line make the problem even more complicated (\cite{golestanian2003roughening,golestanian2004moving}).
Both the time averaging and the spacial homogenization needs to be considered simultaneously. The analysis for these problems will be left for future work.

\section*{Acknowledgments}

The work of Z. Zhang was partially supported by the NSFC grant
(No. 11731006, No. 12071207), the Guangdong Provincial Key Laboratory of Computational Science and Material Design (No. 2019B030301001), and the Guangdong Basic and Applied Basic Research Foundation (2021A1515010359).
The work of X. Xu  partially supported by the NSFC grant (No. 11971469) and by the National Key R\&D Program of China under Grant 2018YFB0704304 and Grant 2018YFB0704300.

\section*{Declaration of Interests}
 The authors report no conflict of interest.
\appendix

\section{Calculate the energy dissipation in a wedge region}
\begin{figure}
\vspace{0.5cm}
\centering
	\includegraphics[width=0.45\textwidth]{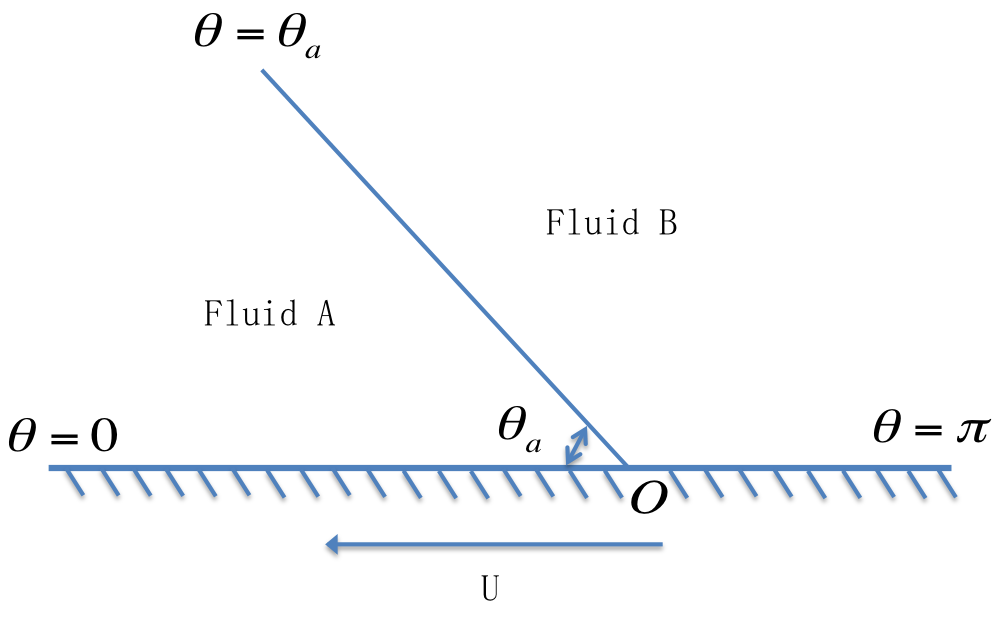}
	\caption{The wedge region near the moving contact point} \label{fig:MCL}
\end{figure}
Since we assume the region is in steady state, $\Phi_{vis}$ can be computed by solving the Stokes equation in the region.
For simplicity, we can change variables and consider a problem as shown in Figure~\ref{fig:MCL}.
We choose a polar coordinate system $(r,\phi)$ near the contact point.
Let the contact point as the origin $O$. The polar axis is in the right direction. In this system, the solid boundary will have
a velocity $U=-v_{ct}$, as shown in Figure~\ref{fig:MCL}. The viscous energy dissipation in the wedge region
can be computed by solving the Stokes equations
\begin{equation}
\left\{\begin{array}{l}
-\mu_A\Delta v_A+\nabla p_A=0, \\
\hbox{div} v_A=0,
\end{array}
\right.\qquad \hbox{in Region A;}
\end{equation}
and
\begin{equation}
\left\{\begin{array}{l}
-\mu_B\Delta v_B+\nabla p_B=0, \\
\hbox{div} v_B=0,
\end{array}
\right.\qquad \hbox{in Region B.}
\end{equation}
We use no slip boundary condition  on $\Gamma_S$. The two-phase flow interface is assumed unchanged with time.
Then the Stokes equation can be solved by the biharmonic equation.

We now calculate the dissipations in the wedge.
To solve the problem, we apply the computations in \cite{cox1986}. Introduce two stream functions $\psi_A$ and $\psi_B$. We
have
\begin{equation}
(v_A)_r=\frac{1}{r}\frac{\partial\psi_A}{\partial\phi}, \qquad (v_A)_{\phi}=-\frac{\partial \psi_A}{\partial r}.
\end{equation}
Here $(v_A)_r$ is the velocity in radial direction, and $(v_A)_{\phi}$ is the velocity in angular direction.
Similar formula hold for $v_B$. Then we have
\begin{equation}
\Delta^2\psi_A=0,\qquad \Delta^2\psi_B=0.
\end{equation}
The equations should satisfies the boundary conditions on the solid surface. We choose the no-slip boundary
condition for the velocity except in the vicinity of the contact line when $r<l$ with the microscopic length $l$ is  a cut-off parameter. When $r>l$,
we have
\begin{align}
&\psi_A=0,\quad \frac{\partial\psi_A}{\partial \phi}=Ur,\qquad \qquad\hbox{on } \phi=0;\\
&\psi_B=0,\quad \frac{\partial\psi_B}{\partial \phi}=-Ur,\qquad \quad\hbox{on } \phi=\pi.
\end{align}
On the interface $\phi=\theta_a$, we have
\begin{equation}
\left\{\begin{array}{l}
\frac{\partial\psi_A}{\partial r}=\frac{\partial\phi_B}{\partial r}=0, \\
\frac{\partial \psi_A}{\partial\phi}=\frac{\partial \psi_B}{\partial\phi},\\
\frac{1}{r^2}\frac{\partial^2\psi_A}{\partial\phi^2}-\frac{\partial^2\psi_A}{\partial r^2}+\frac{1}{r}\frac{\partial\psi_A}{\partial r}
=\lambda
\left(\frac{1}{r^2}\frac{\partial^2\psi_B}{\partial\phi^2}-\frac{\partial^2\psi_B}{\partial r^2}+\frac{1}{r}\frac{\partial\psi_B}{\partial r}\right).
\end{array}
\right.
\end{equation}
where $\lambda=\frac{\mu_A}{\mu_B}$.

The biharmonic equations in the wedge domains can be solved combining the above boundary and interface conditions.
It leads to
\begin{align}
\psi_A=U r\Big( (C_A\phi+D_A)\cos\phi+(E_A\phi+F_A)\sin\phi \Big);\\
\psi_B=U r\Big( (C_B\phi+D_B)\cos\phi+(E_B\phi+F_B)\sin\phi \Big).
\end{align}
Here the two group of coefficients are given by
\begin{align*}
C_A&=\sin\theta_a\Big(-\lambda(\pi\sin\theta_a+\sin^2\theta_a\cos\theta_a+\theta_a(\pi-\theta_a)\cos\theta_a)\\
&\qquad\qquad +\cos\theta_a(\sin^2\theta_a-(\pi-\theta_a)^2)\Big)/\delta;
\\
D_A&=0;\\
E_A&=\sin^2\theta_a\Big(-\lambda(\sin^2\theta_a+\theta_a(\pi-\theta_a))+\sin^2\theta_a-(\pi-\theta_a)^2\Big)/\delta;\\
F_A&=\theta_a\Big(\lambda(\sin^\theta_a+\theta_a(\pi-\theta_a)+\pi\sin\theta_a\cos\theta_a)+(-\sin^2\theta_a+(\pi-\theta_a)^2)\Big)/\delta;
\end{align*}
and
\begin{align*}
C_B&=\frac{-D_B}{\pi}=\sin\theta_a\Big(\lambda\cos\theta_a(\theta_a^2-\sin^2\theta_a)-\pi\sin\theta_a+\sin^2\theta_a\cos\theta_a\\
\qquad\qquad
 &+\theta_a(\pi-\theta_a)\cos\theta_a \Big)/\delta;
\\
E_B&=\sin^2\theta_a\Big(\lambda(\theta_a^2-\sin^2\theta_a)+(\sin^2\theta_a+\theta_a(\pi-\theta_a))\Big)/\delta;\\
F_B&= \Big(\lambda(\sin^2\theta_a-\theta_a^2)(\theta_a-\pi\cos^2\theta_a) -\pi(\pi-\theta_a)\sin\theta_a\cos\theta_a -\theta_a\sin^2\theta_a\\
&\qquad\quad+\pi\sin^2\theta_a\cos^2\theta_a
-(\pi-\theta_a)\theta_a^2)\Big)/\delta.
\end{align*}
Here
$$\delta=\lambda(\theta_a^2-\sin^2\theta_a)(\pi-\theta_a-\sin\theta_a\cos\theta_a)+\big((\pi-\theta_a)^2-\sin^2\theta_a\big)(\theta_a-\sin\theta_a\cos\theta_a).$$

With the formula for $\psi_A$ and $\psi_B$, we can compute the velocities in the two regions. For liquid A,
we have
\begin{align}
&(v_A)_r=U\big((C_A+F_A)\cos\phi+(E_A-D_A)\sin\phi-C_A\phi\sin\phi+E_A\phi\cos\phi\big),\\
&(v_A)_\phi=-U\big( (C_A\phi+D_A)\cos\phi+(E_A\phi+F_A)\sin\phi\big).
\end{align}
Then we have
\begin{equation}
v_A=(v_A)_r \mathbf{r}+(v_A)_\phi \boldsymbol{\tau},
\end{equation}
where $\mathbf{r}$ and $\boldsymbol{\tau}$ are the unit vector in the radial and angular directions, respectively.
We have similar representations for $v_B$. The gradient of the velocity gives
\textcolor{black}{
\begin{align}
\nabla v_A=\frac{2U}{r}(-C_A\sin\phi+E_A\cos\phi) \boldsymbol{\tau}\otimes\mathbf{r}.
\end{align}
Similarly,
\begin{align}
\nabla v_B&=\frac{2U}{r}(-C_B\sin\phi+E_B\cos\phi) \boldsymbol{\tau}\otimes\mathbf{r}.
\end{align}
}
Then the viscous energy dissipations in the wedge regions can be computed by
\begin{align*}
\Psi&=\int_{l_s}^R\int_0^{\theta_a}\mu_A |\nabla v_A|^2 r d\phi d r+\int_{l_s}^R\int_{\theta_a}^\pi \mu_B|\nabla v_B|^2 r d\phi d r\\
&=2\mu_A|\ln\zeta|U^2\Big( C_A^2(\theta_a-\sin\theta_a\cos\theta_a)+C_AE_A(\cos 2\theta_a-1)+ E_A^2(\theta_a+\sin\theta_a\cos\theta_a)\\
&\qquad+\lambda\big(C_B^2(\theta_a-\sin\theta_a\cos\theta_a)+C_BE_B(\cos 2\theta_a-1)+ E_B^2(\theta_a+\sin\theta_a\cos\theta_a)
\big)  \Big),
\end{align*}
where ${\zeta}=D/l$ is the cut-off parameter. Direct calculations lead to
\begin{align}
\Psi&=2\mu_A |\ln\zeta|U^2 \times \nonumber\\
&  \frac{\sin^2\theta_a\Big(\lambda^2(\theta_a^2-\sin^2\theta_a)
+2\lambda(\sin^2\theta_a+\theta_a(\pi-\theta_a))+((\pi-\theta_a)^2-\sin^2\theta_a)\Big) }
{\lambda(\theta_a^2-\sin^2\theta_a)(\pi-\theta_a+\sin\theta_a\cos\theta_a)+((\pi-\theta_a)^2-\sin^2\theta_a)(\theta_a-\sin\theta_a\cos\theta_a)}.
\end{align}

\section{Properties of the averaged dynamics}\label{sec:ode-averaging}

\begin{figure}
\centering
  \includegraphics[width=3.6in]{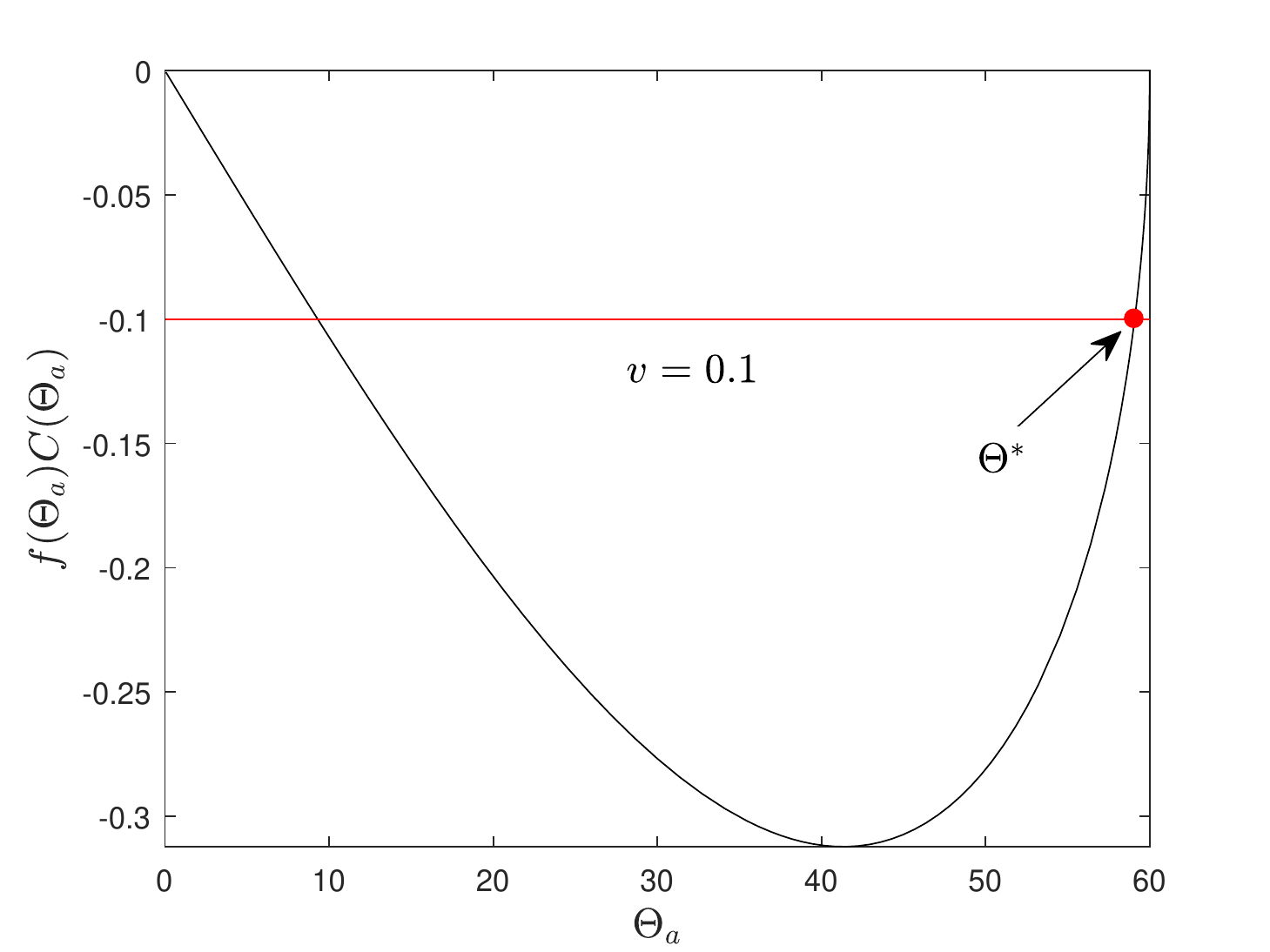}
\caption{Sketch of the function $f(\Theta_a)C(\Theta_a)$ in the case that $v=0.1$, $f$ is given by $\mathcal{F}_1$ in \eqref{eq:F1} and $\varphi$ is given by \eqref{eq:sine} with $\theta_A=60^\circ$ and $\theta_B=120^\circ$. The red point represents the steady apparent contact angle $\Theta^*$ closest to $\theta_A$. It is clear that $f(\Theta_a)C(\Theta_a)$ is non-decreasing at $\Theta^*$.
}
\label{fig:sketch-stable}
\end{figure}

In this section, we study the properties of the averaged dynamics \eqref{eq:slave} and \eqref{eq:ergodic}. Without loss of generality, we assume $v>0$. We also recall that $f$ is a nonnegative function satisfying $f(0)=0$ and $f'(\theta)>0$ for $\theta\in[0,\pi]$, $g$ is a negative function bounded by $-M$ and $-m$. We consider the behavior of the dynamic system for different regimes of $\Theta_a$ starting at $\Theta_a(0)>\theta_B$:
\begin{enumerate}[i)]
  \item When $\Theta_a>\theta_B=\max\limits_{0\leqslant y\leqslant1}\varphi(y)$, the averaged dynamic \eqref{eq:ergodic} is a good approximation. It yields the following properties:
      \begin{enumerate}
        \item $\Theta_a$ monotonically decreases at a speed of at least $mv$ until it arrives at $\theta_B$.

        This is because the harmonic average $C(\Theta_a)=\Big(\int_0^1\frac{\mathrm{d}z}{\cos\varphi(z)-\cos\Theta_a}\Big)^{-1}$ is positive when $\Theta_a>\theta_B$. As a result, we have
      \begin{equation*}
        \dot{\Theta}_a=g(\Theta_a)\Big(f(\Theta_a)\Big(\int_0^1\frac{\mathrm{d}z}{\cos\varphi(z)-\cos\Theta_a}\Big)^{-1}+v\Big)\leq -mv<0.
      \end{equation*}
        \item $\hat{X}_{ct}$ moves in the positive direction with a diminishing velocity.

        In fact $C(\Theta_a)>0$ for $\Theta_a>\theta_B$, and $C(\Theta_a)\rightarrow0$ as $\Theta_a\rightarrow\theta_B$ from the right. This can be shown in a similar way to that in the classical Laplace's method in asymptotic analysis of integral. By expanding $\varphi(z)$ around its local maxima $z_0$ (which is a local minima of $\cos\varphi(z)$), we have
        \begin{align*}
        &\int_0^1\frac{\mathrm{d}z}{\cos\varphi(z)-\cos\Theta_a}\\
        \geqslant&\int_{z_0}^{z_0+\eta}\frac{\mathrm{d}z}{\cos\theta_B+(z-z_0)^2(\cos\varphi)''(z_0)+O((z-z_0)^3)-\cos\Theta_a}\\
        \sim&\frac{\frac{\pi}{2}}{\sqrt{(\cos\varphi)''(z_0)(\cos\theta_B-\cos\Theta_a)}}
        \rightarrow+\infty\quad\mbox{as}\quad \Theta_a\rightarrow\theta_B,
        \end{align*}
        where $\eta$ is a small positive number. As a result, $C(\Theta_a)^{-1}$ diverges to $+\infty$ as $\Theta_a\rightarrow\theta_B$ from the right.

        \item $\Theta_a$ approaches $\theta_B$ exponentially fast.

        In fact, multiplying $\sin\Theta_a$ on both sides of Eq. \eqref{eq:ergodic-theta}, we have
      \begin{align*}
        &\frac{\mathrm{d}}{\mathrm{d}t}(\cos\theta_B-\cos\Theta_a)\\
        =&g(\Theta_a)\sin\Theta_a\Big[f(\Theta_a)\Big(\int_0^1\frac{\mathrm{d}z}{\cos\varphi(y)-\cos\theta_B+\cos\theta_B-\cos\Theta_a}\Big)^{-1}+v\Big]\\
        \leqslant&-m\beta(f(\theta_B)(\cos\theta_B-\cos\Theta_a)+v),
      \end{align*}
      where $\beta=\min\{\sin\Theta_a(0),\sin\theta_B\}$ and we have used the relation $\Theta_a\geqslant\theta_B=\max\limits_{0\leqslant y\leqslant1}\varphi(y)$. An application of Gronwall's inequality implies that
      \begin{equation}\label{eq:exponential-decay}
        \cos\theta_B-\cos\Theta_a\leqslant(\cos\theta_B-\cos\Theta_a(0))e^{-m\beta f(\theta_B) t}-\frac{v}{f(\theta_B)}(1-e^{-m\beta f(\theta_B) t}).
      \end{equation}
      It can be seen that $\Theta_a$ decreases exponentially fast and arrives at $\theta_B$ at some finite time $t_1^*$. Moreover, it is easily estimated that $t_1^*\leqslant\frac1{m\beta f(\theta_B)}\ln(1+\frac{(\cos\theta_B-\cos\Theta_a(0))f(\theta_B)}{v})$. The time period $[0,t_1^*]$ is a transient period for the dynamic of $\Theta_a$ and $\hat{X}_{ct}$.
      \end{enumerate}

  \item When $\Theta_a\in[\theta_A,\theta_B]$, the effective dynamics follows \eqref{eq:slave} which yields the following properties:
      \begin{enumerate}
      \item The effective apparent contact angle $\Theta_a(t)$ decreases monotonically.

      In fact, it is straightforward to show that $\frac{\mathrm{d}}{\mathrm{d}t}\Theta_a\leqslant-vm$ and $\Theta_a(t)\leqslant -vm(t-t_1^*)+\theta_B$. Therefore, $\Theta_a(t)$ must arrive at $\theta_A$ at some finite time $t_2^{*}$ with $t_2^{*}\leqslant t_1^*+\frac{\theta_B-\theta_A}{vm}$.
      \item The effective contact line position $\hat{X}_{ct}$ remains unchanged.
      \end{enumerate}
  \item When $\Theta_a<\theta_A=\min\limits_{0\leqslant y\leqslant1}\varphi(y)$, the dynamics of the effective contact angle and contact point position are given by \eqref{eq:ergodic-theta} and \eqref{eq:ergodic-x}. This dynamic system has the following properties:
      \begin{enumerate}
      \item $\Theta_a$ eventually arrives at a stable steady state $\Theta^*\in(0,\theta_A)$ satisfying $f(\Theta^*)C(\Theta^*)=-v$.

      In fact, $f(\Theta_a)>0$ and $C(\Theta_a)<0$ for $0<\Theta_a<\theta_A$. Since $C(\Theta_a)\rightarrow0$ as $\Theta_a\rightarrow\theta_A$ from the left (similar to the proof in case (i)(b)), we also have $f(\Theta_a)C(\Theta_a)=0$ when $\Theta_a=0$ or $\theta_A$. Then the equation $f(\Theta_a)C(\Theta_a)+v=0$ has at least two roots in $(0,\theta_A)$ for small $v>0$. As a result, the dynamics \eqref{eq:ergodic-theta} admits at least two steady state. As $\Theta_a$ starts from $\theta_A$ in this regime, we are interested in the steady state closest to $\theta_A$. We denote this state as $\Theta^*$.

      $\Theta^*$ is asymptotically stable or semi-stable if the right side of \eqref{eq:ergodic-theta} has non-positive derivative at $\Theta_a=\Theta^*$. Since $g(\Theta_a)<0$, direct calculation shows that this is equivalent to verify the derivative of $f(\Theta_a)C(\Theta_a)$ is non-negative at $\Theta_a=\Theta^*$ (See Figure \ref{fig:sketch-stable}). This can also be proved by contradiction: suppose the derivative of $f(\Theta_a)C(\Theta_a)$ is negative at $\Theta_a=\Theta^*$, then $f(\Theta_a)C(\Theta_a)$ is decreasing nearby $\Theta_a=\Theta^*$, and we can find $\Theta^{**}\in(\Theta^*,\theta_A)$ such that $f(\Theta^{**})C(\Theta^{**})<-v$; but this implies that there must be another root of $f(\Theta_a)C(\Theta_a)+v=0$ in the interval $(\Theta^{**},\theta_A)$ by intermediate value theorem, which contradicts to the assumption that $\Theta^*$ is the closest root to $\theta_A$. Therefore, $\Theta^*$ is a stable steady state.
      \item Before achieving the steady state $\Theta^*$, $\Theta_a$ continues decreasing due to the non-positiveness of the right side of \eqref{eq:ergodic-theta}. Moreover, the effective contact line position $\hat{X}_{ct}$ moves to the negative direction since $-v<f(\Theta_a)C(\Theta_a)<0$.
      \item When $\Theta_a$ arrives at the steady state $\Theta^*$, $\hat{X}_{ct}$ keeps on moving in the negative direction at a constant velocity $\frac{\mathrm{d}\hat{X}_{ct}}{\mathrm{d}t}=-v$.
      \end{enumerate}
\end{enumerate}

Similar results hold in the case of $v<0$. We can write the stable steady effective contact angle as a function of the drag velocity $v$, i.e., $\Theta^*=\Theta^*(v)$. This function has the following properties:
\begin{enumerate}
  \item The steady effective contact angle $\Theta^*(v)$ must be outside the range of the chemical pattern $[\theta_A,\theta_B]$: if $v>0$, $\Theta^*(v)<\theta_A$ and is not far from $\theta_A$, this is called receding contact angle; if $v<0$, $\Theta^*(v)>\theta_B$ and is not far from $\theta_B$, this is called advancing contact angle;
  \item $\Theta^*(v)\rightarrow\theta_A$ as $v\rightarrow0^+$, while $\Theta^*(v)\rightarrow\theta_B$ as $v\rightarrow0^-$. In other words, depending on different directions of the quasi-static motion, the steady effective contact angle approaches the lower bound $\theta_A$ or the upper bound $\theta_B$ of the chemical pattern.

      This is a consequence of implicit function theorem and the smoothness of the function $f(\Theta_a)C(\Theta_a)$ (See also Figure \ref{fig:anglePattern} for a sketch).
  \item In the extreme case $v=0$, $\Theta^*(0)$ can be any value on the interval $[\theta_A,\theta_B]$.
  This can be seen from the averaged dynamics \eqref{eq:slave-theta} and \eqref{eq:ergodic-theta} in three different cases: if the initial contact angle is larger than $\theta_B$, then \eqref{eq:exponential-decay} with $v=0$ shows that the apparent contact angle decays exponentially to $\theta_B$; if the initial contact angle is smaller than $\theta_A$, similar results hold that the apparent contact angle increases exponentially to $\theta_A$; if the initial contact angle is between $\theta_A$ and $\theta_B$, then \eqref{eq:slave-theta} with $v=0$ implies that it is already equilibrium.
\end{enumerate}

From these discussions, we can define the equilibrium apparent contact angle to be any value in the range $[\theta_A, \theta_B]$ when there is chemical roughness on the substrate. The contact line pins when $\Theta_a\in[\theta_A, \theta_B]$; the contact line advances if $\Theta_a>\theta_B$, while the contact line recedes if $\Theta_a<\theta_A$. 

\bibliographystyle{jfm}
\bibliography{literature}

\begin{thebibliography}{81}
\expandafter\ifx\csname natexlab\endcsname\relax\def\natexlab#1{#1}\fi
\def\au#1{#1} \def\ed#1{#1} \def\yr#1{#1}\def\at#1{#1}\def\jt#1{\textit{#1}}
  \def\bt#1{#1}\def\bvol#1{\textbf{#1}} \def\vol#1{#1} \def\pg#1{#1}
  \def\publ#1{#1}\def\arxiv#1{#1}\def\org#1{#1}\def\st#1{\textit{#1}}

\bibitem[Blake(2006)]{blake2006physics}
{\sc \au{Blake, Terence~D}} \yr{2006}  \at{The physics of moving wetting
  lines}.  \jt{Journal of Colloid and Interface Science}  \bvol{299}~(1),
  \pg{1--13}.

\bibitem[Blake \& De~Coninck(2011)]{blake2011dynamics}
{\sc \au{Blake, T.~D.} \& \au{De~Coninck, Jo{\"e}l}} \yr{2011}  \at{Dynamics of
  wetting and kramers' theory}.  \jt{The European Physical Journal Special
  Topics}  \bvol{197}~(1),  \pg{249--264}.

\bibitem[Bonn {\em et~al.\/}(2009)Bonn, Eggers, Indekeu, Meunier \&
  Rolley]{bonn2009wetting}
{\sc \au{Bonn, Daniel}, \au{Eggers, Jens}, \au{Indekeu, Joseph}, \au{Meunier,
  Jacques} \& \au{Rolley, Etienne}} \yr{2009}  \at{Wetting and spreading}.
  \jt{Reviews of Modern Physics}  \bvol{81}~(2),  \pg{739}.

\bibitem[Cassie \& Baxter(1944)]{Cassie44}
{\sc \au{Cassie, A.} \& \au{Baxter, S.}} \yr{1944}  \at{Wettability of porous
  surfaces}.  \jt{Trans. Faraday Soc.}  \bvol{40},  \pg{546--551}.

\bibitem[Chan {\em et~al.\/}(2020)Chan, Yang \& Carlson]{chan2020directional}
{\sc \au{Chan, T.~S.}, \au{Yang, F.} \& \au{Carlson, A.}} \yr{2020}
  \at{Directional spreading of a viscous droplet on a conical fibre}.
  \jt{Journal of Fluid Mechanics}  \bvol{894}.

\bibitem[Choi {\em et~al.\/}(2009)Choi, Tuteja, Mabry, Cohen \&
  McKinley]{Choi09}
{\sc \au{Choi, W.}, \au{Tuteja, A.}, \au{Mabry, J.~M.}, \au{Cohen, R.~E.} \&
  \au{McKinley, G.~H.}} \yr{2009}  \at{A modified cassie-baxter relationship to
  explain contact angle hysteresis and anisotropy on non-wetting textured
  surfaces}.  \jt{J. Colloid Interface Sci.}  \bvol{339},  \pg{208--216}.

\bibitem[Cox(1983)]{cox1983spreading}
{\sc \au{Cox, RG}} \yr{1983}  \at{The spreading of a liquid on a rough solid
  surface}.  \jt{Journal of Fluid Mechanics}  \bvol{131},  \pg{1--26}.

\bibitem[Cox(1986)]{cox1986}
{\sc \au{Cox, R.}} \yr{1986}  \at{The dynamics of the spreading of liquids on a
  solid surface. {Part 1. Viscous flow}}.  \jt{J. Fluid Mech.}  \bvol{168},
  \pg{169--194}.

\bibitem[De~Gennes(1985)]{de1985wetting}
{\sc \au{De~Gennes, Pierre-Gilles}} \yr{1985}  \at{Wetting: statics and
  dynamics}.  \jt{Reviews of Modern Physics}  \bvol{57}~(3),  \pg{827}.

\bibitem[Di {\em et~al.\/}(2016)Di, Xu \& Doi]{DiXuDoi2016}
{\sc \au{Di, Y.}, \au{Xu, X.} \& \au{Doi, M.}} \yr{2016}  \at{Theoretical
  analysis for meniscus rise of a liquid contained between a flexible film and
  a solid wall}.  \jt{Europhys. Lett.}  \bvol{113}~(3),  \pg{36001}.

\bibitem[Doi(2013)]{DoiSoftMatter}
{\sc \au{Doi, M.}} \yr{2013} {\em Soft Matter Physics\/}.  \publ{Oxfort
  University Press}.

\bibitem[Doi(2015)]{Doi15}
{\sc \au{Doi, M.}} \yr{2015}  \at{Onsager priciple as a tool for
  approximation}.  \jt{Chinese Phys. B}  \bvol{24},  \pg{020505}.

\bibitem[Doi {\em et~al.\/}(2019)Doi, Zhou, Di \& Xu]{doi2019application}
{\sc \au{Doi, M.}, \au{Zhou, J.}, \au{Di, Y.} \& \au{Xu, X.}} \yr{2019}
  \at{Application of the onsager-machlup integral in solving dynamic equations
  in nonequilibrium systems}.  \jt{Phys. Rev. E}  \bvol{99}~(6),  \pg{063303}.

\bibitem[Dupont \& Legendre(2010)]{dupont2010numerical}
{\sc \au{Dupont, Jean-Baptiste} \& \au{Legendre, Dominique}} \yr{2010}
  \at{Numerical simulation of static and sliding drop with contact angle
  hysteresis}.  \jt{Journal of Computational Physics}  \bvol{229}~(7),
  \pg{2453--2478}.

\bibitem[Dussan(1979)]{dussan1979spreading}
{\sc \au{Dussan, EB}} \yr{1979}  \at{On the spreading of liquids on solid
  surfaces: static and dynamic contact lines}.  \jt{Annual Review of Fluid
  Mechanics}  \bvol{11}~(1),  \pg{371--400}.

\bibitem[E(2011)]{WeinanBook11}
{\sc \au{E, Weinan}} \yr{2011} {\em Principle of Multiscale Modeling\/}.
  \publ{Cambridge University Press}.

\bibitem[Eggers(2005)]{eggers2005contact}
{\sc \au{Eggers, Jens}} \yr{2005}  \at{Contact line motion for partially
  wetting fluids}.  \jt{Physical Review E}  \bvol{72}~(6),  \pg{061605}.

\bibitem[Extrand(2002)]{Extrand02}
{\sc \au{Extrand, C.~W.}} \yr{2002}  \at{Model for contact angles and
  hysteresis on rough and ultraphobic surfaces}.  \jt{Langmuir}  \bvol{18},
  \pg{7991--7999}.

\bibitem[Gao \& McCarthy(2007)]{Gao07}
{\sc \au{Gao, L.} \& \au{McCarthy, T.~J.}} \yr{2007}  \at{How wenzel and cassie
  were wrong}.  \jt{Langmuir}  \bvol{23},  \pg{3762--3765}.

\bibitem[Gao \& Wang(2012)]{gao2012gradient}
{\sc \au{Gao, Min} \& \au{Wang, Xiao-Ping}} \yr{2012}  \at{A gradient stable
  scheme for a phase field model for the moving contact line problem}.
  \jt{Journal of Computational Physics}  \bvol{231}~(4),  \pg{1372--1386}.

\bibitem[de~Gennes {\em et~al.\/}(2003)de~Gennes, Brochard-Wyart \&
  Quere]{Gennes03}
{\sc \au{de~Gennes, P.G.}, \au{Brochard-Wyart, F.} \& \au{Quere, D.}} \yr{2003}
  {\em Capillarity and Wetting Phenomena\/}.  \publ{Springer Berlin}.

\bibitem[Golestanian(2004)]{golestanian2004moving}
{\sc \au{Golestanian, R.}} \yr{2004}  \at{Moving contact lines on heterogeneous
  substrates}.  \jt{Phil. Trans. Roy. So. London. A}  \bvol{362}~(1821),
  \pg{1613--1623}.

\bibitem[Golestanian \& Rapha{\"e}l(2003)]{golestanian2003roughening}
{\sc \au{Golestanian, R.} \& \au{Rapha{\"e}l, E.}} \yr{2003}  \at{Roughening
  transition in a moving contact line}.  \jt{Phys. Rev. E}  \bvol{67}~(3),
  \pg{031603}.

\bibitem[Guan {\em et~al.\/}(2016{\natexlab{{\em a\/}}})Guan, Wang, Charlaix \&
  Tong]{guan2016asymmetric}
{\sc \au{Guan, D.}, \au{Wang, Y.}, \au{Charlaix, E.} \& \au{Tong, P.}}
  \yr{2016{\natexlab{{\em a\/}}}}  \at{Asymmetric and speed-dependent capillary
  force hysteresis and relaxation of a suddenly stopped moving contact line}.
  \jt{Phys. Rev. Lett.}  \bvol{116}~(6),  \pg{066102}.

\bibitem[Guan {\em et~al.\/}(2016{\natexlab{{\em b\/}}})Guan, Wang, Charlaix \&
  Tong]{guan2016simultaneous}
{\sc \au{Guan, D.}, \au{Wang, Y.}, \au{Charlaix, E.} \& \au{Tong, P.}}
  \yr{2016{\natexlab{{\em b\/}}}}  \at{Simultaneous observation of asymmetric
  speed-dependent capillary force hysteresis and slow relaxation of a suddenly
  stopped moving contact line}.  \jt{Phys. Rev. E}  \bvol{94}~(4),
  \pg{042802}.

\bibitem[Guo {\em et~al.\/}(2013)Guo, Gao, Xiong, Wang, Wang, Sheng \&
  Tong]{guo2013direct}
{\sc \au{Guo, Shuo}, \au{Gao, Min}, \au{Xiong, Xiaomin}, \au{Wang, Yong~Jian},
  \au{Wang, Xiaoping}, \au{Sheng, Ping} \& \au{Tong, Penger}} \yr{2013}
  \at{Direct measurement of friction of a fluctuating contact line}.
  \jt{Physical Review Letters}  \bvol{111}~(2),  \pg{026101}.

\bibitem[Guo {\em et~al.\/}(2019)Guo, Xu, Qian, Di, Doi \& Tong]{guo2019onset}
{\sc \au{Guo, S.}, \au{Xu, X.}, \au{Qian, T.}, \au{Di, Y.}, \au{Doi, M.} \&
  \au{Tong, P.}} \yr{2019}  \at{Onset of thin film meniscus along a fibre}.
  \jt{J. Fluid Mech.}  \bvol{865},  \pg{650--680}.

\bibitem[Gurtin {\em et~al.\/}(1996)Gurtin, Polignone \& Vinals]{gurtin1996two}
{\sc \au{Gurtin, Morton~E}, \au{Polignone, Debra} \& \au{Vinals, Jorge}}
  \yr{1996}  \at{Two-phase binary fluids and immiscible fluids described by an
  order parameter}.  \jt{Mathematical Models and Methods in Applied Sciences}
  \bvol{6}~(06),  \pg{815--831}.

\bibitem[Haley \& Miksis(1991)]{haley1991effect}
{\sc \au{Haley, Patrick~J} \& \au{Miksis, Michael~J}} \yr{1991}  \at{The effect
  of the contact line on droplet spreading}.  \jt{Journal of Fluid Mechanics}
  \bvol{223},  \pg{57--81}.

\bibitem[Huh \& Scriven(1971)]{huh1971hydrodynamic}
{\sc \au{Huh, C.} \& \au{Scriven, L.E.}} \yr{1971}  \at{Hydrodynamic model of
  steady movement of a solid/liquid/fluid contact line}.  \jt{J. colloid
  Interface Sci.}  \bvol{35}~(1),  \pg{85--101}.

\bibitem[Iliev {\em et~al.\/}(2018)Iliev, Pesheva \& Iliev]{iliev2018contact}
{\sc \au{Iliev, Stanimir}, \au{Pesheva, Nina} \& \au{Iliev, Pavel}} \yr{2018}
  \at{Contact angle hysteresis on doubly periodic smooth rough surfaces in
  wenzel's regime: The role of the contact line depinning mechanism}.
  \jt{Physical Review E}  \bvol{97}~(4),  \pg{042801}.

\bibitem[Jacqmin(2000)]{Jacqmin2000}
{\sc \au{Jacqmin, D.}} \yr{2000}  \at{Contact-line dynamics of a diffuse fluid
  interface}.  \jt{J. Fluid Mech.}  \bvol{402},  \pg{57--88}.

\bibitem[Joanny \& Robbins(1990)]{Joanny1990}
{\sc \au{Joanny, J.} \& \au{Robbins, M.}} \yr{1990}  \at{Motion of a contact
  line on a heterogeneous surface}.  \jt{J. Chem. Phys}  \bvol{92}~(2),
  \pg{32063212}.

\bibitem[Joanny \& De~Gennes(1984)]{joanny1984model}
{\sc \au{Joanny, J.~F.} \& \au{De~Gennes, P.-G.}} \yr{1984}  \at{A model for
  contact angle hysteresis}.  \jt{J. Chem. Phys.}  \bvol{81}~(1),
  \pg{552--562}.

\bibitem[Johnson~Jr. \& Dettre(1964)]{Johnson1964}
{\sc \au{Johnson~Jr., R.~E.} \& \au{Dettre, R.~H.}} \yr{1964}  \at{Contact
  angle hysteresis. iii. study of an idealized heterogeneous surfaces}.  \jt{J.
  Phys. Chem.}  \bvol{68},  \pg{1744--1750}.

\bibitem[Man \& Doi(2016)]{ManDoi2016}
{\sc \au{Man, X.} \& \au{Doi, M.}} \yr{2016}  \at{Ring to mountain transition
  in deposition pattern of drying droplets}.  \jt{Phys. Rev. Lett.}
  \bvol{116}~(6),  \pg{066101}.

\bibitem[Marmottant \& Villermaux(2004)]{marmottant2004spray}
{\sc \au{Marmottant, Philippe} \& \au{Villermaux, Emmanuel}} \yr{2004}  \at{On
  spray formation}.  \jt{Journal of Fluid Mechanics}  \bvol{498},
  \pg{73--111}.

\bibitem[Marmur \& Bittoun(2009)]{Marmur09}
{\sc \au{Marmur, A.} \& \au{Bittoun, E.}} \yr{2009}  \at{When wenzel and cassie
  are right: Reconciling local and global considerations}.  \jt{Langmuir}
  \bvol{25},  \pg{1277--1281}.

\bibitem[Miskis \& Davis(1994)]{miskis1994slip}
{\sc \au{Miskis, M.~J.} \& \au{Davis, S.~H.}} \yr{1994}  \at{Slip over rough
  and coated surfaces}.  \jt{Journal of Fluid Mechanics}  \bvol{273},
  \pg{125--139}.

\bibitem[Neumann \& Good(1972)]{Neumann1971}
{\sc \au{Neumann, A.~W.} \& \au{Good, R.~J.}} \yr{1972}  \at{Thermodynamics of
  contact angles: I. heterogeneous solid surfaces}.  \jt{J. Colloid Interface
  Sci.}  \bvol{38},  \pg{341--358}.

\bibitem[Pavliotis \& Stuart(2008)]{pav2008multiscale}
{\sc \au{Pavliotis, G.~A.} \& \au{Stuart, A.~M.}} \yr{2008} {\em Multiscale
  Methods: Averaging and Homogenization\/}.  \publ{New York: Springer}.

\bibitem[Pismen(2002)]{pismen2002mesoscopic}
{\sc \au{Pismen, LM}} \yr{2002}  \at{Mesoscopic hydrodynamics of contact line
  motion}.  \jt{Colloids and Surfaces A: Physicochemical and Engineering
  Aspects}  \bvol{206}~(1),  \pg{11--30}.

\bibitem[Pismen \& Pomeau(2000)]{pismen2000disjoining}
{\sc \au{Pismen, Len~M} \& \au{Pomeau, Yves}} \yr{2000}  \at{Disjoining
  potential and spreading of thin liquid layers in the diffuse-interface model
  coupled to hydrodynamics}.  \jt{Physical Review E}  \bvol{62}~(2),
  \pg{2480}.

\bibitem[Prabhala {\em et~al.\/}(2013)Prabhala, Panchagnula \&
  Vedantam]{Prabhala13}
{\sc \au{Prabhala, B.}, \au{Panchagnula, M.~V.} \& \au{Vedantam, S.}} \yr{2013}
   \at{Three-dimensional equilibrium shapes of drops on hysteretic surfaces}.
  \jt{Colloid Polym. Sci.}  \bvol{291},  \pg{279--289}.

\bibitem[Priest {\em et~al.\/}(2007)Priest, Sedev \&
  Ralston]{priest2007asymmetric}
{\sc \au{Priest, C.}, \au{Sedev, R.} \& \au{Ralston, J.}} \yr{2007}
  \at{Asymmetric wetting hysteresis on chemical defects}.  \jt{Phys. Rev.
  Lett.}  \bvol{99}~(2),  \pg{026103}.

\bibitem[Qian {\em et~al.\/}(2003)Qian, Wang \& Sheng]{QianWangSheng2003}
{\sc \au{Qian, T.}, \au{Wang, X.P.} \& \au{Sheng, P.}} \yr{2003}  \at{Molecular
  scale contact line hydrodynamics of immiscible flows}.  \jt{Phys. Rev. E}
  \bvol{68},  \pg{016306}.

\bibitem[Qian {\em et~al.\/}(2006)Qian, Wang \& Sheng]{QianWangSheng2006}
{\sc \au{Qian, T.}, \au{Wang, X.P.} \& \au{Sheng, P.}} \yr{2006}  \at{A
  variational approach to moving contact line hydrodynamics}.  \jt{J. Fluid
  Mech.}  \bvol{564},  \pg{333--360}.

\bibitem[Raj {\em et~al.\/}(2012)Raj, Enright, Zhu, Adera \& Wang]{Raj12}
{\sc \au{Raj, R.}, \au{Enright, R.}, \au{Zhu, Y.}, \au{Adera, S.} \& \au{Wang,
  E.~N.}} \yr{2012}  \at{Unified model for contact angle hysteresis on
  heterogeneous and superhydrophobic surfaces}.  \jt{Langmuir}  \bvol{28},
  \pg{15777--15788}.

\bibitem[Raphael \& de~Gennes(1989)]{raphael1989dynamics}
{\sc \au{Raphael, Elie} \& \au{de~Gennes, Pierre-Gilles}} \yr{1989}
  \at{Dynamics of wetting with nonideal surfaces. the single defect problem}.
  \jt{The Journal of chemical physics}  \bvol{90}~(12),  \pg{7577--7584}.

\bibitem[Ren \& E(2007)]{ren2007boundary}
{\sc \au{Ren, W.} \& \au{E, W.}} \yr{2007}  \at{Boundary conditions for the
  moving contact line problem}.  \jt{Phys. Fluids}  \bvol{19}~(2),
  \pg{022101}.

\bibitem[Ren \& E(2011)]{RenE2011}
{\sc \au{Ren, W.} \& \au{E, W.}} \yr{2011}  \at{Contact line dynamics on
  heterogeneous surfaces}.  \jt{Phys. Fluids}  \bvol{23},  \pg{072103}.

\bibitem[Ren {\em et~al.\/}(2015)Ren, Trinh \& E]{ren2015distinguished}
{\sc \au{Ren, Weiqing}, \au{Trinh, P.~H.} \& \au{E, Weinan}} \yr{2015}  \at{On
  the distinguished limits of the navier slip model of the moving contact line
  problem}.  \jt{Journal of Fluid Mechanics}  \bvol{772},  \pg{107--126}.

\bibitem[Renardy {\em et~al.\/}(2001)Renardy, Renardy \&
  Li]{renardy2001numerical}
{\sc \au{Renardy, Michael}, \au{Renardy, Yuriko} \& \au{Li, Jie}} \yr{2001}
  \at{Numerical simulation of moving contact line problems using a
  volume-of-fluid method}.  \jt{Journal of Computational Physics}
  \bvol{171}~(1),  \pg{243--263}.

\bibitem[Schwartz \& Eley(1998)]{schwartz1998simulation}
{\sc \au{Schwartz, Leonard~W} \& \au{Eley, Richard~R}} \yr{1998}
  \at{Simulation of droplet motion on low-energy and heterogeneous surfaces}.
  \jt{Journal of Colloid and Interface Science}  \bvol{202}~(1),
  \pg{173--188}.

\bibitem[Schwartz \& Garoff(1985)]{Schwarts1985}
{\sc \au{Schwartz, L.~W.} \& \au{Garoff, S.}} \yr{1985}  \at{Contact angle
  hysteresis on heterogeneous surfaces}.  \jt{Langmuir}  \bvol{1},
  \pg{219--230}.

\bibitem[Seppecher(1996)]{seppecher1996moving}
{\sc \au{Seppecher, Pierre}} \yr{1996}  \at{Moving contact lines in the
  {Cahn-Hilliard} theory}.  \jt{International Journal of Engineering Science}
  \bvol{34}~(9),  \pg{977--992}.

\bibitem[Seveno {\em et~al.\/}(2009)Seveno, Vaillant, Rioboo, Adao, Conti \&
  De~Coninck]{seveno2009dynamics}
{\sc \au{Seveno, David}, \au{Vaillant, Alexandre}, \au{Rioboo, Romain},
  \au{Adao, H}, \au{Conti, J} \& \au{De~Coninck, Jo{\"e}l}} \yr{2009}
  \at{Dynamics of wetting revisited}.  \jt{Langmuir}  \bvol{25}~(22),
  \pg{13034--13044}.

\bibitem[Shikhmurzaev(1993)]{shikhmurzaev1993moving}
{\sc \au{Shikhmurzaev, Y.~D.}} \yr{1993}  \at{The moving contact line on a
  smooth solid surface}.  \jt{Inter. J. Multiphase Flow}  \bvol{19}~(4),
  \pg{589--610}.

\bibitem[Sibley {\em et~al.\/}(2015)Sibley, Nold \& Kalliadasis]{Sibley15}
{\sc \au{Sibley, D.~N.}, \au{Nold, A.} \& \au{Kalliadasis, S.}} \yr{2015}
  \at{The asymptotics of the moving contact line:cracking an old nut}.
  \jt{J.~Fluid Mech.}  \bvol{764},  \pg{445--462}.

\bibitem[Snoeijer \& Andreotti(2013)]{snoeijer2013moving}
{\sc \au{Snoeijer, J.~H.} \& \au{Andreotti, B.}} \yr{2013}  \at{Moving contact
  lines: scales, regimes, and dynamical transitions}.  \jt{Annual Rev. Fluid
  Mech.}  \bvol{45},  \pg{269--292}.

\bibitem[Spelt(2005)]{spelt2005level}
{\sc \au{Spelt, Peter~DM}} \yr{2005}  \at{A level-set approach for simulations
  of flows with multiple moving contact lines with hysteresis}.  \jt{Journal of
  Computational Physics}  \bvol{207}~(2),  \pg{389--404}.

\bibitem[Sui {\em et~al.\/}(2014)Sui, Ding \& Spelt]{sui2014numerical}
{\sc \au{Sui, Y.}, \au{Ding, H.} \& \au{Spelt, P.}} \yr{2014}  \at{Numerical
  simulations of flows with moving contact lines}.  \jt{Annual Rev. Fluid
  Mech.}  \bvol{46},  \pg{97--119}.

\bibitem[Sui \& Spelt(2013{\natexlab{{\em a\/}}})]{sui2013efficient}
{\sc \au{Sui, Yi} \& \au{Spelt, Peter~DM}} \yr{2013{\natexlab{{\em a\/}}}}
  \at{An efficient computational model for macroscale simulations of moving
  contact lines}.  \jt{Journal of Computational Physics}  \bvol{242},
  \pg{37--52}.

\bibitem[Sui \& Spelt(2013{\natexlab{{\em b\/}}})]{sui2013validation}
{\sc \au{Sui, Yi} \& \au{Spelt, Peter~DM}} \yr{2013{\natexlab{{\em b\/}}}}
  \at{Validation and modification of asymptotic analysis of slow and rapid
  droplet spreading by numerical simulation}.  \jt{Journal of Fluid Mechanics}
  \bvol{715},  \pg{283--313}.

\bibitem[Tanner(1979)]{tanner1979spreading}
{\sc \au{Tanner, L.H.}} \yr{1979}  \at{The spreading of silicone oil drops on
  horizontal surfaces}.  \jt{Journal of Physics D: Applied Physics}
  \bvol{12}~(9),  \pg{1473}.

\bibitem[Wang {\em et~al.\/}(2008)Wang, Qian \& Sheng]{QianWangSheng2008}
{\sc \au{Wang, X.P.}, \au{Qian, T.} \& \au{Sheng, P.}} \yr{2008}  \at{Moving
  contact line on chemically patterned surfaces}.  \jt{J. Fluid Mech.}
  \bvol{605},  \pg{59--78}.

\bibitem[Wenzel(1936)]{Wenzel36}
{\sc \au{Wenzel, R.~N.}} \yr{1936}  \at{Resistance of solid surfaces to wetting
  by water}.  \jt{Ind. Eng. Chem.}  \bvol{28},  \pg{988--994}.

\bibitem[Whyman {\em et~al.\/}(2008)Whyman, Bormashenko \& Stein]{Whyman08}
{\sc \au{Whyman, G.}, \au{Bormashenko, E.} \& \au{Stein, T.}} \yr{2008}
  \at{The rigorous derivative of young, cassie-baxter and wenzel equations and
  the analysis of the contact angle hysteresis phenomenon}.  \jt{Chem. Phy.
  Letters}  \bvol{450},  \pg{355--359}.

\bibitem[Xu(2016)]{xu2016modified}
{\sc \au{Xu, X.}} \yr{2016}  \at{Modified wenzel and cassie equations for
  wetting on rough surfaces}.  \jt{SIAM J. Appl. Math.}  \bvol{76}~(6),
  \pg{2353--2374}.

\bibitem[Xu {\em et~al.\/}(2016)Xu, Di \& Doi]{xu2016variational}
{\sc \au{Xu, X.}, \au{Di, Y.} \& \au{Doi, M.}} \yr{2016}  \at{Variational
  method for contact line problems in sliding liquids}.  \jt{Phys. Fluids}
  \bvol{28},  \pg{087101}.

\bibitem[Xu \& Wang(2020)]{xu2020theoretical}
{\sc \au{Xu, Xianmin} \& \au{Wang, Xiaoping}} \yr{2020}  \at{Theoretical
  analysis for dynamic contact angle hysteresis on chemically patterned
  surfaces}.  \jt{Physics of Fluids}  \bvol{32}~(11),  \pg{112102}.

\bibitem[Xu \& Wang(2011)]{XuWang2011}
{\sc \au{Xu, X.} \& \au{Wang, X.~P.}} \yr{2011}  \at{Analysis of wetting and
  contact angle hysteresis on chemically patterned surfaces}.  \jt{SIAM J.
  Appl. Math.}  \bvol{71},  \pg{1753--1779}.

\bibitem[Xu \& Wang(2013)]{XuWang2013}
{\sc \au{Xu, X.} \& \au{Wang, X.~P.}} \yr{2013}  \at{The modified cassie's
  equation and contact angle hysteresis}.  \jt{Colloid Polym. Sci.}
  \bvol{291},  \pg{299--306}.

\bibitem[Xu {\em et~al.\/}(2019)Xu, Zhao \& Wang]{XuZhaoWang2019}
{\sc \au{Xu, X.}, \au{Zhao, Y.} \& \au{Wang, X.-P.}} \yr{2019}  \at{Analysis
  for contact angle hysteresis on rough surfaces by a phase field model with a
  relaxed boundary condition}.  \jt{SIAM J. Appl. Math.}  \bvol{79},
  \pg{2551--2568}.

\bibitem[Young(1805)]{Young1805}
{\sc \au{Young, T.}} \yr{1805}  \at{An essay on the cohesion of fluids}.
  \jt{Philos. Trans. R. Soc. London}  \bvol{95},  \pg{65--87}.

\bibitem[Yue(2020)]{yue2020thermodynamically}
{\sc \au{Yue, Pengtao}} \yr{2020}  \at{Thermodynamically consistent phase-field
  modelling of contact angle hysteresis}.  \jt{Journal of Fluid Mechanics}
  \bvol{899}.

\bibitem[Yue \& Feng(2011)]{yue2011can}
{\sc \au{Yue, P.} \& \au{Feng, J.~J.}} \yr{2011}  \at{Can diffuse-interface
  models quantitatively describe moving contact lines?}  \jt{The European
  Physical Journal Special Topics}  \bvol{197}~(1),  \pg{37--46}.

\bibitem[Zhang \& Yue(2020)]{zhang2020level}
{\sc \au{Zhang, Jiaqi} \& \au{Yue, Pengtao}} \yr{2020}  \at{A level-set method
  for moving contact lines with contact angle hysteresis}.  \jt{Journal of
  Computational Physics}  \pg{p. 109636}.

\bibitem[Zhang \& Ren(2019)]{zhang2019distinguished}
{\sc \au{Zhang, Zhen} \& \au{Ren, Weiqing}} \yr{2019}  \at{Distinguished limits
  of the navier slip model for moving contact lines in stokes flow}.  \jt{SIAM
  Journal on Applied Mathematics}  \bvol{79},  \pg{1654--1674}.

\bibitem[Zhou \& Doi(2018)]{zhou2018dynamics}
{\sc \au{Zhou, J.} \& \au{Doi, M.}} \yr{2018}  \at{Dynamics of viscoelastic
  filaments based on onsager principle}.  \jt{Phys. Rev. Fluids}  \bvol{3}~(8),
   \pg{084004}.

\bibitem[Zhou \& Sheng(1990)]{zhou1990dynamics}
{\sc \au{Zhou, Min-Yao} \& \au{Sheng, Ping}} \yr{1990}  \at{Dynamics of
  immiscible-fluid displacement in a capillary tube}.  \jt{Physical Review
  Letters}  \bvol{64}~(8),  \pg{882}.

\end{thebibliography}

\end{document}